\documentclass[%
 reprint,
 superscriptaddress,
 amsmath,amssymb,
 aps,
 prx,
]{revtex4-2}

\usepackage{url}
\usepackage{mathtools}

\usepackage[T1]{fontenc}
\usepackage[utf8]{inputenc}
\usepackage{graphicx}
\usepackage{dcolumn}
\usepackage{bm}

\usepackage{physics}
\usepackage{setspace}
\usepackage{algpseudocode}
\usepackage{amsmath}
\usepackage{amssymb}
\usepackage{array}
\usepackage{dashrule}

\usepackage{physics}
\usepackage{stackengine}

\usepackage{tcolorbox}
\tcbuselibrary{listings, breakable}

\usepackage{amsmath,accents}

\newcommand{\dashbar}[1]{\widetilde{#1}}

\newcounter{algorithm} 
\newcommand{\algcaption}[1]{%
  \refstepcounter{algorithm}%
  \noindent\rule{\textwidth}{0.8pt}
  \textbf{Algorithm~\thealgorithm. #1}
  \noindent\rule{\textwidth}{0.8pt}
}

\usepackage{hyperref}
\begin{document}

\title{%
  Programmable Quantum Matter:\texorpdfstring{\\}{ }
  Heralding Large Cluster States in Driven Inhomogeneous Spin Ensembles
}

\author{Pratyush Anand}
\thanks{These authors contributed equally to this work.}
\affiliation{Department of Electrical Engineering and Computer Science, Massachusetts Institute of Technology, Cambridge, MA 02139, USA}
\affiliation{Research Laboratory of Electronics, Massachusetts Institute of Technology, Cambridge, MA 02139, USA}

\author{Louis Follet}
\thanks{These authors contributed equally to this work.}
\affiliation{Department of Electrical Engineering and Computer Science, Massachusetts Institute of Technology, Cambridge, MA 02139, USA}
\affiliation{Research Laboratory of Electronics, Massachusetts Institute of Technology, Cambridge, MA 02139, USA}

\author{Odiel Hooybergs}
\thanks{These authors contributed equally to this work.}
\affiliation{Department of Electrical Engineering and Computer Science, Massachusetts Institute of Technology, Cambridge, MA 02139, USA}
\affiliation{Research Laboratory of Electronics, Massachusetts Institute of Technology, Cambridge, MA 02139, USA}
\affiliation{Department of Physics, ETH Zürich, 8093 Zürich, Switzerland}
\affiliation{Quantum Center, ETH Zürich, 8093 Zürich, Switzerland}

\author{Dirk R. Englund}%
\affiliation{Department of Electrical Engineering and Computer Science, Massachusetts Institute of Technology, Cambridge, MA 02139, USA}
\affiliation{Research Laboratory of Electronics, Massachusetts Institute of Technology, Cambridge, MA 02139, USA}


\begin{abstract}
Atom-like emitters in solids have emerged as promising platforms for quantum sensing and information processing. Among the major challenges are inhomogeneities in emitter fine structure, which complicates quantum control. Here, we introduce a framework that leverages this emitter diversity to simplify the experimental resources needed to create optically heralded spin cluster states across $N_{q}$ emitters from the conventional order $O(N_{q})$ to $O(1)$ within ensembles of $N_{q}\sim 10-100$. Specifically, the optimized pulse sequence simultaneously corrects parameter variations (pulse-length error and frequency detuning error), achieving single-qubit gate fidelities exceeding 99.99\% for errors (normalized relative to Rabi drive strength) up to 0.3, while maintaining fidelities above 99\% even for errors as large as 0.4. Applying this optimized pulse sequence in the form of a Carr-Purcell-Meiboom-Gill (CPMG) based dynamical decoupling protocol to the dominant noise spectral density of silicon-vacancy centers in diamond, our approach enhances ensemble-average coherence times by more than a factor of $7$ relative to interleaved bang-bang based CPMG. For state-of-the-art dilution refrigeration systems, we further estimate sharply reduced heating when driving a global resonant optimal dynamical decoupling across $N_{q}$ silicon-vacancy spins, potentially resolving the current trade-off between spin coherence and scaling to $N_{q} \gg 1$. We further introduce a modified single-photon entanglement protocol with an efficient algorithm for deterministic entanglement compilation. Depending on the decoupling window, our method yields order $O(10^{2}\text{--}10^{4})$ more entanglement links than bang-bang sequences, with theoretical guarantees of order $\Omega(N_q)$ unique links—improvable via control-parameter tuning. Our approach thus offers enhanced fidelity, scalability, and robustness. Together, these techniques provide foundational tools—including global unitary control, phase denoising, remote entanglement, and compilation—for scalable quantum computing architectures based on heterogeneous spin ensembles.
\end{abstract}

\maketitle 

\section{Introduction}


\begin{figure*}[htb]
    \centering
    \includegraphics[width=1\textwidth]{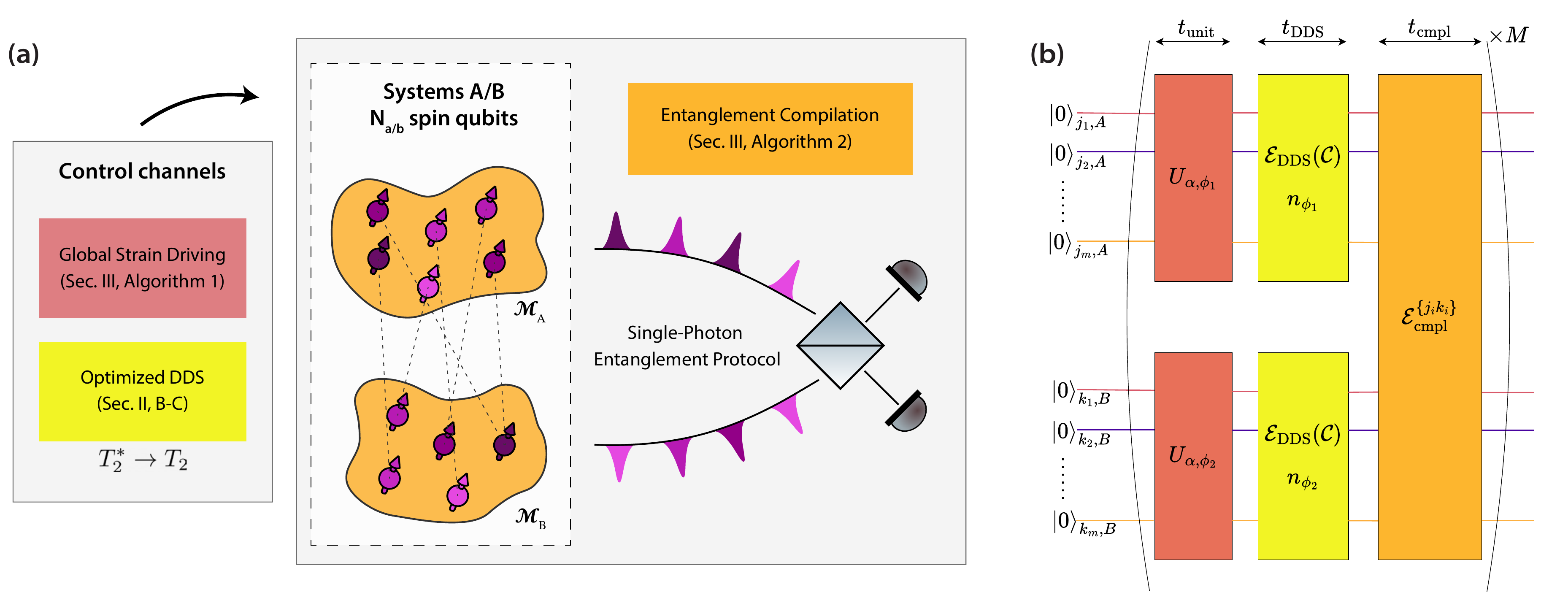}
    \caption{\textbf{Architecture and protocol for mechanical bipartite graph cluster-state generation.}
    \textbf{(a) Architecture overview.} Devices \(\mathcal{M}_{A/B}\) (Systems A/B, \(N_{a,b}\) spin qubits) are arbitrary mechanical structures that house heterogeneous color-center ensembles with non-uniform spin frequencies and spin–strain susceptibilities (represented by arbitrary color shades), operating in noisy environments.
    Each ensemble is controlled by a single global strain drive that provides programmable qubit addressing (Sec.~\ref{sec_ApplicationQuantumComputation}, Algorithm~\ref{alg:1}) and is operated with optimized dynamical decoupling sequences (DDS; Sec.~\ref{sec_theoretical_framework} B–C), yielding robust quantum operations and extended spin coherence. 
    A single-photon entanglement protocol generates heralded Bell pairs between the two ensembles; these pairs are compiled (Sec.~\ref{sec_ApplicationQuantumComputation}, Algorithm~\ref{alg:2}) into a two-dimensional bipartite-graph cluster state. For clarity, only a representative subset of entanglement links is shown with dashed lines.
    \textbf{(b) Quantum circuit diagram of the full protocol}. The protocol starts with two sets of qubits in system A and B respectively: $\{\ket{0}_{j_{i},A}\}, \{\ket{0}_{k_{i},B}\}$ initialized in their respective ground states. A global initialization unitary $U_{\alpha,\phi_{1,2}}$ (in red) of time duration $t_{\text{unit}}$ prepares the initial state in a coherent superposition. Further, the global dynamical decoupling channel $\mathcal{E}_{\text{DDS}}$ (in yellow) reduces the phase noise of the systems, taking a time of $t_{\text{dds}}$. After this step, the compiler schedules pair-wise entanglement in the sequence: $\{(j_{i},k_{i})\}$, represented by the block $\mathcal{E}_{\text{cmpl}}$ (in orange) which takes time $t_{\text{cmpl} }$. The whole block is repeated for $M$ attempts to increase the success rate of entanglement. The inner architecture of these blocks are described in detail in SI Sec.~III \cite{googleSI_Programmable_Quantum_Matter_2025Google}.}
    \label{fig:overview}
\end{figure*}

Among atom-like emitters, diamond color centers now support high-coherence spin–photon interfaces and optically heralded remote entanglement~\cite{beukers2024remote,Sukachev_2017,arjona2022photonic,knaut2024entanglement,iwasaki2017tin,parker2024diamond,rugar2021quantum}. Building on the quantum network milestones and recent modular frameworks for photon-mediated entanglement, these platforms provide a concrete path toward programmable quantum network nodes~\cite{barrett2005efficient,bernien2013heralded,hensen2015loophole,humphreys2018deterministic,pompili2021realization,doi:10.1126/science.add9771,beukers2024remote, PRXQuantum.5.010303}. Scaling such nodes imposes several key requirements:\\
\textbf{R1:} spin coherence must be preserved throughout repeated, probabilistic optical attempts, necessitating dynamical decoupling that remains effective across heterogeneous ensembles;\\
\textbf{R2:} control  must be uniform and high fidelity — both single- and two-qubit gates should tolerate amplitude and detuning variations with minimal per-emitter calibration \cite{Dong_2010}.\\
\textbf{R3:} optically heralded entanglement must produce indistinguishable photons while protecting the memory state, as in single-photon or Barrett–Kok–type protocols~\cite{Hermans_2023,barrett2005efficient,bernien2013heralded,pompili2021realization}.\\
\textbf{R4:} compilation and scheduling should orchestrate many parallel entanglement attempts with guarantees on link count and uniqueness \cite{maronese2021quantumcompiling,wang2024atomique,li2024dynamic,patil2022entanglement,kaur2023entanglement,humphreys2018deterministic}.\\
\textbf{R5:} all of the above requirements must be thermally feasible within cryogenic power budgets during extended control and decoupling windows \cite{krinner2019engineering}.

Today’s implementations fall short on several fronts. Standard bang–bang (CPMG/XY) decoupling \cite{Viola_1998} is detuning-sensitive and relies on high peak power, leading to rapid performance loss and heating in inhomogeneous ensembles. For the number of qubits being $N_q$, per-emitter interleaving of controls scales as order $O(N_q)$, compressing interpulse spacings below thermalization times and eroding $T_2$ and $T_2^*$ \cite{Sukachev_2017,nguyen2019integrated}. Entanglement scheduling via optical switching introduces insertion loss and synchronization complexity that directly suppress heralding rates, while photon-frequency inhomogeneity often forces filtering or matching that further reduces throughput unless compensated at the control level.

Here, we replace lossy optical-switch scheduling with \emph{resonant, composite drive engineering}. A single optimized waveform provides global unitary control and dynamical decoupling across heterogeneous emitters—reducing control overhead from $O(N_q)$ to $O(1)$—while maintaining high fidelity under detuning spread and lowering thermal load. Combined with a deterministic entanglement compiler, this enables optically heralded links with orders-of-magnitude higher throughput and guaranteed coverage.

As seen from Fig. ~\ref{fig:overview} we propose to use two spatially separated spin ensembles to create a bipartite cluster state which acts as a resource for measurement-based quantum computing \cite{jozsa2005introductionmeasurementbasedquantum,RevModPhys.79.135}. This state is set of entanglement links $\{j-k\}$ where $j, k$ are qubit labelings in devices $\mathcal{M}_{A}$ and $\mathcal{M}_{B}$ respectively. This approach, however, transforms the challenge from a simple entanglement generation and scheduling \cite{pompili2022experimental,li2024dynamic,ni2025entanglement} limitation into a complex, multi-parameter optimization problem.
The central task becomes minimizing the global cost function $J(\theta)$, defined as the average two-qubit error over all generated links:
\begin{equation}
    J(\theta) = \left\langle \epsilon_{jk}(\theta) \right\rangle_{j,k}
\end{equation}
Optimization is performed on the vector of experimental parameters $\theta = \{ P_{\mathrm{pulse}}, P_{\mathrm{DDS}}, P_{\mathrm{comp}} \}$, which includes the temporal shape of control pulses, the architecture of the dynamical decoupling sequences \cite{Viola_1998,Lucero_2010}, and the scheduling of entanglement attempts respectively. The error for each link, $\epsilon_{jk}$, is a function of both coherent control errors and decoherence during the entanglement protocol, as detailed in Eq.~\ref{Eq_e_ff} . Crucially, the fidelity decoherence for each link is determined by the compilation strategy $P_{\mathrm{comp}}$, which demands an efficient compilation strategy to build a graph state in PSPACE \cite{jain2009qippspace} and polynomial time. The value of $J(\theta)$ is therefore determined by the interplay between single-qubit gate fidelities, thermally limited coherence times, and the probabilistic efficiency of the underlying single-photon entanglement protocol \cite{Hermans_2023}.
Our work systematically deconstructs and optimizes the components of $J(\theta)$ to deliver a scalable and robust solution.

To solve this optimization problem, we take advantage of the diversity of color centers. This inherent heterogeneity provides the unique spectral labels necessary for qubit addressability under a global drive, which resolves the challenge of scheduling remote entanglement attempts that would otherwise scale super-exponentially. Our integrated approach, summarized in Fig.~\ref{fig:overview}, therefore delivers four key components: (i) high-fidelity global unitary control (\textbf{R2, R5}), (ii) broadband dynamical decoupling (\textbf{R1, R2}), (iii) a single-photon remote entanglement protocol (\textbf{R3}), and (iv) an efficient compilation algorithm to generate a bipartite cluster state \cite{Nielsen_2006} (\textbf{R4}).


We illustrate it on diamond-based group-IV defect centers \cite{bradac_quantum_2019, PhysRevLett.112.036405, iwasaki_germanium-vacancy_2015, PhysRevLett.118.223603, PhysRevLett.119.253601, PhysRevX.11.041041, PhysRevX.15.021011}, specifically the negatively charged silicon-vacancy center (SiV$^-$) which are optically active \cite{PhysRevB.51.16681, PhysRevLett.77.3041, Sukachev_2017} and feature addressable—via microwave or mechanical strain—electronic spin states \cite{pingault_coherent_2017, PhysRevB.97.205444} (serving as control qubits) and nearby nuclear spin states \cite{PhysRevLett.122.190503, doi:10.1126/science.add9771} (serving as memory qubits). These defect centers have been previously demonstrated as a quantum network node \cite{bhaskar_experimental_2020, doi:10.1126/science.add9771, PRXQuantum.5.010303}, single photon sources \cite{PhysRevLett.113.113602, PhysRevLett.129.053603} (serving as communication qubits), and have also been proposed for blind quantum computing \cite{doi:10.1126/science.adu6894}. Having a well-studied \cite{PhysRevB.97.205444} strain dependence of its qubit characteristics makes this substrate programmable up to device constraints. Therefore, the SiV$^-$ center contains all the major components required by a quantum computational substrate in our proposed framework.


In Section \ref{sec_theoretical_framework}, we introduce the proposed system architecture, consisting of an ensemble of SiV$^-$ color centers in diamond interfaced with a mechanical structure. First, we establish how the mechanical strain of the structure can be used to drive arbitrary single-qubit gates on a single color center (Section \ref{sec_system_description}). Next, we show how a concatenated composite pulse sequence can be designed using gradient ascent pulse engineering (GRAPE) \cite{khaneja_optimal_2005} (Section \ref{sec_error_correcting_pulses}). This optimal control pulse can simultaneously correct pulse length errors and off-resonance errors \cite{PhysRevA.80.032303}. Hence it can be applied globally to perform an arbitrary single-qubit gate on all color centers at once. The use of these global optimal control pulses as rephasing $\pi$ pulses allows for simultaneous dynamical decoupling of all color centers, resulting in an effective decoherence time $T_2$ instead of $T_2^*$ for all color centers. In Section \ref{sec_dynamical_decoupling} a frequency-domain analysis of this effect is given using the filter function formalism and the performance of the GRAPE based Carr-Purcell-Meiboom-Gill (CPMG) sequence \cite{PhysRev.94.630, meiboom_modified_1958} is compared against the conventional bang-bang based CPMG sequence. In Section \ref{sec_ApplicationQuantumComputation} we elaborate on how this system can be useful for quantum computing purposes. We set up a formalism to create a bipartite-graph cluster states by linking two systems using a single-photon protocol for entanglement generation. A statistical definition of the system's quantum volume \cite{Cross_2019} is used to show the superior performance of our GRAPE optimal control pulse sequence over conventional bang-bang algorithms. Finally, we describe in detail how the full system can be compiled.

\section{Theoretical framework}
\label{sec_theoretical_framework}

\subsection{System description}
\label{sec_system_description}

Consider a set of $N_q$ group-IV color centers in diamond. The ground states (GS) and excited states (ES) manifold of each color center $CC_i$ are described by the Hamiltonian $\hat{\mathcal{H}
 }_i^{\text{GS/ES}}$ that includes the strain, spin-orbit (SO) and Zeeman interaction: 
 \begin{align}
 \hat{\mathcal{H}}_i^{\text{GS/ES}} = \hat{\mathcal{H}}_i^{\text{strain}} + \hat{\mathcal{H}}_i^{\text{SO(GS/ES)}} + \hat{\mathcal{H}}_i^{\text{Zeeman}}
 \end{align}

While our protocol is generally applicable to group-IV color centers due to their shared $D_{3d}$ symmetry, we illustrate it using the transverse-oriented \(\mathrm{SiV}^{-}\) center as a well-characterized model system. Its explicit Hamiltonian, which accounts for the defect's orbital structure and its interaction with the magnetic field and lattice strain, is detailed in Sec.~I.A of the Supplements \cite{googleSI_Programmable_Quantum_Matter_2025Google}.

Let us focus on the GS manifold and use its bottom 2 energy levels as a qubit with energy spacing $\hbar \omega_i$. An arbitrary single-qubit gate of type $\hat{\mathcal{U}}^  {\text{ideal}}(\theta, \phi)=\exp{-i\frac{\theta}{2}(\cos{\phi}\hat{\sigma}_x + \sin{\phi}\hat{\sigma}_y)}$ can be implemented using the strain-driving Hamiltonian $\hat{\mathcal{H}}_i^{\text{drive}}$ \cite{PhysRevB.97.205444}.

\begin{align}
    \hat{\mathcal{H}}_i^{\text{drive}} = 
    \begin{bmatrix}
    0 & 0 & \epsilon_{E_{gx},i}^{ac} & 0 \\
    0 & 0 & 0 & \epsilon_{E_{gx},i}^{ac} \\
    \epsilon_{E_{gx},i}^{ac} & 0 & 0 & 0 \\
    0 & \epsilon_{E_{gx},i}^{ac} & 0 & 0 
  \end{bmatrix}
  \cos(\omega_d t_i + \varphi)
\end{align}

The strain oscillations are driven at frequency $\omega_d$ and phase offset $\varphi$, where these are chosen to satisfy $\omega_d = \omega_i$ and $\varphi=\pi - \phi$. The evolution time $t_i$ should be picked equal to $\frac{\theta}{\Omega_i}$. The effective Rabi frequency $\Omega_i$ is determined by fitting a squared sinusoid to the probability of finding the color center in its ground state after initialization in the first excited state. Further details on the realization of strain-driven arbitrary single-qubit gates are provided in Sec.~I.B of Supplements \cite{googleSI_Programmable_Quantum_Matter_2025Google}.

In our framework we are using a global strain drive, i.e. strain pulses that act on all $N_q$ color centers. Hence, the strain drive frequency is detuned differently from all color centers: $\omega_d = \omega_i + \Delta_i$. This occurs due to different local strain environments $\epsilon_{E_{gx},i}^{dc}$. Furthermore, the effective Rabi frequency can vary for all color centers: $\Omega_i = \Omega (1 + \epsilon_i)$. These variations encompass 2 effects: different strain modulation amplitudes $\epsilon_{E_{gx},i}^{ac}$ due to different locations in the device and different spin-strain susceptibilities for different local strain environments. As a result, the dynamic evolution for a time $t=\frac{\theta}{\Omega}$ of each color center $i$ under $\hat{\mathcal{H}}_i^{\text{drive}}$ is given by the unitary operator $\hat{\mathcal{U}}_{\epsilon_i , f_i}^{\text{real}}(\theta, \phi)$.

\begin{align*}
    &\hat{\mathcal{U}}_{\epsilon_i , f_i}^{\text{real}}(\theta, \phi) = \exp{-\frac{i}{\hbar}\hat{\mathcal{H}}^{\text{drive}}_i t} \\
    &\cong \exp{-i\frac{\theta}{2} [ (1+\epsilon_i)(\cos{\phi}\hat{\sigma}_{x,i} + \sin{\phi}\hat{\sigma}_{y,i}) - i f_i \hat{\sigma}_{z,i} ] }
\end{align*}

Here $\cong$ denotes the corresponding action on the bottom 2 energy levels. By comparing $\hat{\mathcal{U}}_{\epsilon_i , f_i}^{\text{real}}(\theta, \phi)$ with $\hat{\mathcal{U}}^{\text{ideal}}(\theta, \phi)$ it is clear that each color center experiences an amplitude error $\epsilon_i$ and an off-resonance error $f_i=\frac{\Delta_i}{\Omega}$.

The total system Hamiltonian for our set of $N_q$ color centers and global strain drive is given by:

\begin{align}
    \hat{\mathcal{H}}^{\text{total}} = \sum_{i=1}^{N_q} \left( \hat{\mathcal{H}}_i^{\text{GS}} + \hat{\mathcal{H}}_i^{\text{drive}}\right)
    \label{eq:syst_Hamiltonian}
\end{align}

\subsection{Error-Correcting Pulses}
\label{sec_error_correcting_pulses}
\begin{figure*}[htb]
    \centering
    \includegraphics[width=0.9\textwidth]{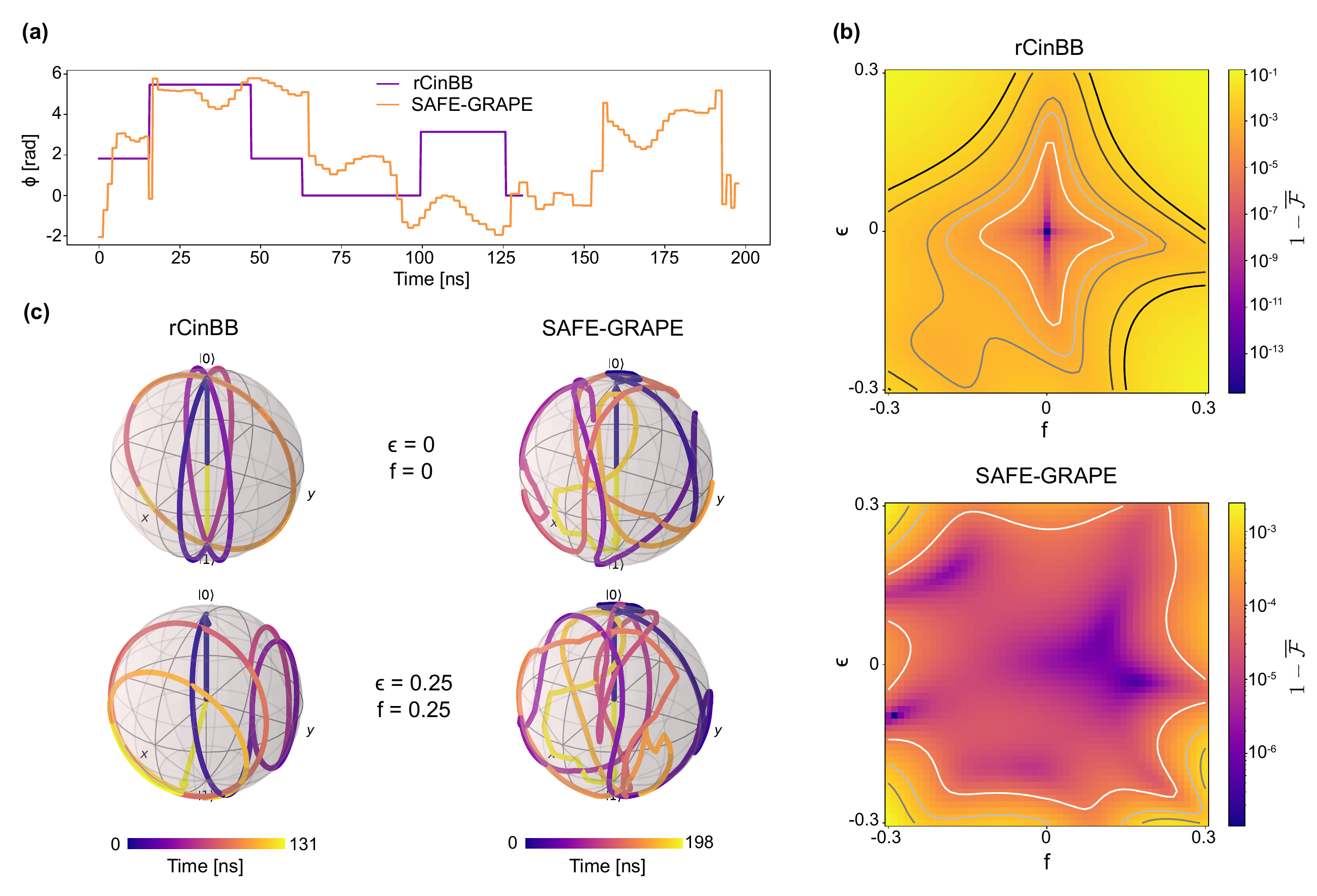}
    \caption{\textbf{SAFE-GRAPE optimal control pulse sequence for implementing \boldmath$\hat{\mathcal{U}}^{\text{ideal}}(\theta=\pi, \phi=0)$.} \textbf{(a)} Control pulse sequence before (rCinBB) and after (SAFE-GRAPE) optimization for a Rabi drive strength $\Omega=200$ Mrad/s. \textbf{(b)} Average gate infidelity heatmap for rCinBB (top) and SAFE-GRAPE (bottom) pulse sequence. Contours denote infidelities of $10^{-4}, 5\times10^{-4}, 10^{-3}, 5\times10^{-3}$ and $ 10^{-2}$, increasing from white to black. Note that both plots have their own color bar. \textbf{(c)}  Action of rCinBB (left) and SAFE-GRAPE (right) pulse sequence on the Bloch sphere for initial state $\ket{0}$ in case of $\epsilon=f=0$ (top) and $\epsilon=f=0.25$ (bottom). On each Bloch sphere, the purple and yellow arrows point at the initial and final state respectively. }
    \label{fig:GRAPE}
\end{figure*}

\subsubsection{Composite Pulses}

The application of a single strain pulse to a set of $N_q$ color centers introduces systematic control errors $\{\epsilon_i, f_i\}_{i\in [1...N_q]}$.
To mitigate these errors, robust pulse sequences from the nuclear magnetic resonance (NMR) field are adapted. Composite pulses, denoted as $\hat{\mathcal{U}}_{CP}(\theta, \phi)$, are constructed as sequences of elementary single-qubit gates $\hat{\mathcal{U}}(\theta_i, \phi_i)$:  
\begin{align}
\hat{\mathcal{U}}_{CP}(\theta, \phi) = \prod_{i=1}^{N_p} \hat{\mathcal{U}}(\theta_i, \phi_i)
\end{align}
where $N_p$ is the number of pulses in the sequence. Specifically, the BB1 sequence \cite{WIMPERIS1994221} addresses amplitude errors, while CORPSE \cite{PhysRevA.67.042308} mitigates off-resonance errors. For simultaneous robustness against both errors, we employ the reduced CinBB (rCinBB) pulse \cite{Bando2012Concatenated}, a concatenated composite pulse integrating CORPSE within the BB1 framework (see SI Sec.~II.B \cite{googleSI_Programmable_Quantum_Matter_2025Google}). The operator for the rCinBB pulse is expressed as:
$\hat{\mathcal{U}}_{rCinBB}(\theta,\phi) = \prod_{i=1}^{6} \hat{\mathcal{U}}(\theta_i, \phi_i)$,
with parameters given in SI Table S2 \cite{googleSI_Programmable_Quantum_Matter_2025Google}.  



\subsubsection{SAFE-GRAPE}

More accurate error mitigation can be achieved by Simultaneous Amplitude and Frequency Error-correcting GRadient Ascent Pulse Engineering (SAFE-GRAPE) of composite pulses. Define the composite search space $\boldsymbol{\Omega}$ as the region of errors that should be mitigated, e.g. $\boldsymbol{\Omega} = \{(\epsilon,f) | \epsilon_{\text{min}}<\epsilon<\epsilon_{\text{max}} \wedge f_{\text{min}}<f<f_{\text{max}} \}$. The goal is to find a set $\{\theta_i , \phi_i \}_{i\in [1...N_p]}$ such that:

\begin{align}
    \forall (\epsilon , f) \in \boldsymbol{\Omega}: \prod_{i=1}^{N_p} \hat{\mathcal{U}}_{\epsilon , f}^{\text{real}}(\theta_i, \phi_i) \approx \hat{\mathcal{U}}^{\text{ideal}}(\theta,\phi)
\end{align}

The SAFE-GRAPE algorithm discretizes $\boldsymbol{\Omega}$ to $\boldsymbol{\Omega}^*$, e.g. $\boldsymbol{\Omega^*} = \{ \underbrace{\epsilon_{min}, ...,\epsilon_{max}}_{N_{\epsilon}} \} \times \{ \underbrace{f_{min}, ...,f_{max}}_{N_{f}}\}$. For each point in the discretized composite search space, we calculate the average gate infidelity $1-\overline{\mathcal{F}}(\epsilon,f)$ between the composite pulse sequence and the target operation. The SAFE-GRAPE algorithm then minimizes the total loss function $\mathcal{L}$, which is given by:

\begin{widetext}
\begin{align}
    \mathcal{L} = \sum_{(\epsilon , f) \in \boldsymbol{\Omega^*}} \underbrace{ \left[ 1 - \frac{1}{2}\text{Tr}\left[ \left( \hat{\mathcal{U}}^{\text{ideal}}(\theta,\phi) \right)^\dagger \prod_{i=1}^{N_p} \hat{\mathcal{U}}_{\epsilon , f}^{\text{real}}(\theta_i, \phi_i) \right] \right] }_{1-\overline{\mathcal{F}}(\epsilon,f)} \mathcal{W}(\epsilon,f) 
\end{align}
\end{widetext}

where we introduced the additional weight factor $\mathcal{W}(\epsilon,f)$. If all the points in the composite search space are equally important, $\mathcal{W}(\epsilon,f)=1$ is a good choice. Alternatively, the center $(\bar{\epsilon}, \bar{f})$ of the composite search space can be prioritized using a Gaussian weight factor $\mathcal{W}(\epsilon,f)=\mathcal{N} \exp \left( - \frac{(\epsilon - \bar{\epsilon})^2}{2 \sigma_{\epsilon}^2} - \frac{(f - \bar{f})^2}{2 \sigma_{f}^2} \right) $, with the appropriate normalization constant $\mathcal{N}$.

At the start of the SAFE-GRAPE algorithm we define the hyperparameters of the model. $\theta$ and $\phi$ define the target unitary $\hat{\mathcal{U}}^{\text{ideal}}(\theta, \phi)$. By selecting different values for $\theta$ and $\phi$ it is possible to implement an arbitrary single-qubit gate. $N_p$ sets the number of pulses in the composite pulse sequence. $\epsilon_{min},\epsilon_{max},N_{\epsilon},f_{min},f_{max}$ and $N_f$ define the discretized composite search space $\boldsymbol{\Omega^*}$. The relative importance of points within $\boldsymbol{\Omega^*}$ is set by the definition of $\mathcal{W}(\epsilon,f)$. The effective Rabi drive strength $\Omega$ is fixed, so all off-resonance errors $f_i$ are constant for the respective color centers with fixed detuning $\Delta_i$. Since $t=\frac{\theta}{\Omega}$, we can now represent the composite pulse sequence $\prod_{i=1}^{N_p} \hat{\mathcal{U}}_{\epsilon , f}^{\text{real}}(\theta_i, \phi_i)$ by the set of trainable parameters $\{t_i , \phi_i \}_{i\in [1...N_p]}$. These are initialized based on the rCinBB composite pulse sequence. We use a sigmoid re-parametrization in order to force each $t_i$ to lie in $[t_{min}, t_{max}]$, where $t_{min}$ and $t_{max}$ are also hyperparameters of the model:

\begin{align}
    t_i \xrightarrow{} \tilde{t}_i = t_{min} + (t_{max} - t_{min})\frac{1}{1+\exp(-t_i)}
\end{align}

We implement the SAFE-GRAPE algorithm in \texttt{PyTorch} \cite{paszke2019pytorchimperativestylehighperformance} and use the built-in L-BFGS optimizer \cite{Liu1989}. SI Sec.~II.C \cite{googleSI_Programmable_Quantum_Matter_2025Google} lists all the parameter values used in the SAFE-GRAPE algorithm. 

As an example, we implement SAFE-GRAPE for a $\pi$-X gate  $\hat{\mathcal{U}}^{\text{ideal}}(\theta=\pi, \phi=0)$. The resulting optimal control pulse sequence is shown and compared to the rCinBB pulse sequence in Fig.~\ref{fig:GRAPE}(a). To illustrate the effect of both pulse sequences, we show their action on the ground state in Fig.~\ref{fig:GRAPE}(c). When no amplitude error or off-resonance error is present, both rCinBB and SAFE-GRAPE perform well, achieving a fidelity of 100\% and $>99.999\%$ respectively. Once amplitude and off-resonance errors are introduced, rCinBB fidelity decreases more rapidly. In case of $\epsilon=f=0.25$, the respective fidelities drop to 92.7\% and 99.987\%. SAFE-GRAPE still manages to flip the initial $\ket{0}$ state close to the $\ket{1}$ state on the Bloch sphere, whereas for rCinBB the yellow arrow indicating the final state is clearly deviating from the $\ket{1}$ state. Hence, at the cost of higher bandwidth requirements and a 51\% increase in pulse duration w.r.t. the rCinBB pulse sequence, SAFE-GRAPE produces a more robust pulse with $<10^{-4}$ average gate infidelity for most of the composite search area (Fig.~\ref{fig:GRAPE}(b)). The average gate infidelity represents the fidelity between the state after evolution by the composite pulse sequence and the target final state, averaged over all possible input states. Here we have calculated the fidelity based on perfect control pulses, such that the resulting infidelity maps reflect best the differences between the two composite control pulse sequences. 

\subsection{Dynamical Decoupling: Frequency Domain Picture}
\label{sec_dynamical_decoupling}
\begin{figure*}[htb]
    \centering
    \includegraphics[width=0.9\textwidth]{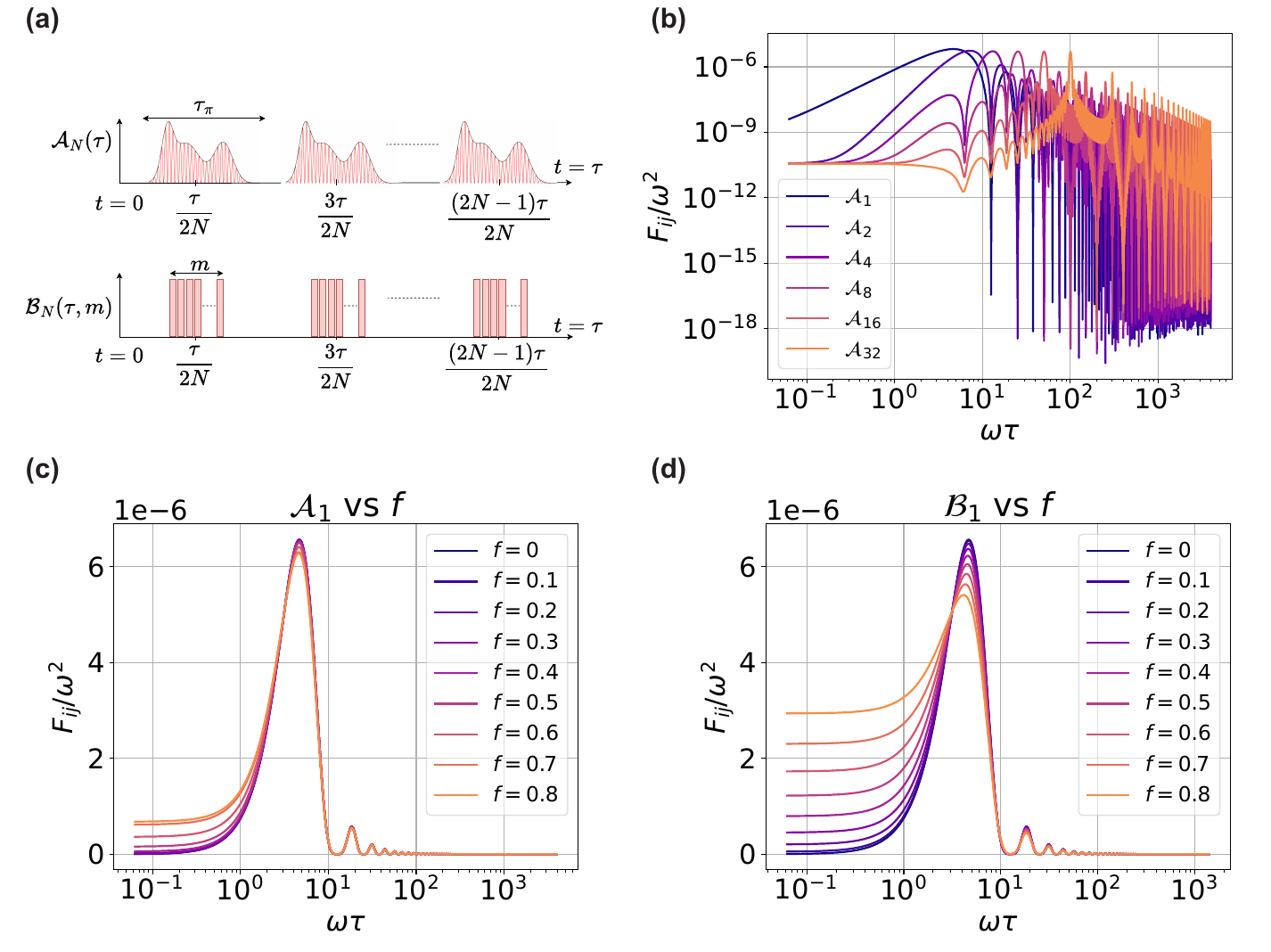}
     \caption{\textbf{Filter function simulations}. \textbf{(a)} Schematic for sequences $\mathcal{A}_{N}(\tau)$ and $\mathcal{B}_{N}(\tau,m)$. \textbf{(b)} As $N$ increases, the modified filter function $F_{ij}/\omega^{2}$ for $\mathcal{A}_{N}$ reduces, representing increased noise suppression. \textbf{(c)} $F_{ij}/\omega^{2}$ for $\mathcal{A}_{1}$ rises for increased pulse detuning $f$. \textbf{(d)} There is a much larger increment in $F_{ij}/\omega^{2}$ for $\mathcal{B}_{1}(m=1)$ for the same change in $f$. This shows that sequence $\mathcal{A}_{N}$ is more robust against off-resonance errors and hence better for implementing a dynamical decoupling sequence than sequence $\mathcal{B}_{N}$. ($f = 0$ for plot \textbf{(b)}; $\epsilon = 0$ for all plots.)}
    \label{fig:Filter_Function}
\end{figure*}
\begin{figure*}[htb]
    \centering
    \includegraphics[width= \linewidth]{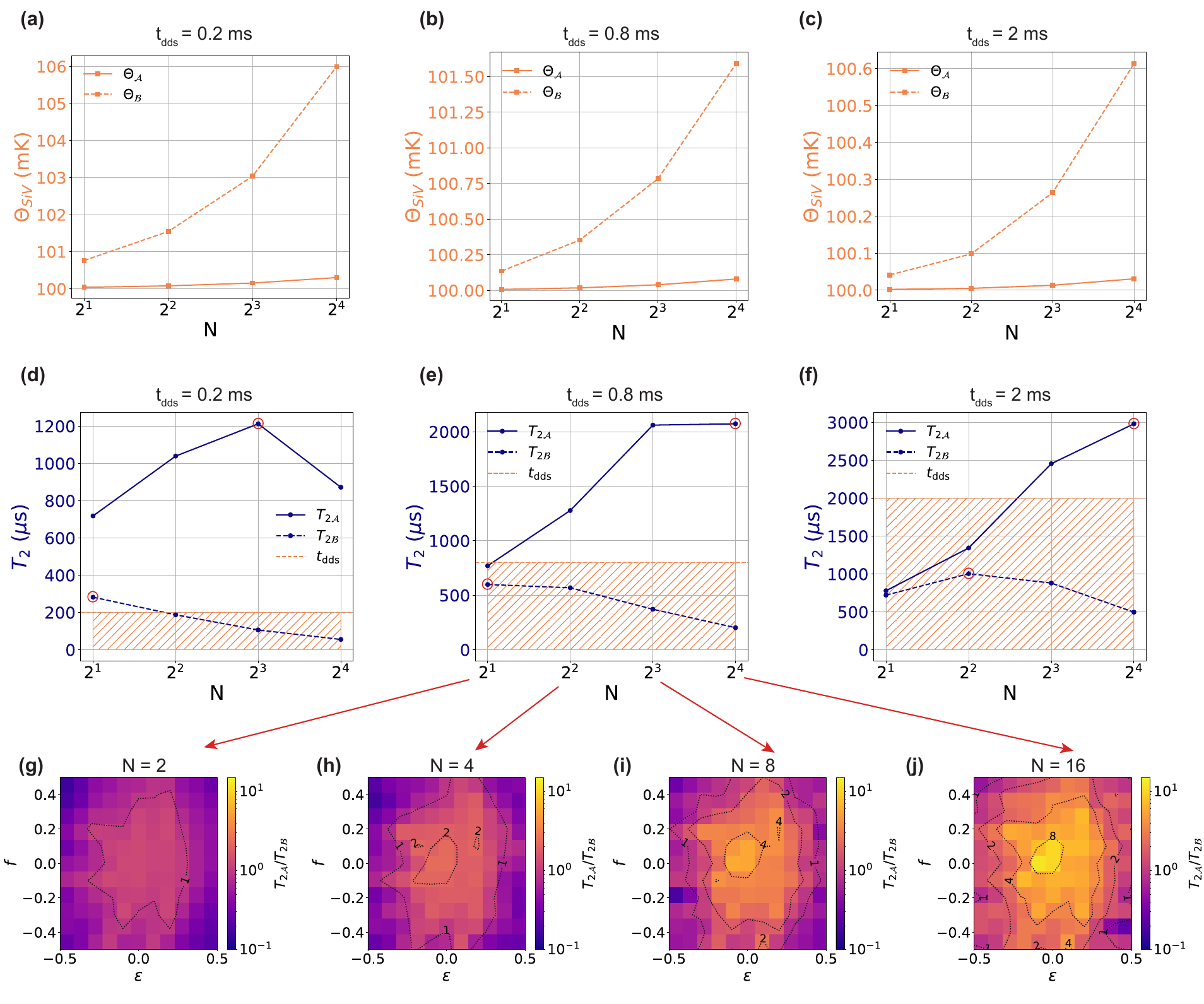}
    \caption{\textbf{Temperature and $T_{2}$ simulations}. \textbf{(a-c)} Variation of SiV$^{-}$ temperature $\Theta_{\text{SiV}}$ with the CPMG index $N$. Sequence $\mathcal{B}$ consistently leads to a higher rise in temperature compared to $\mathcal{A}$. As $t_{\text{dds}}$ is increased, the rise in temperature reduces for both sequences due to the reduction in the duty cycle of the active heat load. \textbf{(d-f)} Variation of $T_{2}$ (for the central qubit, i.e. $\epsilon=f=0$) with $N$. The orange shaded region represents the region $T_{2}<t_{\text{dds}}$, signifying that the decoupling protocol is futile in that region. Red circles represent the optimal CPMG index (i.e. $N$ for which $T_{2}$ is maximum), signifying the interplay between heat-induced decoherence and improved coherence due to dynamical decoupling. Furthermore, $T_{2}$ for sequence $\mathcal{A}$ is consistently larger than that for $\mathcal{B}$. As $t_{\text{dds}}$ increases, $T_{2}$ values for both sequences increase but sequence $\mathcal{B}$ gets completely submerged in the shaded region \textbf{(e,f)}. This implies that decoupling via $\mathcal{B}$ does not work for larger values of $t_{\text{dds}}$. \textbf{(g-j)} For $t_{\text{dds}}=~0.8$ ms, we plot the heat-maps for $T_{2\mathcal{A}}/T_{2\mathcal{B}}$ on the composite ($\epsilon$-$f$) search space. Here the dotted contours show islands where the ratio is above a marked threshold. As $N$ increases, the contour landscape becomes more pronounced with maximum enhancement above $8$ for $N=16$. ($m$ is set to 121 for $\mathcal{B}$ in the above simulations.)}
    \label{fig:T2_sims}
\end{figure*}
In the previous section, we talked about a GRAPE based composite pulse sequence to implement a $\pi$ rotation which corrects both off-resonance errors and amplitude errors. Correcting these errors ensures that all the qubits in the ensemble experience the same $\pi$ pulse upto a given fidelity as seen from the heat-map in Fig.~\ref{fig:GRAPE}. We propose to concatenate this optimized pulse for many cycles to implement a dynamical decoupling sequence on all qubits simultaneously. In order to quantify these sequences, we use the filter function formalism \cite{Biercuk_2011}, and thereby estimate the enhancement in $T_{2}$ time of the system of qubits.

In the presence of a decohering environment, the Hamiltonian in Eq.~\ref{eq:syst_Hamiltonian}, changes to:
\begin{align}
    \hat{\mathcal{H}}^{\text{total}} = \sum_{i=1}^{N_q} \left(\hat{\mathcal{H}}_i^{\text{GS}} + \hat{\mathcal{H}}_i^{\text{drive}} + \hat{\mathcal{H}}_i^{\text{noise}}\right)
\end{align}
where $\hat{\mathcal{H}}_i^{\text{noise}}$ is given by \cite{Hangleiter_2021, Cerfontaine_2021}:
\begin{align}
    \hat{\mathcal{H}}_i^{\text{noise}} = \sum_{k}s_{k,i}(t)b_{k,i}(t)\hat{\mathbf{B}}_{k,i}
\end{align}
Here $\hat{\mathbf{B}}_{k,i}$ is the set of noise operators via which qubit $i$ couples to the environment, $b_{k,i}(t)$ is the classically fluctuating noise variable for qubit $i$ and noise mode $k$, $s_{k,i}(t)$ is the deterministic time-dependent sensitivity for the noise operator $\hat{\mathbf{B}}_{k,i}$.
In order to simplify our analysis, we assume that the qubit ensemble only experiences dephasing noise (i.e. $k=1$, $\hat{\mathbf{B}}_{i}=\hat{\sigma}_{z,i}$) and we further assume that the noise mode is always coupled to the qubit with a constant sensitivity $s_{i}$, which can be absorbed in the noise variable $b_{i}(t)$. For the filter-function formalism, we consider the power spectral density of the fluctuating variables given by:
 \begin{align}
     \mathcal{S}_{i}(\omega) = \int_{0}^{\infty}e^{-j\omega\tau}\langle~b_{i}(t+\tau)~b_{i}(t)\rangle~d\tau
 \end{align}
 Furthermore, we define the coherence of the $i^{th}$ qubit spin state after time $\tau$ as \cite{Biercuk_2011}:
 \begin{equation}
     W_{i}(\tau) = |\overline{\langle\sigma_{y,i}\rangle}| \equiv e^{-\chi_{i}(\tau)}
 \end{equation}
Here the angled brackets represent the quantum-mechanical expectation value and the overline represents the average over multiple measurement outcomes on qubit $i$. Using these definitions, we have the following relation \cite{Biercuk_2011}:
\begin{equation}
    \chi_{ij}(\tau) = \frac{2}{\pi}\int_{0}^{\infty}\frac{\mathcal{S}_{i}(\omega)}{\omega^2}F_{ij}(\omega\tau)~d\omega
\end{equation}
We introduce the filter function (spectral dependence) $F_{ij}(\omega\tau)$ of the experiment sequence $j$ being performed on qubit $i$, which in our case will be a dynamical decoupling sequence. We further represent $\chi(\tau)$ as follows:
\begin{equation}
    \chi_{ij}(\tau) \equiv \bigg(\frac{\tau}{T_{2,ij}}\bigg)^{z_{ij}}
\end{equation}
where $T_{2,ij}$ is the coherence time of qubit $i$, after implementing an experimental sequence $j$, and $z_{ij}$ is the corresponding scaling, which depends on the noise spectrum $\mathcal{S}_{i}$. We further simplify our analysis by assuming that all qubits experience the same noise spectrum and are equally sensitive (i.e. $s_{i} = s$, $\mathcal{S}_{i}(\omega) = \mathcal{S}(\omega)~\forall~i$). We consider a double-exponential noise spectrum as seen in the SI of SiV$^{-}$ paper \cite{Sukachev_2017}:
\begin{equation}
    \mathcal{S}(\omega) = c_{0} e^{-\omega^2/\omega_{0}^2} + c_{1} e^{-\omega^2/\omega_{1}^2}
\end{equation}
where we assume the same values from the paper: $c_{0} = 10^6~s^{-1}$, $c_{1} = 10^9~s^{-1}$, $\omega_{0} = 1.8\times10^{3}~s^{-1}$, $\omega_{1} = 50~s^{-1}$. For the composite pulse sequence generated by optimizing the cost function $\mathcal{L}$ in Eq. 8, we numerically evaluate the filter-function using the \texttt{filter\_function} package \cite{hangleiter_2024_11192918} available in \texttt{Python}.
 
For the comparison, we consider two sequence forms (Fig.~\ref{fig:Filter_Function}(a)): \{$\mathcal{A}_{i}(\tau)$, $\mathcal{B}_{i}(\tau, m)$\} for $i\in\{2,4,8,16\}$, where the sequence $\mathcal{A}_{i}(\tau)$ is composed of $i$ cycles of SAFE-GRAPE optimized pulses uniformly spaced over a measurement window $\tau$, and $\mathcal{B}_{i}(\tau, m)$ represents the $m$-times (addressing $m$ qubits individually) concatenated CPMG sequence with $i$ cycles over the measurement window $\tau$ in the bang-bang regime.    

Fig.~\ref{fig:Filter_Function}(b) shows the numerically simulated filter functions \( F_{ij}/\omega^2 \) versus \( \omega \tau \) on a log-log scale for sequences \( \mathcal{A}_N \), where \( N \) is the number of \(\pi\) pulses for $\epsilon=f=0$. As \( N \) increases from 1 to 32, the low-frequency part of the filter (e.g., \( \omega \tau \lesssim 10 \)) is increasingly suppressed, with filter values dropping by over 7 orders of magnitude—from around \( 10^{-5} \) for \( \mathcal{A}_1 \) to below \( 10^{-12} \) for \( \mathcal{A}_{32} \). Strong oscillations appear in the range \( \omega \tau \sim 10^2 \) to \( 10^6 \), becoming denser and more structured as \( N \) increases, which reflects enhanced frequency selectivity.

At very high frequencies (\( \omega \tau \gtrsim 10^7 \)), the curves converge to a baseline around \( 10^{-21} \)–\( 10^{-23} \), suggesting that additional pulses beyond a certain point offer diminishing returns in this regime. Overall, the filter functions show that higher-order CPMG sequences provide broader and deeper suppression of low-frequency noise, with tunable filtering properties governed by the pulse number.

The plots in Fig.~\ref{fig:Filter_Function}(c) compare the filter functions \( F_{ij}/\omega^2 \) for sequences \( \mathcal{A}_1 \) as the parameter \( f \) varies from 0 to 0.8. In the low-frequency regime ($\omega\tau << 1$) for \( \mathcal{A}_1 \), the function shifts from $4\times10^{-9}$ ($f=0$) to $1.8\times10^{-8}$ ($f=0.3$), suggesting an increment by $\sim$4.5 times, while the peak amplitude remains stable (within 0.1\%) at approximately \( 6.5 \times 10^{-6} \) across values of \( f\in(0, 0.3) \). In contrast, \( \mathcal{B}_1 \) at low frequencies, shows a change in the function from $3\times10^{-9}$ ($f=0$) to $4.5\times10^{-7}$ (
$f=0.3$), which is an increment by $\sim$150 times. Further, the peak amplitude reduces by 3\% (for $f=0$ to 0.3) and even undergoes a leftward shift and narrowing of the filter bandwidth.

These differences indicate that \( \mathcal{B}_1 \) is more sensitive to variations in \( f \), with its filtering strength and spectral selectivity degrading significantly as \( f \) increases. In contrast, \( \mathcal{A}_1 \) maintains consistent filtering performance, making it $\sim30$ times more robust under fluctuations in $f$ (from 0 to 0.3). Thus, based on the stability of peak height, location, and shape, \( \mathcal{A}_1 \) is quantitatively more resilient to changes in \( f \) than \( \mathcal{B}_1 \). This approach is system agnostic, and shows robustness of sequence $\mathcal{A}$ over $\mathcal{B}$ irrespective of the dominant noise spectrum of hardware.

We now focus on SiV$^{-}$ based system placed on the cold-plate stage (100 mK) of a dilution refrigerator. Implementing a pulse sequence ($\mathcal{A}$ or $\mathcal{B}$) leads to active and passive heat-loads on the sample stage, which depends on multiple parameters like: dilution fridge geometry and cooling cycle, thermal properties of cryogenic coax cable, sample footprint, specific heat capacity of the sample and it's thermal anchoring with the stage. Suppose for sequence $\mathcal{A}_{N}$, pulse incoming times are $t_{j,A}=\{ \frac{(2j-1)t_{\text{dds}}}{2N} \}$, where $1\le j \le N$, and for sequence $\mathcal{B}_{N}(m)$, the incoming times are $t_{jk,B}=\{ \frac{(2j-1)t_{\text{dds}}}{2N} + (k-1)t_{\pi,B}\}$, where $1\le j \le N$ and $1\le k \le m$. We assume a realistic set of device parameters and derive the following expressions for time-dependent temperature of SiV$^{-}$ system $\Theta_{\text{SiV}}$ (see SI Sec.~III.F Eq. (89) \cite{googleSI_Programmable_Quantum_Matter_2025Google}):
\begin{subequations}
\begin{align}
\Theta_{\mathrm{SiV}}(t)
&= \overline{\Theta}_{\mathrm{CP}}
  + \sum_{j=1}^{N}\Bigl(
      \Theta_{\mathrm{SiV}}(t,\,t_{j,A})
      - \overline{\Theta}_{\mathrm{CP}}
    \Bigr)
  \\
&\quad \text{for }\mathcal{A}_{N}, \nonumber\\
\Theta_{\mathrm{SiV}}(t)
&= \overline{\Theta}_{\mathrm{CP}}
  + \sum_{j=1}^{N}\sum_{k=1}^{m}\Bigl(
      \Theta_{\mathrm{SiV}}(t,\,t_{jk,B})
      - \overline{\Theta}_{\mathrm{CP}}
    \Bigr)
  \\
&\quad \text{for }\mathcal{B}_{N}(m). \nonumber
\end{align}
\end{subequations}
Here $\overline{\Theta}_{\text{CP}}$ takes into account the slow temperature rise in the cold-plate as in SI Eq.~(83), and $\Theta_{\text{SiV}}(t, t_{0})$ is given by:
\begin{equation}
\begin{split}
  \Theta_{\mathrm{SiV}}(t,t_{0})
  &= \Theta_{\mathrm{CP}}
    + P_{\mathrm{th}}\bigl(
       e^{-\frac{t-t_{0}}{\tau_{\mathrm{th,SiV}}}}
       - e^{-9\,\frac{t-t_{0}}{\tau_{\mathrm{th,SiV}}}}
      \bigr) \\
  &\quad\;\times \operatorname{ReLU}(t - t_{0}) \,.
\end{split}
\end{equation}
where $P_{\text{th}}$ is a normalization constant (estimated in SI Eq.~(87)), and $\tau_{\text{th, SiV}}$ is the thermalization time-scale of the sample. Thus, while the dynamical decouping pulses are applied, it simultaneously increases the temperature of the sample which impacts the $T_{2}$ and $T_{1}$ time of the system. Assuming that $1/T_{2(1)}$ of SiV$^{-}$ has a linear dependence with $\Theta_{\text{SiV}}$ in the low temperature and low strain regime \cite{pingault_coherent_2017, Jahnke_2015}, SI Eq.~(91-92) yields the following effective $T_{2}$ times for a measurement window $t_{\text{dds}}$ in case of both sequences:
\begin{widetext}
\begin{subequations}
    \begin{align}
       T_{2}(\Theta_{\text{SiV}}(t_{\text{dds}},\mathcal{A}_{N})) &= \frac{T_{2}(0.1~\text{K},\mathcal{A}_{N})}{1 + T_{2}(0.1~\text{K},\mathcal{A}_{N})\cdot(\Theta_{\text{SiV}}(t_{\text{dds}},\mathcal{A}_{N})-0.1)\cdot3\cdot10^{6}}\\[1.5ex]
       T_{2}(\Theta_{\text{SiV}}(t_{\text{dds}},\mathcal{B}_{N})) &= \frac{T_{2}(0.1~\text{K},\mathcal{B}_{N})}{1 + T_{2}(0.1~\text{K},\mathcal{B}_{N})\cdot(\Theta_{\text{SiV}}(t_{\text{dds}},\mathcal{B}_{N})-0.1)\cdot3\cdot10^{6}}
    \end{align}
\end{subequations}
\end{widetext}
This implies that two competing effects are in play: (a) heat load due to the pulses decohering the system and (b) dynamical decoupling improving the coherence of the system. Fig.\ref{fig:T2_sims}(a-c) show the rise of SiV$^{-}$ temperature $\Theta_{\text{SiV}}$ with CPMG index $N$, since the heat-load increases with $N$. For these simulations, we take $m=121$ in $\mathcal{B}(m,\tau)$ to address the 121 qubits in the $\epsilon$-$f$ space. Further, as $t_{\text{dds}}$ increases from 0.2 - 2 ms, the rise in $\Theta_{\text{SiV}}$ reduces for both $\mathcal{A}$ and $\mathcal{B}$, due to the reduction of the duty-cycle of the incoming pulses while $t_{\text{dds}}$ increases. Since, $\mathcal{B}$ leads to a larger heat-load into the sample compared to $\mathcal{A}$, we observe that for $t_{\text{dds}}= 0.2$ ms $\Theta_{\text{SiV}}$ increases by $\sim$5.2\% for $\mathcal{B}$ in contrast to a negligible rise of $\sim$0.3\% for $\mathcal{A}$. Since, $\mathcal{B}$ consistently leads to larger SiV$^{-}$ temperatures compared to $\mathcal{A}$, it leads to lower $T_{2}$ due to larger phonon-induced decoherence, as shown in Fig.\ref{fig:T2_sims}(d-f). Further, as $t_{\text{dds}}$ increases from 0.2 - 2 ms, the $T_{2}$ value for both $\mathcal{A}$ and $\mathcal{B}$ increases, since the effective thermal load reduces with increase in $t_{\text{dds}}$. We also observe that for a particular value of $t_{\text{dds}}$, there exists an optimal value of $N$ (encircled in red in Fig. \ref{fig:T2_sims}(d-f)) which maximizes $T_{2}$. The existence of this optimal value is a demonstration of the interplay between the two effects of heat-induced decoherence and increased coherence due to decoupling sequences. Table \ref{tab:t2_comparison1} and \ref{tab:t2_comparison2} compare the coherence times \( T_{2\mathcal{A}} \) and \( T_{2\mathcal{B}} \) for two sequences across various pulse numbers \( N \) for two different values of $t_{\text{dds}}$. The numbers reported here are the average and standard deviation over all grid points in the composite search space. Sequence \( \mathcal{A} \) consistently produces longer coherence times than sequence \( \mathcal{B} \), with enhancement factors extending to 7. Table \ref{tab:t2_comparison2} shows that for larger values of $t_{\text{dds}}$ the enhancement ratio reduces because the heat-load due to $\mathcal{B}$ approaches the heat-load due to $\mathcal{A}$ as seen from Fig. \ref{fig:T2_sims}(a-c). 

\begin{table}[h]
\centering
\caption{Comparison of coherence times and mean enhancement for $t_{\text{dds}}=0.2$ ms. Values shown are (mean $\pm$ standard deviation) over the grid points.
}
\begin{tabular}{|c| c| c |c|}
\hline
$N$ & $T_{2\mathcal{A}}$ (ms) & $T_{2\mathcal{B}}$ (ms) & Enhancement \\
\hline
\rule{0pt}{2.5ex}
2   & 0.43$^{\pm0.18}$  & 0.28  & 1.5 \\
\hline
\rule{0pt}{2.5ex}
4   & 0.51$^{\pm0.27}$  & 0.19  & 2.7 \\
\hline
\rule{0pt}{2.5ex}
8   & 0.49$^{\pm0.29}$  & 0.10  & 4.7 \\
\hline
\rule{0pt}{2.5ex}
16   & 0.39$^{\pm0.19}$  & 0.06  & 7.2 \\
\hline
\end{tabular}
\label{tab:t2_comparison1}
\end{table}

\begin{table}[h]
\centering
\caption{Comparison of coherence times and mean enhancement for $t_{\text{dds}}=0.8$ ms. Values shown are (mean $\pm$ standard deviation) over the grid points.
}
\begin{tabular}{|c| c| c |c|}
\hline
$N$ & $T_{2\mathcal{A}}$ (ms) & $T_{2\mathcal{B}}$ (ms) & Enhancement \\
\hline
\rule{0pt}{2.5ex}
2   & 0.45$^{\pm0.19}$  & 0.59  & 0.75 \\
\hline
\rule{0pt}{2.5ex}
4   & 0.58$^{\pm0.34}$  & 0.57  & 1.0 \\
\hline
\rule{0pt}{2.5ex}
8   & 0.65$^{\pm0.47}$  & 0.37  & 1.8 \\
\hline
\rule{0pt}{2.5ex}
16   & 0.62$^{\pm0.44}$  & 0.20  & 3.0 \\
\hline
\end{tabular}
\label{tab:t2_comparison2}
\end{table}
The orange shaded regions in Fig.\ref{fig:T2_sims}(d-f) represents the region for which $T_{2}<t_{\text{dds}}$, indicating that any pulse sequence within that region is not useful for dynamical decoupling. For lower values of $t_{\text{dds}}= 0.2$ ms, $\mathcal{B}$ is within this region for $N>2$ whereas $\mathcal{A}$ always stays outside. Increasing $t_{\text{dds}}$ completely submerges $\mathcal{B}$ in this region, implying that $\mathcal{B}$ cannot be used for dynamical decoupling at all for larger values of $t_{\text{dds}}$. Sequence $\mathcal{A}$ is now only partially submerged, implying that $\mathcal{A}$ not just shows enhancement in $T_{2}$, but due to a lower heat-load it also performs useful dynamical decoupling. Fig \ref{fig:T2_sims}(g-j) shows the 2D-color map of the enhancement ratio for $t_{\text{dds}}=0.8$ ms and various $N$. The contours represent fixed enhancement ratios. As $N$ increases, we observe larger contour landscapes with some enhancement values extending even beyond 8.

This section shows that the dynamical decoupling sequence implemented by the SAFE-GRAPE optimized unitaries outperforms traditional bang-bang sequences in terms of: heat-load ($\Theta_{\text{SiV}}$ increases by $\sim$5.2\% for $\mathcal{B}$ and $\sim$0.3\% for $\mathcal{A}$), $T_{2}$ enhancement (reaching over 7), and feasibility ($\mathcal{B}$ stops to decouple noise for larger values of $t_{\text{dds}}$). In the Supplementary section, we further change the number of qubits $m$ from $10^{2}$ to $10^{4}$ and observe that as the number of qubits scale, protocol $\mathcal{B}$ goes completely within the orange shaded region, implying that the approach $\mathcal{B}$ is non-scalable for dynamical-decoupling. Hence we have created a phase-noise reduced platform, which can be used to implement the single-photon entanglement protocol of the next section. 

\section{Application to Quantum Computation}
\label{sec_ApplicationQuantumComputation}


\subsection{Entanglement Operations and Cluster State Generation}
\begin{figure*}[htb]
    \centering
    \includegraphics[width=\linewidth]{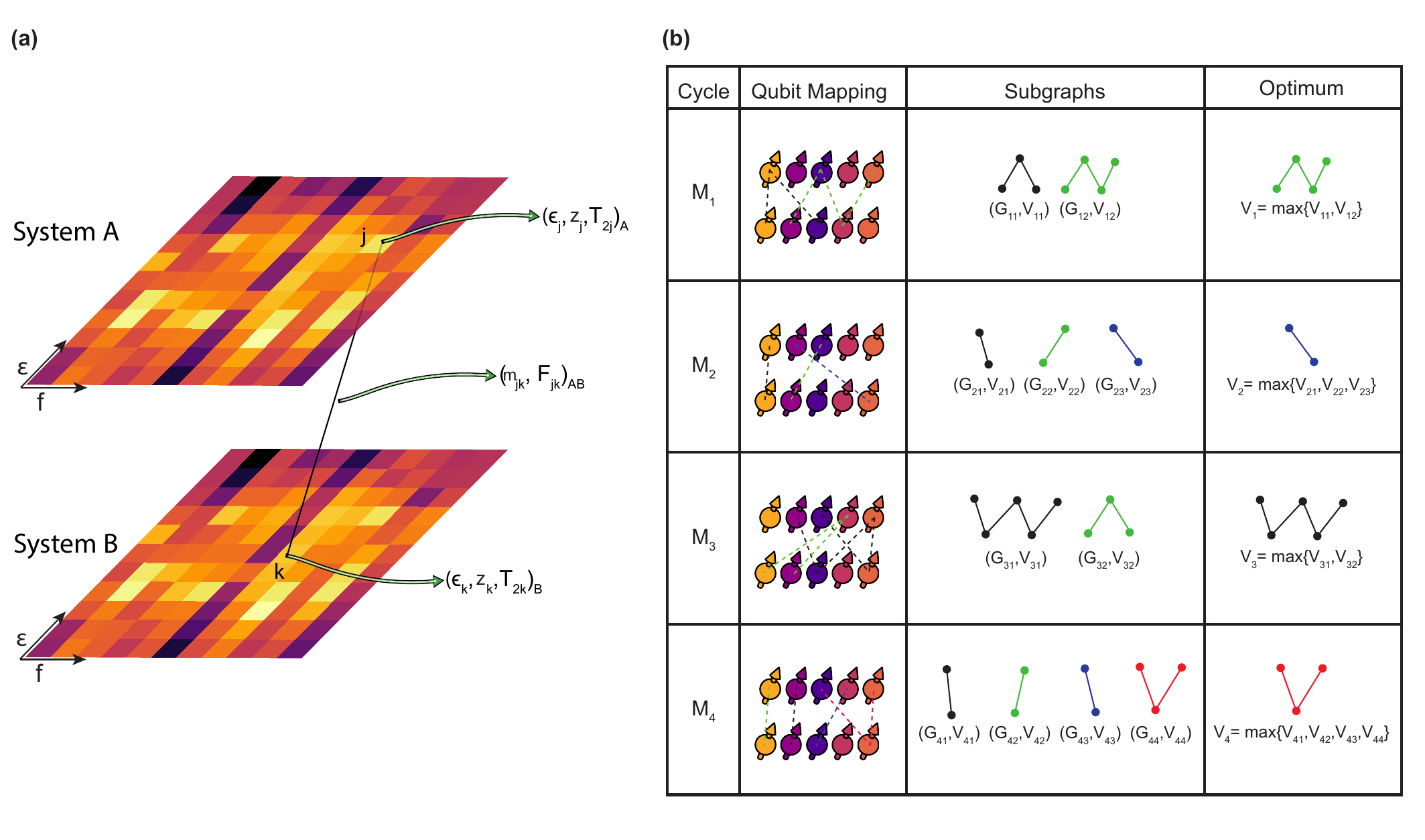}
    \caption{\textbf{Entanglement links statistics.}\textbf{(a)} Checkerboard representation ($\epsilon$-$f$ space) of two systems, in which qubit $j$ (system A) and $k$ (system B) is highlighted (colors only represent inhomogeneity and have no physical significance); an entanglement attempt on $j$-$k$, leads to a success after waiting for $m_{jk}$ attempts with a fidelity $F_{jk}$. \textbf{(b)} A graphical representation of the qubits over 4  different experimental cycles, where M$_{i}$ shows the $i^{\text{th}}$ experimental cycle. For any experimental cycle, the policy produces different configurations of subgraphs G$_{ij}$ with quantum volume V$_{ij}$ corresponding to the $j^{\text{th}}$ cluster of the $i^{\text{th}}$ experimental cycle.}
    \label{fig:Graph_building}
\end{figure*}

\begin{figure*}
    \centering
    \includegraphics[width=0.8\linewidth]{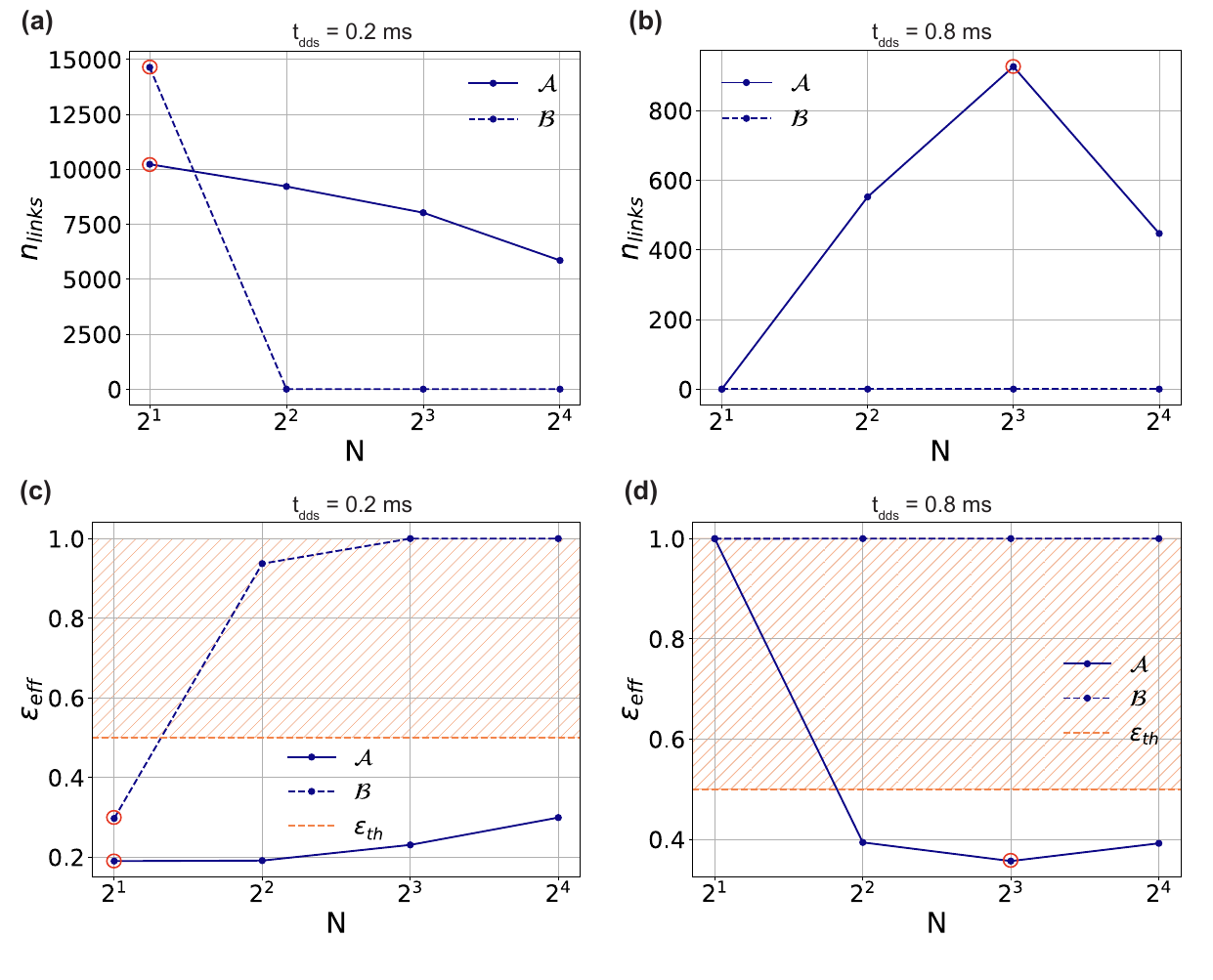}
    \caption{\textbf{Entanglement simulations}. Simulations over two systems (each system containing 121 points on the $\epsilon$-$f$ space), with $121^{2} (=14641)$ possible links. Here $n_{\text{links}}$ is the number of links which give effective error $\epsilon_{\text{eff}}<0.5$ and the red circle represents the optimal value of $N$ which maximizes $n_{\text{links}}$ and minimizes $\epsilon_{\text{eff}}$. For these simulations $t_{\text{cmpl}} = 10~\mu$s. \textbf{(a)} For a low value of $t_{\text{dds}}= 0.2$ ms, $\mathcal{B}$ starts off ($N = 2$) with performing better than $\mathcal{A}$, but for $N>2$, $n_{\text{links}}$ for $\mathcal{B}$ drops to zero as also confirmed from Fig. \ref{fig:T2_sims}(d) where sequence $\mathcal{B}$ is completely submerged in shaded region for $N>2$. On the contrary $n_{\text{links}}$ for $\mathcal{A}$ is consistent around $\sim 5000-10000$.  \textbf{(b)} For a larger value of $t_{\text{dds}} = 0.8$ ms, $n_{\text{links}}$ for $\mathcal{B}$ drops to zero for all $N$ which is confirmed from Fig. \ref{fig:T2_sims}(e) where $\mathcal{B}$ is completely submerged in the shaded region, whereas $\mathcal{A}$ reaches an optimum value of $\sim$900 for $N=8$, and thus performs better than $\mathcal{B}$. \textbf{(c-d)} Variation of $\epsilon_{\text{eff}}$ with $N$ (with the orange region showing the region $\epsilon_{\text{eff}}>0.5$). For $t_{\text{dds}}=0.2$ ms, $\mathcal{B}$ gets submerged in the orange region for $N>2$ whereas $\mathcal{A}$ is consistently outside the shaded region, always having a lower $\epsilon_{\text{eff}}$ than $\mathcal{B}$. For $t_{\text{dds}}= 0.8$ ms, $\mathcal{B}$ is always submerged in the shaded region. $\mathcal{A}$ starts off from the shaded region, but already for $N>2$ it exits it, achieving an optimal value for $N=8$. For large $t_{\text{dds}}$, $\mathcal{A}$ consistently outperforms $\mathcal{B}$.}
    \label{fig:QV_sims}
\end{figure*}
We implement a single-photon protocol \cite{Hermans_2023} for entanglement generation and start from the following initial state:
\begin{equation}
    \psi_{in}(\alpha, \phi) = \sqrt{\alpha}\ket{0} + \sqrt{1-\alpha}~\text{e}^{i\phi}\ket{1}
\end{equation}
as seen in Fig. \ref{fig:overview}(b). The checkerboard configuration in $\epsilon$-$f$ space, from Fig.~\ref{fig:Graph_building} is taken as an example where two qubits $j$ (system A) and $k$ (system B) are highlighted. We take a statistical approach to describe the resulting quantum volume. In order to generate a first successful entanglement link between qubits $j$ and $k$, one has to wait for a sufficient number of attempts $m_{jk}$. $m_{jk}$ is a random variable following a geometric distribution $P_{jk}(m) = 2\alpha\eta(1-2\alpha\eta)^{m-1}$, where $\eta$ is the single photon detection efficiency (see SI Sec.~III Eq. (32) \cite{googleSI_Programmable_Quantum_Matter_2025Google}). For our simulations, we take $(\alpha,\eta)=(10^{-4},10^{-2})$. This means that the process of cluster-state generation creates a distribution $\mathcal{D}_{m}$ of waiting times, and a distribution $\mathcal{D}_{\mathcal{F}}$ of the fidelity for the entanglement link generated between each of the total $N_q^2$ pairs:
\begin{equation}
\begin{split}
\mathcal{D}_{m} &= \{m_{jk}\}_{\substack{1 \leq i \leq N_q \\ 1 \leq j \leq N_q}} \\
\mathcal{D}_{\mathcal{F}} &= \{\mathcal{F}_{jk}\}_{\substack{1 \leq i \leq N_q \\ 1 \leq j \leq N_q}}
\end{split}
\end{equation}

From Sec.~II.C we get for each qubit $j$ a triplet ($\epsilon_{j},z_{j},T_{2j}$), which are the gate error, $T_2$ scaling and $T_{2}$ time.
SI Sec.~III.E yields the following expression for the lower bound on the ensemble averaged entanglement fidelity for the pair $jk$ as a function of the experimental time duration $\tau~(=t_{\text{dds}}+t_{\text{cmpl}})$\cite{googleSI_Programmable_Quantum_Matter_2025Google}:
\begin{widetext}
\begin{equation}
  \rule{0pt}{3ex} 
  \rule[-1.5ex]{0pt}{0pt} 
  \displaystyle
  \overline{\mathcal{F}_{jk}(\alpha,\tau)} \ge \bigg(1-\text{e}^{-\frac{\tau}{T_{1}}}\alpha - (1- \text{e}^{-\frac{\tau}{T_{ 
 1}}})~p_{\text{th}}(\tau)\bigg)\bigg(\frac{1 +\text{e}^{-((\tau/T_{2j})^{z_{j}}+(\tau/T_{2k})^{z_{k}})} - \sqrt{1 - \text{e}^{-2(\tau/T_{2j})^{z_{j}}}}\sqrt{1 - \text{e}^{-2(\tau/T_{2k})^{z_{k}}}}}{2}\bigg)
\end{equation}
\end{widetext}
Here the overline represents an ensemble average over multiple measurement attempts leading to averaging over the noise bath, 
and $p_{\text{th}}$ is the thermodynamic steady-state probabilities given by the following Boltzmann distribution:
\begin{equation}
    p_{\text{th}} = \frac{1}{1 + \text{e}^{-\hbar\omega / (k_B \Theta_{\text{SiV}})}}
\end{equation}
where $\hbar\omega$ is the energy of the qubit, and $k_{B}$ is the Boltzmann constant. All three parameters: $T_{2}, T_{1}, p_{\text{th}}$ depend on $\tau$ through the following mapping: $\{T_{2}, T_{1}, p_{\text{th}}\}\to\Theta_{\text{SiV}}\to~\tau$. Each measurement cycle, leads to a bipartite graph $\mathcal{G}_{\tau}$ with edges given by the set $\mathcal{A}_{\tau}$. Suppose that $\mathcal{G}_{\tau}$ is a union of $\mathcal{K}_{\tau}$ disconnected subgraphs, then we have the following expressions:
\begin{equation}
\begin{split}
\mathcal{G}_{\tau} &= \{ (j,k) \mid m_{jk} < M \,\}\\
\mathcal{G}_{\tau} &= \bigoplus_{i=1}^{\mathcal{K}_{\tau}} \mathcal{G}^i_{\tau}
\end{split}
\end{equation}
$M$ is the total number of attempts (Fig.~\ref{fig:overview}(b)), $\mathcal{G}^i_{\tau}$ is the $i^{th}$ connected subgraph of $\mathcal{G}_{\tau}$. We represent the set of edges and nodes in $\mathcal{G}^i_{\tau}$ by $\mathcal{E}^i_{\tau}$, $\mathcal{N}^i_{\tau}$ respectively. Each subgraph $\mathcal{G}^i_{\tau}$ acts as a quantum computational substrate with its quantum volume defined as \cite{Cross_2019, Moll2018_QST_VQA}:
\begin{equation}
\log_{2}(\text{V}^i_\tau) = 
\underset{2 \le n \le |\mathcal{N}^i_{\tau}|}{\text{argmax}}\, 
\min\left(n, \frac{1}{n\epsilon_{\text{eff}}}\right) 
\le 
\left\lfloor \frac{1}{\sqrt{\epsilon_{\text{eff}}}} \right\rfloor
\end{equation}
Here $\epsilon_{\text{eff}}$ is the average two-qubit gate error over all possible links ($j$-$k$) such that $\overline{\epsilon_{jk}}<0.5$ given by:
\begin{equation}
\begin{split}
  \epsilon_{\mathrm{eff}}
  &= \frac{1}{n_{\mathrm{links}}\bigl(t_{\mathrm{dds}},\,t_{\mathrm{cmpl}}\bigr)} \\[0.5ex]
  &\quad\times
    \sum_{\substack{(j,k)\in\mathcal{G}_{\tau} \\ \overline{\epsilon_{jk}}<0.5}}
      \overline{\epsilon_{jk}}\bigl(
        \alpha,\,
        t_{\mathrm{dds}},\,
        t_{\mathrm{cmpl}},\,
        N
      \bigr)
\end{split}
\label{eq:eps_eff}
\end{equation}
Here, $n_{\text{links}}(t_{\text{dds}},t_{\text{cmpl}})$ is the number of links such that $\overline{\epsilon_{jk}}<0.5$, $t_{\text{dds}}$ is the time window for decoupling sequences, $t_{\text{cmpl}}$ is the compilation time (Fig.~\ref{fig:overview}(b)), $N$ is the CPMG pulse index and $\epsilon_{jk}(\alpha,\tau)$ is an ensemble averaged two-qubit gate error on the link $j$-$k$ over experimental duration $\tau$. The expression for $\epsilon_{jk}$ and its temporal average $\overline{\epsilon_{jk}}$ is derived in SI Sec.~III.E-F ~\cite{googleSI_Programmable_Quantum_Matter_2025Google}:

\par\vspace{0.75\baselineskip} 

\begin{widetext}
  \begin{subequations}
  \label{Eq_e_ff}
\begin{align}
\epsilon_{jk}(\alpha, \tau,N) &=
 1 - \left(1 - \text{e}^{-\frac{\tau}{T_{1}}}\alpha 
- (1 - \text{e}^{-\frac{\tau}{T_{1}}})\,p_{\text{th}}\right)  \notag \\
&\quad  \times 
\left(\frac{1 +\text{e}^{-((\tau/T_{2j})^{z_{j}}+(\tau/T_{2k})^{z_{k}})} - \sqrt{1 - \text{e}^{-2(\tau/T_{2j})^{z_{j}}}}\sqrt{1 - \text{e}^{-2(\tau/T_{2k})^{z_{k}}}}}{2}\right) \notag \\
&\quad  + (2N+1)\epsilon_{1\text{-qubit}} \label{eq:eps_jk1}\\[1.5em]
\overline{\epsilon_{jk}}(\alpha, t_{\text{dds}}, t_{\text{cmpl}}, N) &= \frac{1}{t_{\text{cmpl}}}\int^{t_{\text{dds}}+t_{\text{cmpl}}}_{t_{\text{dds}}}\epsilon_{jk}(\alpha, \tau,N)~\text{d}\tau \label{eq:eps_jk2}
\end{align}
\end{subequations}
\end{widetext}
The first term in Eq. (\ref{eq:eps_jk1}) is the error due to entanglement (which we approximate as CNOT gate error) and the second term is the local unitary (single-qubit) error. Eq. (\ref{eq:eps_jk1}) follows from the universality theorem \cite{Nielsen_Chuang_2010} stating that any two qubit gate can be represented as a combination of a CNOT gate and single-qubit unitary. Since the entanglement process occurs stochastically in the time-window $t=(t_{\text{dds}}, t_{\text{dds}}+t_{\text{cmpl}})$, we take the temporal average of $\epsilon_{jk}(\alpha, \tau,N)$ over this window. We can then estimate $\epsilon_{\text{eff}}$ by plugging Eq. (\ref{eq:eps_jk1}) and (\ref{eq:eps_jk2}) in Eq. (\ref{eq:eps_eff}). We report two figures of merits for the cluster-state generation: $\epsilon_{\text{eff}}(t_{\text{dds}}, t_{\text{cmpl}}, N)$ and $n_{\text{links}}$.
\begin{table}[h]
\centering
\caption{Comparison of entanglement link statistics for  $t_{\text{dds}}=0.2$ ms and $t_{\text{cmpl}}=10~\mu$s.
}
\begin{tabular}{|c| c| c | c | c|}
\hline
$N$ & $n_{\text{links},\mathcal{A}}$ & $n_{\text{links},\mathcal{B}}$ & $\epsilon_{\text{eff},\mathcal{A}}$ & $\epsilon_{\text{eff},\mathcal{B}}$\\
\hline
\rule{0pt}{2.5ex}
2   & 10231  & 14641 & 0.19 & 0.29\\
\hline
\rule{0pt}{2.5ex}
4   & 9221  & 0 & 0.19 & 0.94\\
\hline
\rule{0pt}{2.5ex}
8   & 8029  & 0 & 0.23 & 1.0\\
\hline
\rule{0pt}{2.5ex}
16   & 5857  & 0 & 0.30 & 1.0\\
\hline
\end{tabular}
\label{tab:qv_comparison1}
\end{table}
\begin{table}[h]
\centering
\caption{Comparison of entanglement link statistics for  $t_{\text{dds}}=0.8$ ms and $t_{\text{cmpl}}=10~\mu$s.
}
\begin{tabular}{|c| c| c | c | c|}
\hline
$N$ & $n_{\text{links},\mathcal{A}}$ & $n_{\text{links},\mathcal{B}}$ & $\epsilon_{\text{eff},\mathcal{A}}$ & $\epsilon_{\text{eff},\mathcal{B}}$\\
\hline
\rule{0pt}{2.5ex}
2   & 0  & 0 & 1.0 & 0.999\\
\hline
\rule{0pt}{2.5ex}
4   & 552  & 0 & 0.39 & 0.9999\\
\hline
\rule{0pt}{2.5ex}
8   & 926  & 0 & 0.36 & 1.0\\
\hline
\rule{0pt}{2.5ex}
16   & 447  & 0 & 0.39 & 1.0\\
\hline
\end{tabular}
\label{tab:qv_comparison2}
\end{table}

We perform the entanglement error simulations over two systems (for $t_{\text{dds}}$ = $\{$0.2 ms, 0.8 ms$\}$ and $t_{\text{cmpl}}=10~\mu$s), each containing 121 grid-points (in $\epsilon$-$f$ space), implying that the total number of links are $121^2~(= 14641)$. For low values of $t_{\text{dds}}=0.2$ ms, we observe from Fig. \ref{fig:QV_sims}(a) that $n_{\text{links}}$ for $\mathcal{B}$ ($N=2$) starts off high at 14641 but sharply reduces to 0 for $N>2$, which is supported by Fig. \ref{fig:T2_sims}(d), where $\mathcal{B}$ is submerged in the shaded region for $N>2$. On the contrast, $\mathcal{A}$ varies from roughly 10000 to 6000 with increase in $N$. Thus, for $t_{\text{dds}}= 0.2$ ms, $n_{\text{links}}$ of $\mathcal{A}$ exceeds that of $\mathcal{B}$ by approximately $O(10^3\text{--}10^4)$. For $t_{\text{dds}}=0.8$ ms $n_{\text{links}}$ for $\mathcal{B}$ is zero for all $N$, whereas $n_{\text{links}}$ achieves an optimal value at $\sim~900$ for $\mathcal{A}_8$. This is shown in Fig. \ref{fig:QV_sims}(a-b). Fig. \ref{fig:QV_sims}(c-d) shows the variation of $\epsilon_{\text{eff}}$ with $N$, where the orange shaded area represents the region for which $\epsilon_{\text{eff}}>0.5$. All sequences within this region give $n_{\text{links}}=0$. For $t_{\text{dds}}=0.2$ ms $\mathcal{B}$ starts off outside this region but enters it for $N>2$, whereas $\mathcal{A}$ is consistently outside the shaded region up to $N =16$. For $t_{\text{dds}}=0.8$ ms, $\mathcal{B}$ is always inside the shaded region, whereas $\mathcal{A}$ starts from this region, but exits it for $N>2$. Moreover it attains an optimal value for $N=8$. The existence of an optimal $N$ (encircled in red) is again a signature of two competing effects as discussed previously. Thus, for $t_{\text{dds}}= 0.8$ ms, $n_{\text{links}}$ for $\mathcal{A}$ exceeds that of $\mathcal{B}$ by approximately $O(10^2\text{--}10^3)$. Table \ref{tab:qv_comparison1} and \ref{tab:qv_comparison2} summarize these values for the two sequences.

This section shows that the SAFE-GRAPE based entanglement protocol outperforms conventional bang-bang CPMG in terms of: the number of entanglement links (exceeding by $O(10^2\text{--}10^4)$, depending on the temporal window of the dynamical decoupling sequences) and feasibility ($\mathcal{B}$ is unable to create links with $\overline{\epsilon_{jk}}<0.5$ for larger values of $t_{\text{dds}}$).

\subsection{Programmability and Compilation}

\begin{figure*}
    \centering
    \includegraphics[width=\linewidth]{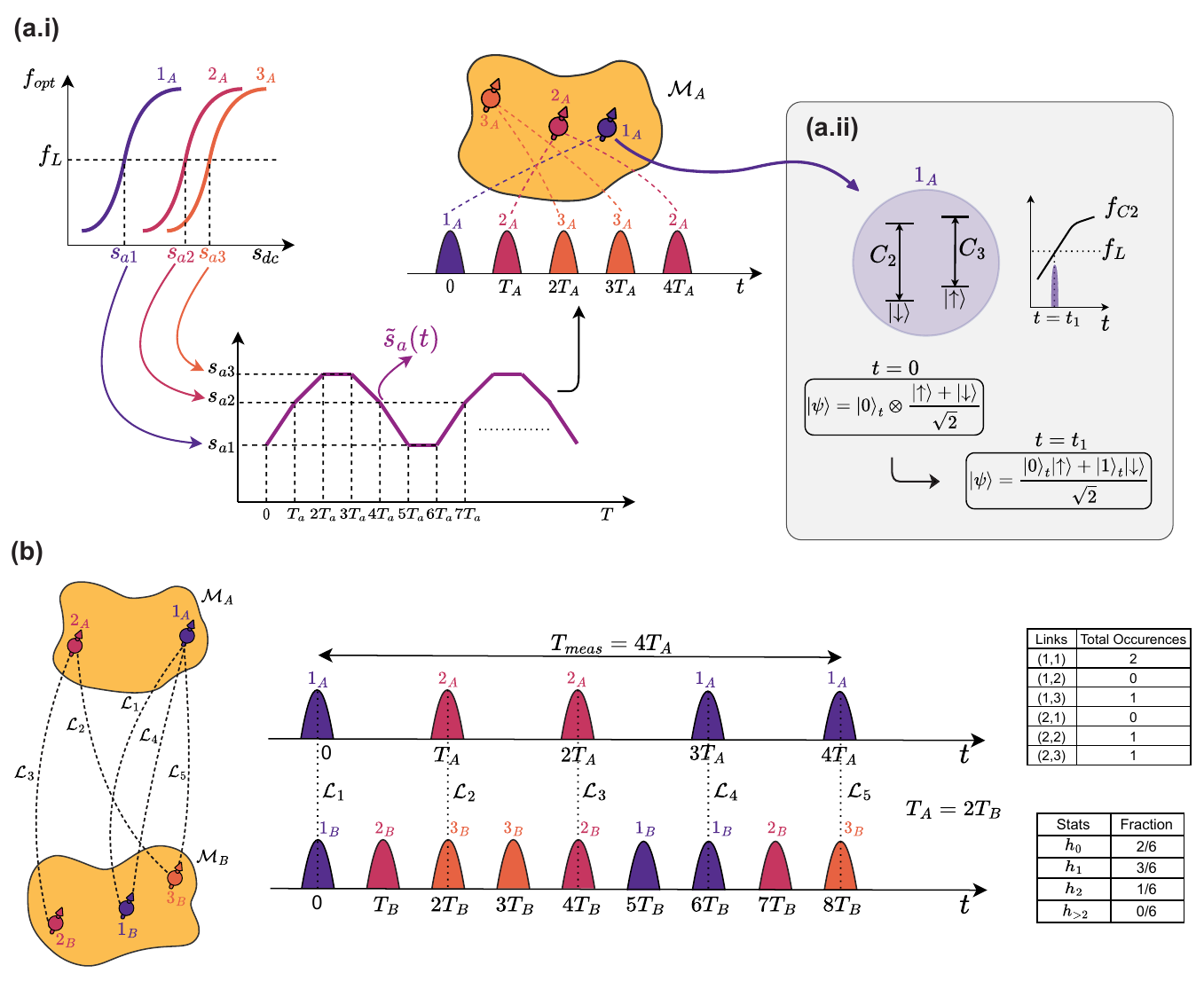}
    \caption{\textbf{Programmability and entanglement compilation}. \textbf{(a.i)} Schematic for Algorithm~\ref{alg:1} illustrated for three qubits (in purple, crimson and orange colors), where one first starts with the curve $f_{opt}$ (optical frequency) vs $s_{dc}$ (DC-strain), then maps it onto the time plot to construct the global strain drive $\tilde{s}_{a}(t)$, which leads to a train of spin-photon entanglements (photonic time-bin qubits are represented as colored pulses where the color matches that of the qubit with which it is entangled. \textbf{(a.ii)} Spin-photon entanglement generation per color-center at the timestep where laser frequency $f_{L}$ matches its optical transition frequency  \textbf{(b)} Train of entangled time-bin qubits arriving at the beam-splitter from two systems in case of $T_{A}$ = $2T_{B}$. Every temporal coincidence between two pulse trains means an entanglement attempt between the respective remote qubits (Bell-state projection). For example at $t=T_{A}=2T_{B}$, photonic qubit at $T_{A}$ (entangled with qubit $2_{A}$) temporally overlaps with photonic qubit at $2T_{B}$ (entangled with qubit $3_{B}$). Hence, this is an entanglement attempt between qubits $2_{A}$ and $3_{B}$ (represented in dotted line by link $\mathcal{L}_{2}$). For a measurement window of $4T_{A}$ 4 unique links are attempted, leaving behind 2 links for all-to-all connectivity. The table on the right shows the link statistics and the number $h_{j}$ represents the fraction of total possible links that occur $j$-times in a given window.}
    \label{fig:Programmability}
\end{figure*}

The task of programmability and compilation of color-center qubits relies on the assumption (see Sec.~\ref{sec:strain_window}) that there exists a strain regime where there is a one-to-one mapping $\mathbb{F}_{aj}$ from the strain applied on system A, to the optical frequency of qubit $j$ as follows:
\begin{equation}
    \begin{split}
        f_{j}(t) &= \mathbb{F}_{aj}(s_{a}(t))\\ 
        f_{j} &= \mathbb{F}_{aj}(s_{a})
    \end{split}
\end{equation}
where $f_{j}(t)$ is the time-dependent optical transition frequency of qubit $j$, due to the slow modulation in the strain $s_{a}(t)$. Assuming that we are in the strain regime where $\mathbb{F}_{aj}$ is invertible, we have the following:
\begin{equation}
    s_{a} = \mathbb{F}_{aj}^{-1}(f_{j})
\end{equation}
To define the strain which leads to the optical frequency equal to the laser frequency $f_{L}$, we adapt the following definition (Fig. \ref{fig:Programmability} (a.i)):
\begin{equation}
    s_{aj} = \mathbb{F}_{aj}^{-1}(f_{L})
\end{equation}
which leads to a set of strains $\{s_{aj}\}_{\substack{1\le j \le N_{a}}}$, where $N_{a(b)}$ is the number of qubits in system A(B). Without loss of generality, we can assume that the qubits are labeled such that the sequence is strictly monotonic:
\begin{equation}
    s_{a1} < s_{a2} <  \cdots < s_{aj} < \cdots < s_{aN_{a}}
\end{equation}
Fig \ref{fig:Programmability}(a.ii) shows the schematic for a laser interacting with a 4-level system to generate spin-photon entanglement, where the photonic qubit is encoded in the time-bin basis. The task of generating spin-photon entanglement at timestamps: \{$0, T_{a}, 2T_{a}, \cdots, (N_{a}-1)T_{a}$\} can be solved by finding a global drive $\tilde{s}_{a}(t)$ for system A which satisfies the following condition:
\begin{equation}
    \tilde{s}_{a}((k-1)\cdot T_{a}) = s_{ak}\hspace{20pt} \forall k\in\{1,N_{a}\}
\end{equation}
under the following constraints:
\begin{equation}
    \begin{split}
        &(i) \hspace{20pt} T_{a} > 1/\gamma_{opt}\\
        &(ii) \hspace{20pt} \frac{d\tilde{s}_{a}(t)}{dt} < s_{max}
    \end{split}
\end{equation}
The first constraint penalizes overlap between the emitted photons from different qubits, while the second constraint avoids sidebands due to strain modulation. We propose a simplistic solution based on Algorithm~\ref{alg:1}.

\begin{figure*}[t]
\algcaption{Computation of \( \tilde{s}_{a}(t) \)}
\begin{spacing}{1.2}
\label{alg:1}
\begin{algorithmic}[1]
    \State \textbf{Given sequences:} \( \{\mathbb{F}_{aj}\}_{1\le j\le N_{a}}, \{kT_{a}\}_{0\le k\le (N_{a}-1)} \)
    \State \textbf{Evaluate:} \( s_{aj} \gets \mathbb{F}_{aj}^{-1}(f_L) \)
    \State \textbf{Assume (w.l.o.g.):} The sequence \(\{ s_{aj} \}\) is strictly monotonic.
    \State \(\beta_j \gets \dfrac{s_{a,j+1} - s_{a,j}}{T_a}\)
    \State \(\beta^* \gets \max\{\beta_j\}\)
    \State \(\kappa \gets \dfrac{\beta^*}{s_{\max}}\)
    \State \( T_a \gets {T_a}(\operatorname{ReLU}(\kappa-1)+1) \)
    \For{\(1 \le j \le N_{a}-1\)}
        \If{\((j-1)T_a \le t \le jT_a\)}
            \State \( \tilde{s}_{a}(t) \gets \dfrac{s_{a,j+1} - s_{a,j}}{T_a}\, t + \Bigl(j\, s_{a,j} + (1-j)\, s_{a,j+1}\Bigr)\)\Comment{Linear interpolation}
        \EndIf
    \EndFor
    \For{\( N_{a}T_{a} \le t \le (2N_{a}-1)T_{a}\)}
        \State \( \tilde{s}_{a}(t) \gets \tilde{s}_{a}((2N_{a}-1)T_{a}-t) \)\Comment{Palindromic extension of $\tilde{s}_{a}(t)$}
    \EndFor
    \For{\( t \ge 2N_{a}T_{a} \)}
        \State \( q \gets \left\lfloor \frac{t}{2N_{a}T_{a}} \right\rfloor\ \)
        \State \( \tilde{s}_{a}(t) \gets \tilde{s}_{a}(t - 2N_{a}T_{a}q) \)\Comment{Periodic extension of $\tilde{s}_{a}(t)$}
    \EndFor
    \State \Return \( \tilde{s}_{a}(t)  \) 
\end{algorithmic}
\end{spacing}
\noindent\rule{\textwidth}{0.8pt}
\end{figure*}

We use the same algorithm to compute the global drive for system B, $\tilde{s}_{b}(t)$ corresponding to a mapping sequence: $\{\mathbb{F}_{bj}\}_{1\le j\le N_{b}}$ and time sequence $\{kT_{b}\}_{0\le k\le (N_{b}-1)}$. Because of the way our global drive is constructed, spin-photon entanglement is established at the following time sequences $\mathcal{T}_{A(B)} \equiv \{kT_{a(b)}\}_{k\in\mathbb{W}}$. Further, the spin-photon entanglement at timestep $\mathcal{T}_{A(B)}(j)$ is with the spin-qubit labeled $\theta_{ja(b)}$, where qubit sequence $\mathcal{Q}_{A(B)} \equiv \{\theta_{ja(b)}\}$ is also termed as a triangular-wave sequence which has the following compact form:
\begin{equation}
\begin{aligned}
    \theta_{ja(b)} =\ & 1 + \min\Bigl\{ j \bmod (2N_{a(b)}), \\
    & (2N_{a(b)} - 1) - \bigl(j \bmod (2N_{a(b)})\bigr) \Bigr\}
\end{aligned}
\label{eq:triangularwave}
\end{equation}
where $j$ is in the set of whole numbers.

The second stage of compilation builds on the two time sequences $\mathcal{T}_{A}$ and $\mathcal{T}_{B}$. This stage is for creating entanglement link attempts efficiently between system A and B, by overlapping the two timing sequences. We propose Algorithm~\ref{alg:2} for this stage. Depending on the global strain drive, we can get different qubit mapping sequences $\mathcal{Q}_{A(B)}$. Because of the construction of Algorithm \ref{alg:1} (pseudocode: lines 14 and 18), where the global drive is forced to be palindromic and periodic in nature, the qubit sequence it creates has a triangular wave-form structure as described in Eq. (\ref{eq:triangularwave}). This can be seen pictorially in Fig. \ref{fig:Programmability}(b). Algorithm \ref{alg:2} takes as an input these two qubit mapping sequences $\{\mathcal{Q}_{A},\mathcal{Q}_{B}\}$, and periodic timing sequences $\{\mathcal{T}_{A},\mathcal{T}_{B}\}$. It finds the scaling $m_{scal}$ (= $\frac{T_{A}}{T_{B}}$), for which the number of unique entanglement links is maximized. This is demonstrated with an example in Fig. \ref{fig:Programmability}(b). We define the metric $h_{j}$ as fraction of total possible links that occur $j$ times in a given window. We see that for $T_{meas}=4T_{A}=8T_{B}$, system A generates time-bin qubits $\{1_{A}, 2_{A}, 2_{A}, 1_{A}, 1_{A}\}$, whereas system B generates $\{1_{B}, 2_{B}, 3_{B}, 3_{B}, 2_{B}, 1_{B}, 1_{B}, 2_{B}, 3_{B}\}$. We see that temporal overlaps only lead to 5 entanglement attempts: $\{1_{A}-1_{B}, 2_{A}-3_{B}, 2_{A}-2_{B}, 1_{A}-1_{B}, 1_{A}-3_{B}\}$.\newline 
Out of these 5 attempts we get the following statistics:\\ 
\begin{itemize}
    \item links that occur exactly once:  $\{2_{A}-3_{B}, 2_{A}-2_{B}, 1_{A}-3_{B} \}\to h_{1} = 3/6$
    \item links that occur exactly twice:  $\{1_{A}-1_{B}\}\to h_{2} = 1/6$
    \item links that never occur:  $\{1_{A}-2_{B}, 2_{A}-1_{B}\}\to h_{0} = 2/6$
\end{itemize}
These links statistics are also summarized in Fig. \ref{fig:Programmability}(b). Because of the construction of Algorithm \ref{alg:1} and \ref{alg:2}, the fraction of attempted links $\sum_{i\ge1}h_{i}$ satisfies the following inequality:
\begin{equation}
    \sum_{i\ge1}h_{i} \ge \frac{\text{max}\{ N_{a},N_{b}\}}{N_{a}N_{b}}
\end{equation}
The above relation implies that using the combination of Algorithms~\ref{alg:1} and \ref{alg:2}, the number of unique entanglement links is bounded by order $\Omega(N_q)$. One can further improve this bound by sweeping on the value of $m_{scal}$.

In Fig. \ref{fig:Compiler_sims} we show a more thorough analysis of our compilation algorithm. For all combinations of $(N_a,N_b)$ qubits in the respective systems (illustrated for $N_b\le N_a\le10)$ we compute the optimal $m_{scal}$ that minimizes the fraction of missing links $h_0$ (Fig. \ref{fig:Compiler_sims}(a-b)). Fig. \ref{fig:Compiler_sims}(c) shows the minimum required measurement window to achieve the minimal $h_0$ in each of these scenarios. In Fig. \ref{fig:Compiler_sims}(d) we show for a specific configuration how the link statistics evolve over time. In this specific case, we find that after time $t=20T_A$ all possible links between qubits of both systems have been attempted. 

\begin{figure*}[t]
\algcaption{Entanglement Compiler}
\begin{spacing}{1.2}
\label{alg:2}
\begin{algorithmic}[1]
    \State \textbf{Given:} $\mathcal{Q}_{A}, \mathcal{Q}_{B}$, $T_{a}, T_{b}, N_{a}, N_{b}$
    \State \( E_{max} \gets N_{a}N_{b}    \) 
    \State \textbf{Assume:} $T_{a}=T_{b} = T_{0}$
    \For{\(1 \le m_{scal} \le \text{min}(N_{a}, N_{b})\)}
        \State \( T_{a} \gets m_{scal}T_{0}\)
        \State \( j_{\max} \gets \mathcal{J}(N_{a},N_{b},m_{scal})  \)\Comment{For derivation of $\mathcal{J}$, see Supplements}
        \For{\(0 \le j \le (j_{\max})\)}
            \State attempt entanglement link between qubits $\mathcal{Q}_{A}(j)$ and $\mathcal{Q}_{B}(m_{scal}j)$.
        \EndFor
        \State \( E(m_{scal}, N_{a}, N_{b}) \gets \text{total number of unique links made between} ~\mathcal{Q}_{A} ~\text{and}~ \mathcal{Q}_{B}\)
    \EndFor
\State \( m^{*}_{scal} \gets \text{min}\{\operatorname*{argmin}_{m_{scal}} E(m_{scal}, N_{a}, N_{b})\} \)
\State \( E^{*} \gets E(m^{*}_{scal}, N_{a}, N_{b}) \)\\
\Return \( m^{*}_{scal}, E^{*}  \)
\end{algorithmic}
\end{spacing}
\noindent\rule{\textwidth}{0.8pt}
\end{figure*}

\begin{figure*}
    \centering
    \includegraphics[width=0.7\linewidth]{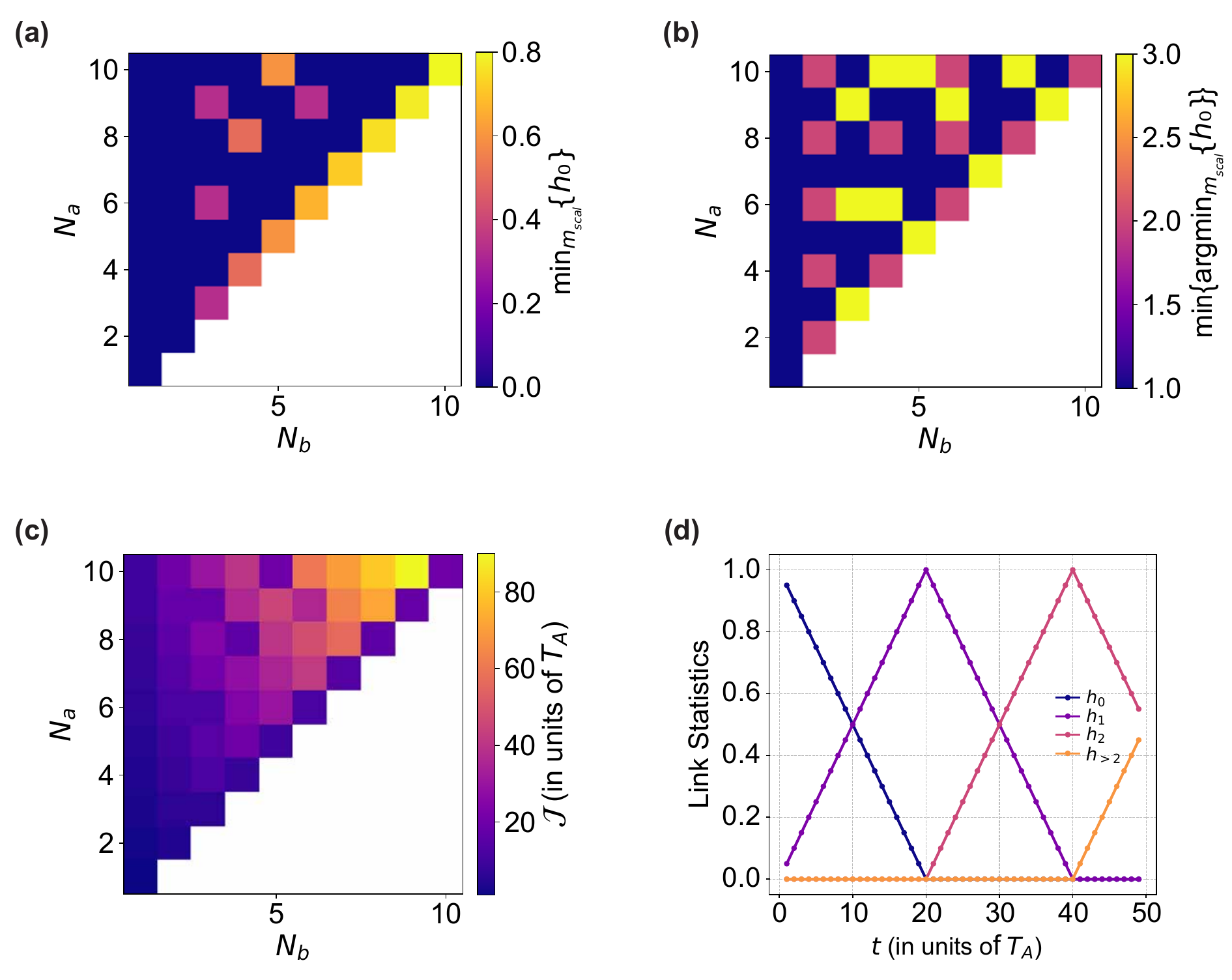}
    \caption{\textbf{Simulation for entanglement compilation (Algorithm~\ref{alg:2})}. Here the number $h_{j}$ represents the fraction of total possible links that occur $j$-times in an indefinite time-window, and $m_{scal}$ refers to the scaling between $T_{A}$ and $T_{B}$ (i.e. $T_{A} = m_{scal}T_{B}$). \textbf{(a)} Color map over the number of qubits in the two systems ($N_{a}$ vs $N_{b}$), where the color quantifies the minimum $h_{0}$ (i.e. the fraction of links which never get entangled) obtained over any integer scaling $m_{scal}$. Dominant blue squares suggest there are multiple configurations $(N_{a}, N_{b}, m_{scal})$ for which $h_{0}=0$, implying all-to-all connectivity. \textbf{(b)} The scaling $m_{scal}$ which minimizes $h_{0}$. \textbf{(c)} the minimum measurement window $t=\mathcal{J}T_{A}$, which minimizes $h_{0}$. \textbf{(d)} The trend for link statistics $h_{j}$ as the measurement window $t$ increases for the configuration $(N_{a}, N_{b}, m_{scal}) = (5,4,1)$.} 
    \label{fig:Compiler_sims}
\end{figure*}

\subsection{Working strain window for Algorithm \ref{alg:1}}
\label{sec:strain_window}

Algorithm~\ref{alg:1} requires a control parameter that
drives every optical qubit \emph{monotonically} and \emph{injectively} past a
global laser frequency \(f_L\).
We demonstrate that such a window exists on a state‑of‑the‑art solid‑state
platform \cite{PhysRevB.97.205444} by simulating the spin–conserving \(\mathrm{C}_2\) transition of
several \(\mathrm{SiV}^{-}\) centers in a nanomechanical structure, although the argument applies to any
emitter whose lowest‑order response to strain is linear.

For center \(i\) we decompose the strain that couples to the diamond
\(E_{gx}\) mode into three additive terms
\begin{equation}
\epsilon_{E_{gx},i}(t)=
\underbrace{\epsilon_{E_{gx},i}^{\text{bias}}}_{\text{fabrication}}
+
\underbrace{\epsilon_{E_{gx}}^{\text{dc}}}_{\text{piezo set–point}}
+
\underbrace{\epsilon_{E_{gx}}^{\text{ac}}(t)}_{\text{fast control}}
\end{equation}

so that in the low‑strain limit
(\(|\epsilon_{E_{gx}}|\lesssim10^{-4}\)) the optical frequency is

\begin{equation}
f_{\text{opt},i}
      =f_0+\Delta d\;\epsilon_{E_{gx},i}
\end{equation}
with
\begin{equation}
\Delta d=d_{es}-d_{gs}\simeq0.5~\mathrm{PHz/strain}
\end{equation}

The static biases \(\epsilon_{E_{gx},i}^{\text{bias}}\) are drawn from the
Gaussian distribution \(\mathcal N(0,\sigma_{\epsilon})\) with
\(\sigma_{\epsilon}=6\times10^{-5}\), derived from reported inhomogeneous C-transition spread in nanofabricated structures \cite{PhysRevB.97.205444, googleSI_Programmable_Quantum_Matter_2025Google}. For simplicity, we assume only transverse-oriented SiV centers with $\epsilon_{yy}\neq0,\
\epsilon_{xx}=\epsilon_{xz}=0$, simulated by diagonalizing the full
spin–orbit–Zeeman Hamiltonian with parameters detailed in SI Table~S1 \cite{googleSI_Programmable_Quantum_Matter_2025Google}.

\begin{figure}[tb]
  \centering
  \includegraphics[width=\columnwidth]{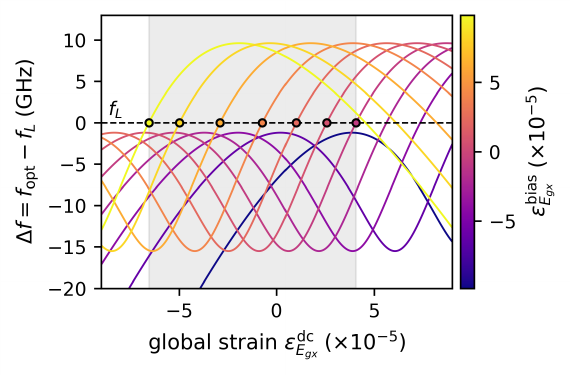}
  \caption{\textbf{Common monotonic strain window.}
  Calculated detuning \(\Delta f=f_{\text{opt}}-f_L\) for eleven
  \(\mathrm{SiV}^{-}\) centres whose bias strains (colour bar) span the
  \(\sigma_{\epsilon}=6\times10^{-5}\) distribution of
  Ref.\,\cite{PhysRevB.97.205444}, and arbitrary laser frequency.  Black circles mark laser crossings detected
  within \(|\Delta f|<0.2~\mathrm{GHz}\) ($C_2$ transition linewidth) and with positive local slope.  The shaded region extends from
  the earliest to the latest of these crossings and is truncated at the first
  additional (negative‑slope) intersection of the outer‑bias centre,
  guaranteeing that the mapping
  \(f_{\text{opt},i}=\mathbb{F}_{ai}(\epsilon_{E_{gx}}^{\text{dc}})\) is one‑to‑one for
  all qubits inside the band.}
  \label{fig:strain_window}
\end{figure}

Figure~\ref{fig:strain_window} shows the detuning
\(\Delta f=f_{\text{opt}}-f_L\) versus the piezo set‑point
\(\epsilon_{E_{gx}}^{\text{dc}}\).  We identify a strain regime (grey band) where every center
crosses \(f_L\) exactly once with a positive slope, in order to fulfil the assumptions of Algorithm~\ref{alg:1} for the
entire ensemble.  If a larger spectral spacing between crossings is required, one may simply omit the most crowded emitters without affecting the existence of the monotonic window.

\section{Conclusion and Outlook}

We introduce a hardware-agnostic control-and-compilation framework that turns fabrication-induced inhomogeneity into a programmable resource for scalable quantum information processing. A single, globally applied strain drive—optimized via SAFE-GRAPE (Simultaneous Amplitude and Frequency Error-correcting GRadient Ascent Pulse Engineering) on a composite-pulse basis—implements uniform high-fidelity single-qubit gates across heterogeneous SiV$^{-}$ ensembles. This approach corrects for significant variations in Rabi frequencies and spectral detunings to yield single-qubit gate fidelities exceeding 99.99\% for normalized (wrt. Rabi strength) errors up to 0.3. Simulation results demonstrate that SAFE-GRAPE-based CPMG sequence \( \mathcal{A} \) consistently outperforms alternative bang-bang based CPMG sequence \( \mathcal{B} \) across multiple performance metrics and operating regimes. In the low-frequency regime, \( \mathcal{A} \) exhibits stable filtering behavior under variations in a detuning parameter \(   f \), with a modest 4.5-fold increase in low-frequency response and $<$0.1\% variation in peak amplitude, compared to a 150-fold increase and 3\% amplitude degradation for \( \mathcal{B} \). This makes \( \mathcal{A}\sim 30 \) times more robust to parameter fluctuations, independent of the underlying noise spectrum.

In the decoupling regime, SAFE-GRAPE achieves a higher thermal robustness, with the heat-load parameter $\Theta_{\text{SiV}}$ increasing by approximately $5.2\%$ for protocol $\mathcal{B}$ and $0.3\%$ for protocol $\mathcal{A}$, while also delivering a significant coherence-time improvement, with $T_{2}$ enhancements reaching more than a factor of seven. Furthermore, unlike the bang-bang protocol, which ceases to suppress noise for large values of $t_{\text{dds}}$, SAFE-GRAPE maintains its feasibility, thereby enabling a phase-noise-reduced operational platform suitable for implementing the single-photon entanglement protocol.

In the entanglement generation stage, the SAFE-GRAPE based protocol outperforms the bang-bang CPMG approach by achieving a markedly higher number of successful entanglement links, exceeding by $O(10^2\text{--}10^4)$ depending on the temporal window of the decoupling sequences. This advantage is further reinforced by improved operational feasibility, as protocol $\mathcal{B}$ in the bang-bang case fails to generate high-fidelity links at large $t_{\text{dds}}$, whereas SAFE-GRAPE continues to operate effectively.  

Finally, the theoretical analysis indicates that by employing the combination of Algorithm \ref{alg:1} and Algorithm \ref{alg:2}, the number of unique entanglement links generated by the system is guaranteed to be bounded by order $\Omega(N_q)$. This bound is not fundamental and can be improved further by sweeping over the scaling parameter $m_{scal}$, offering a path towards even greater scalability. Collectively, these results establish SAFE-GRAPE as a robust and scalable control strategy for both noise suppression and high-rate entanglement generation, paving the way for its integration into large-scale quantum networking architectures. We also provide an end-to-end framework, summarized as an effective quantum circuit diagram in Fig.~\ref{fig:overview}(b), that delivers global control, dynamical decoupling, remote entanglement, and efficient compilation to build a bipartite cluster state.

Looking forward, the SAFE-GRAPE-optimized control pulses can be further improved by incorporating more advanced machine-learning-driven, in-situ optimization techniques to adapt to dynamic changes in the qubit environment. On the hardware front, the performance can be enhanced by engineering nanomechanical structures with optimized mode shapes \cite{raniwala2025spin}, enabling more efficient strain transfer and access to higher-frequency control regimes, although practical hurdles like charge-state instability will require parallel mitigation strategies.

The general nature of our framework ensures its applicability extends beyond the SiV$^{-}$ center in diamond. It can be readily adapted to other promising strain-sensitive platforms, such as other group-IV color centers in diamond (SnV$^{-}$ \cite{iwasaki2017tin,parker2024diamond,rugar2021quantum}, GeV$^{-}$ \cite{iwasaki2015germanium}) or defect centers in silicon \cite{prabhu_individually_2023, saggio_cavity-enhanced_2024, dhaliah2022first, buzzi2025spectral, PRXQuantum.5.010102, PRXQuantum.4.020308} or silicon carbide (silicon vacancy V$_{\text{Si}}$ \cite{sorman2000silicon}), thereby providing a general method for harnessing inhomogeneity in various solid-state systems.

Furthermore, the demonstrated global dynamical decoupling sequences can be repurposed for quantum sensing \cite{PhysRevLett.134.120802,ulanowski2025cavity} and single-photon detection applications \cite{https://doi.org/10.48550/arxiv.2401.10455} (see SI Sec.~V \cite{googleSI_Programmable_Quantum_Matter_2025Google}). Protecting the entire ensemble from decoherence simultaneously enhances its collective sensitivity to external fields, turning it into a parallelized quantum sensor.

Ultimately, this work lays the technical groundwork for measurement-based quantum processors built upon programmable quantum matter. By treating heterogeneity as a resource, our approach provides a practical roadmap towards networked demonstrations of multi-thousand-qubit cluster states, a critical step on the path to fault-tolerant quantum information systems.

\section{Code Availability}
The simulations were performed using \texttt{QuTiP} \cite{johansson2012qutip, johansson2013qutip, lambert2024qutip5quantumtoolbox} and \texttt{PyTorch} \cite{paszke2019pytorchimperativestylehighperformance} in Python. The codes can be found in our GitHub repository \cite{Git_Codes_2025}. 

\section{Acknowledgements}
The authors thank Dr. Matthew Trusheim, Dr. Hanfeng Wang and Dr. Mahmoud Jalali Mehrabad for proofreading the manuscript and offering constructive feedback. We also thank Hamza Raniwala and Dr. Ian Christen for contributing to the initial discussions of this project. We thank Dr. Ethan G. Arnault for helpful discussions regarding the thermal modeling. P.A. would like to thank Amit and Deepali Sinha Foundation Presidential Fellowship from MIT, MITRE Quantum
Moonshot Program, and NVIDIA Academic Grant Program. L.F. acknowledges support from the Air Force Office of Scientific Research (AFOSR) under Award No. GR108261. O.H. acknowledges support from the National Science Foundation (NSF) Engineering Research Center for Quantum Networks (CQN) awarded under cooperative agreement number 1941583.

\section{Author contributions}
D.R.E., P.A., L.F., and O.H. conceived the project. P.A. performed the simulations and theoretical work for filter-functions, thermal model, entanglement fidelity and compilation algorithms. L.F. performed simulations in strain modeling, thermal modeling, and entanglement protocol and contributed to the theory for the global unitary control and compilation algorithm. O.H. defined and implemented the SAFE-GRAPE algorithm for global unitary control and performed simulations for strain driving of SiV$^{-}$ centers. P.A., L.F., O.H. prepared the manuscript. All authors discussed the results and revised the manuscript. D.R.E. supervised the project.

\section{Competing interests}
The authors declare no competing interests.

\bibliography{bib}

\clearpage
\clearpage
\onecolumngrid 

\begin{center}
  \textbf{\large Supplementary Information for}\\[1ex]
  \textbf{\large ``Programmable Quantum Matter: Heralding Large Cluster States in Driven Inhomogeneous Spin Ensembles''}
\end{center}

\addcontentsline{toc}{section}{Supplementary Information}

\addtocontents{toc}{\protect\SIonlyTOCstart}

\begingroup
\makeatletter
  \newif\ifSIonly@print
  \SIonly@printfalse
  \def\SIonlyTOCstart{\SIonly@printtrue}

  \let\SIold@section\l@section
  \def\l@section#1#2{\ifSIonly@print \SIold@section{#1}{#2}\fi}

  \@ifundefined{l@subsection}{}{%
    \let\SIold@subsection\l@subsection
    \def\l@subsection#1#2{\ifSIonly@print \SIold@subsection{#1}{#2}\fi}
  }
  \@ifundefined{l@subsubsection}{}{%
    \let\SIold@subsubsection\l@subsubsection
    \def\l@subsubsection#1#2{\ifSIonly@print \SIold@subsubsection{#1}{#2}\fi}
  }

  \def\contentsname{Contents (Supplementary Information)}

  \tableofcontents
\makeatother
\endgroup

\setcounter{section}{0}
\setcounter{equation}{0}
\setcounter{figure}{0}
\setcounter{table}{0}

\renewcommand{\thesection}{S\arabic{section}}
\renewcommand{\thefigure}{S\arabic{figure}}
\renewcommand{\thetable}{S\arabic{table}}

\renewcommand{\theHsection}{S\arabic{section}}
\renewcommand{\theHfigure}{S\arabic{figure}}
\renewcommand{\theHtable}{S\arabic{table}}

\clearpage

\section{Strain Driving of Group-IV Color Centers in Diamond}




\subsection{Monotonic Strain Window for the \(\mathrm{SiV}^{-}\) Center}

As stated in the main text, Algorithm 1 requires a control parameter that drives every optical qubit monotonically and injectively past a global laser frequency $f_L$. This section details the simulation that confirms
such an operational window exists for an ensemble of negatively charged silicon-vacancy (\(\mathrm{SiV}^{-}\)) centers, a
state-of-the-art solid-state platform. The simulation models the spin-conserving $C_2$ optical transition of an
inhomogeneous \(\mathrm{SiV}^{-}\) ensemble under an applied quasi-static strain.

\subsubsection{Simplified Model for the \(\mathrm{SiV}^{-}\) Center}

\begin{figure}[h!]
    \centering
    \includegraphics[width=0.5\textwidth]{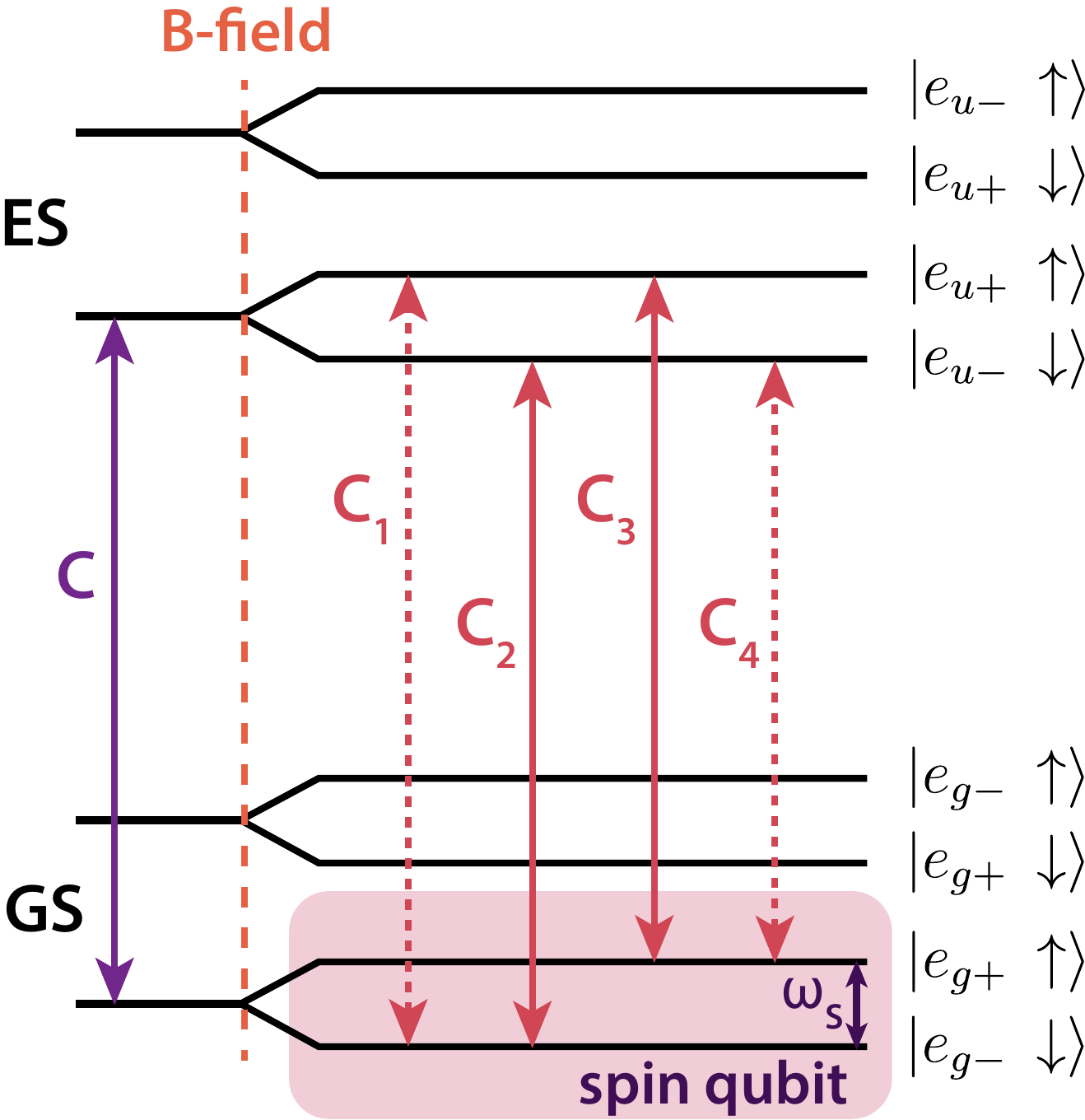}
    \caption{\textbf{Energy level diagram for the \(\mathrm{SiV}^{-}\) center.} The diagram illustrates the fundamental energy level structure of the \(\mathrm{SiV}^{-}\) center. The ground (GS) and excited (ES) state manifolds are split by the spin-orbit interaction. An external magnetic field further splits each orbital branch into spin-up ($\uparrow$) and spin-down ($\downarrow$) sublevels via the Zeeman effect. The two lowest-energy ground state sublevels form the spin qubit, with transition frequency $\omega_s$. The simulation specifically calculates the frequency of the spin-conserving $C_2$ optical transition.}
    \label{fig:siv_levels}
\end{figure}

We adopt a set of simplifying assumptions consistent with experimental work performed by Meesala, \textit{et al.} \cite{PhysRevB.97.205444}: the model considers only transverse-oriented emitters where the strain tensor is assumed to be dominated by the $\epsilon_{yy}$ component. The energy level diagram for the \(\mathrm{SiV}^{-}\), showing the ground state (GS), excited state (ES), and the relevant optical transitions (\(\mathrm{C}_1\)-\(\mathrm{C}_4\)), is shown in Figure~\ref{fig:siv_levels}.


Under the additional assumption of the specific strain environment, the Hamiltonian governing these levels simplifies significantly. The SO interaction is characterized by the SO coupling strength $\lambda_{\text{SO}}^{\text{GS/ES}}$. The Zeeman interaction describes the coupling of the orbital and spin angular momenta to a uniform magnetic field $\Vec{B}=B_x \Vec{1}_x + B_z \Vec{1}_z$ through their respective gyromagnetic ratios $\gamma_L$ and $\gamma_S$. Finally, this yields the following Hamiltonian, written in the basis of spin-orbit eigenstates $\{ |e_{-} \downarrow\rangle , |e_{+} \uparrow\rangle , |e_{+} \downarrow\rangle , |e_{-} \uparrow\rangle \}$ for a given manifold (GS/ES) and center $i$:
\begin{align}
 & \hat{\mathcal{H}}_i^{\text{GS/ES}} \nonumber \\
 & =
 \begin{pmatrix}
    - \frac{\lambda_{\text{SO}}}{2} - \gamma_L B_z - \gamma_S B_z & 0 & \epsilon_{E_{gx}} & \gamma_S B_x \\
    0 & - \frac{\lambda_{\text{SO}}}{2} + \gamma_L B_z + \gamma_S B_z & \gamma_S B_x & \epsilon_{E_{gx}} \\
    \epsilon_{E_{gx}} & \gamma_S B_x & \frac{\lambda_{\text{SO}}}{2} + \gamma_L B_z - \gamma_S B_z & 0 \\
    \gamma_S B_x & \epsilon_{E_{gx}} & 0 & \frac{\lambda_{\text{SO}}}{2} - \gamma_L B_z + \gamma_S B_z 
 \end{pmatrix}
\end{align}
Here, the primary effect of strain is captured by the off-diagonal $\epsilon_{E_{gx}}$ term, approximated by $\epsilon_{E_{gx}}=- d_{gs/es}\, \epsilon_{yy}$.

\subsubsection{Simulation and Parameters}
To accurately model the inhomogeneous ensemble, we must define a statistical distribution of the static fabrication biases, $\epsilon_{E_{gx},i}^{\text{bias}}$. We derive the the standard deviation, $\sigma_\epsilon$, of this distribution from experimentally measured inhomogeneous broadening of the \(\mathrm{SiV}^{-}\) C-transition, which is reported by Meesala, \textit{et~al.}~\cite{PhysRevB.97.205444} to have a standard deviation of $\sigma_f \approx 31$~GHz for centers in nanofabricated devices. In the low-strain limit, the optical frequency shift is approximately linear with strain:
\begin{equation}
    \Delta f \approx (d_{es} - d_{gs}) \epsilon_{E_{gx}} \equiv \Delta d \cdot \epsilon_{E_{gx}}
\end{equation}

Using the susceptibility parameters from Table~\ref{tab:simparams}, the differential susceptibility is $\Delta d = (1.8 - 1.3) \times 10^{15}$~Hz/strain = 0.5~PHz/strain. This allows us to estimate the standard deviation of the underlying strain distribution:
\begin{equation}
    \sigma_{\epsilon} = \frac{\sigma_f}{\Delta d} = \frac{31 \times 10^9 \text{ Hz}}{0.5 \times 10^{15} \text{ Hz/strain}} \approx 6 \times 10^{-5}
\end{equation}
Based on this calculation, we adopt a rounded value of $\sigma_\epsilon = 6 \times 10^{-5}$ for the simulation. The static biases $\epsilon_{E_{gx},i}^{\text{bias}}$ are therefore drawn from the Gaussian distribution $\mathcal{N}(0, \sigma_\epsilon)$. The frequency of the \(\mathrm{C}_2\) transition is then computed for each center by numerically diagonalizing the simplified Hamiltonian for the total strain $\epsilon_{E_{gx},i} = \epsilon_{E_{gx},i}^{\text{bias}} + \epsilon_{E_{gx}}^{\text{dc}}$ and parameter values from Table \ref{tab:simparams}. The results, plotted as the detuning $\Delta f$ in Figure 10 of the main text, confirm that a common monotonic window exists for the simulated inhomogeneous ensemble.

\begingroup
 \renewcommand{\thetable}{S\arabic{table}}
 \setcounter{table}{0}
 \begin{table}[h!]
    \centering
    \vspace{4pt}
    \begin{tabular}{|l|c|l|}
      \hline
      \textbf{Quantity} & \textbf{Symbol} & \textbf{Value} \\ \hline
      spin–orbit split (ground)   & \(\lambda^{\text{gs}}_{\text{SO}}\) & \(46~\mathrm{GHz}\) \\
      spin–orbit split (excited)  & \(\lambda^{\text{es}}_{\text{SO}}\) & \(255~\mathrm{GHz}\) \\
      orbital \(g\)--factor       & \(g_L\)                         & \(1.4\) \\
      spin \(g\)--factor          & \(g_S\)                         & \(14\) \\
      magnetic field              & \(B\)                           & \(0.17~\mathrm{T}\) (in the \(xz\) bisector) \\
      strain coefficient (GS)     & \(d_{gs}\)                      & \(1.3\times10^{15}~\mathrm{Hz/strain}\) \\
      strain coefficient (ES)     & \(d_{es}\)                      & \(1.8\times10^{15}~\mathrm{Hz/strain}\) \\
      bias strain distribution, std. dev. & \(\sigma_\epsilon\)             & \(6\times10^{-5}\) \\
      assumed strain components   & ---                             & \(\epsilon_{yy}\neq0,\; \epsilon_{xx}=\epsilon_{xz}=0\) \\
      \hline
    \end{tabular}
    \caption{Parameters used in the \(\mathrm{SiV}^{-}\)-specific simulation. Adapted from \cite{PhysRevB.97.205444}.}
    \label{tab:simparams}
 \end{table}
\endgroup

\subsection{Strain Driving Simulations}
\label{sec:straindriving}

\begin{figure}[h!]
    \centering
    \includegraphics[width=\textwidth]{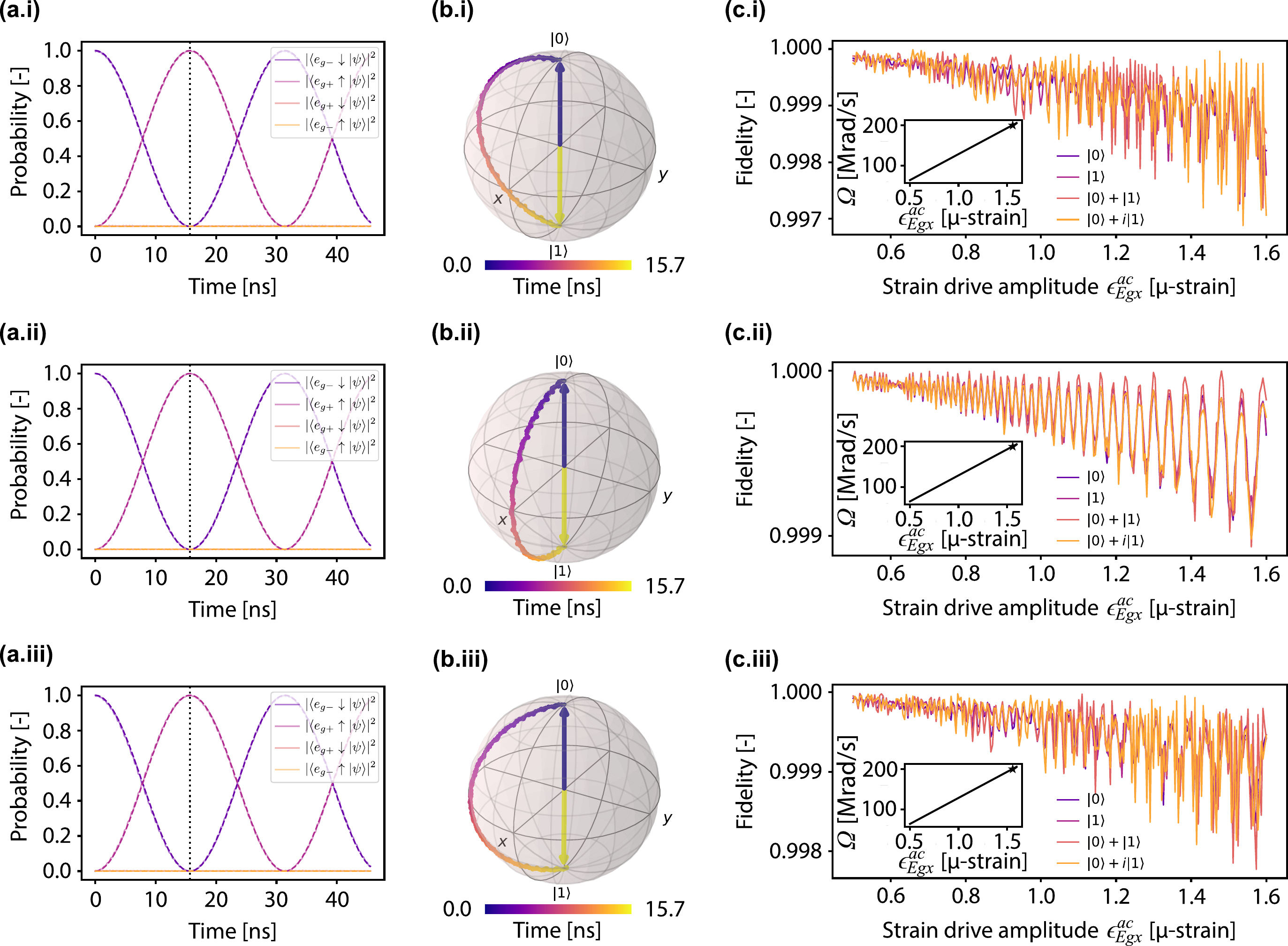}
    \caption{\textbf{Arbitrary single-qubit gates $\mathbf{\hat{\mathcal{U}}(\theta,\phi)}$ implemented by strain driving.} \textbf{(a)} System evolution under strain drive Hamiltonian $\hat{\mathcal{H}}^{\text{drive}}$ for initial state $|0\rangle = |e_{g-}\downarrow\rangle$. Rabi oscillations between the eigenstates of the bottom two energy levels can be observed. The colored dashed lines correspond to the best ideal Rabi oscillations fit. The strain drive amplitude $\epsilon_{Egx}^{ac}$ is chosen to realize a Rabi drive strength $\Omega$ of 200 Mrad/s. The black vertical dotted line indicates the time at which a $\pi$ rotation is achieved. \textbf{(b)} Bloch sphere representation of the time evolution during this $\pi$ rotation gate. The arrows mark the initial and final state. \textbf{(c)} State fidelity between the state after evolution under the strain drive Hamiltonian and the ideal final state, for different initial states. \textbf{(inset of (c))} Optimal fit of the Rabi drive strength for a set strain drive amplitude. The star indicates the point that realizes $\Omega$ of 200 Mrad/s. Note that arbitrary single-qubit gates can be implemented. Here the results are shown for \textbf{(i)} rotations around the x-axis ($\hat{\mathcal{U}}(\theta,\phi=0)$) implemented by $\hat{\mathcal{H}}^{\text{drive}}(\varphi=\pi)$, \textbf{(ii)} rotations around the y-axis ($\hat{\mathcal{U}}(\theta,\phi=\pi/2)$) implemented by $\hat{\mathcal{H}}^{\text{drive}}(\varphi=\pi/2)$ and \textbf{(iii)} rotations around the xy-bisector ($\hat{\mathcal{U}}(\theta,\phi=\pi/4)$) implemented by $\hat{\mathcal{H}}^{\text{drive}}(\varphi=3\pi/4)$.},
    \label{fig:strain_driving}
\end{figure}

Strain driving \(\hat{\mathcal{H}}^{\text{drive}}(\varphi=\pi-\phi)\) can be used to implement arbitrary single-qubit gates $\hat{\mathcal{U}}^{\text{ideal}}(\theta,\phi)$ on the spin qubit defined by the \(\mathrm{SiV}^{-}\) $\hat{\mathcal{H}}^{\text{GS}}$ lowest two eigenstates. The parameter $\theta$ is determined by the Rabi drive strength $\Omega$ and the evolution time. The parameter $\phi$ is controlled using the phase $\varphi$ of the strain drive. Fig~\ref{fig:strain_driving} illustrates three cases of those arbitrary gates: rotation on the Bloch sphere around the x-axis (\ref{fig:strain_driving}.i), y-axis (\ref{fig:strain_driving}.ii) and xy-bisector (\ref{fig:strain_driving}.iii). Here, the static strain $\epsilon_{Egx}^{\text{dc}}$ is set to $4\times10^{-6}$ and the applied magnetic field $\vec{B}$ to $0.25\text{T}\times\frac{1}{\sqrt{2}}(\vec{1}_x+\vec{1}_z)$. In all three cases, fidelities above 99.7\% are achieved for various initial states  (FIG. \ref{fig:strain_driving} (c)). FIG. \ref{fig:strain_driving} (a) shows that Rabi oscillations are induced by strain driving after initializing the qubit in its ground state. The strain drive amplitude $\epsilon_{Egx}^{\text{ac}}$ is picked as $1.560\times10^{-6}$, $1.560\times10^{-6}$ and $1.553\times10^{-6}$ respectively, in order to realize $\Omega=200$ Mrad/s. The slight population of the two upper levels contributes to the imperfect fidelity of the Rabi oscillations between the two lowest energy levels. 

\section{Concatenated Composite Pulses}

\subsection{Concept}

\begin{figure}[h]
    \centering
    \includegraphics[width=\linewidth]{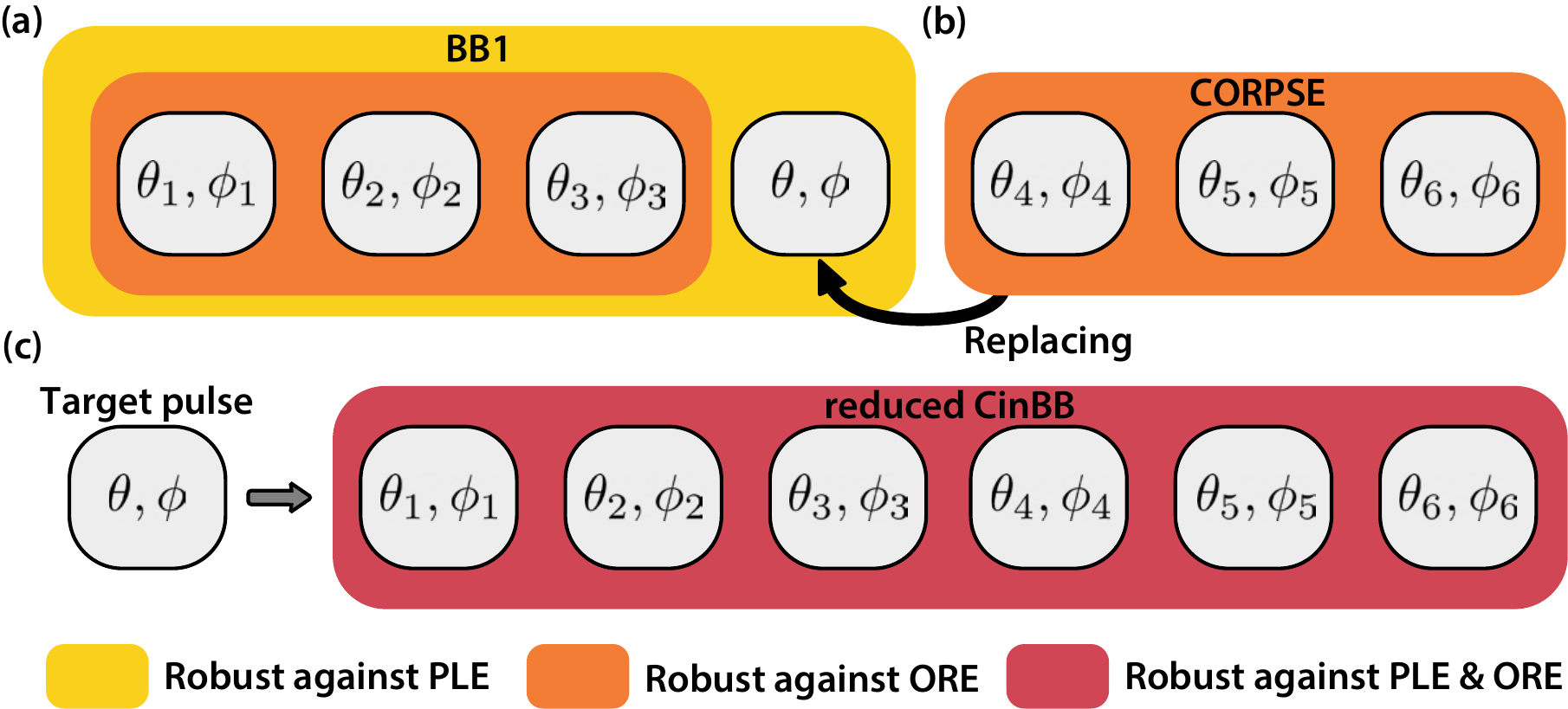}
    \caption{\textbf{Schematic construction of the six‐pulse \textit{r}CinBB composite pulse sequence.}
             (a) Four‐pulse BB1 outer sequence cancels PLE.
             (b) Three‐pulse CORPSE block cancels ORE and is REP‐PLE.
             (c) Substituting only the fourth BB1 rotation yields the
             six‐pulse rCinBB, first‐order robust to both PLE and ORE.}
    \label{fig:schematic_rCinBB}
  \end{figure}

Composite-pulse control replaces an elementary single-qubit gate
\(\hat{\mathcal{U}}_{\mathrm{ideal}}(\theta,\phi)\), defined as
\[
\hat{\mathcal{U}}_{\mathrm{ideal}}(\theta,\phi)= \exp\!\left[-\tfrac{i}{2}\theta\left(\cos\phi\,\hat\sigma_x + \sin\phi\,\hat\sigma_y\right)\right],
\]
with a short sequence of rotations
\(R(\theta_i,\phi_i)\)
whose combined propagator suppresses systematic errors to first order.
In a concatenated composite pulse (CCCP) \cite{Bando2012Concatenated} an \emph{outer} composite
sequence that cancels one error (here the amplitude / pulse-length error, PLE) is
nested with an \emph{inner} sequence that cancels the complementary
off-resonance error (ORE) while preserving the residual effect of PLE to
first order, the residual-error-preserving (REP) property.
Because the inner block then behaves like an effective single rotation
for PLE, the concatenation is first-order robust to both PLE and ORE.
The construction is summarised schematically in Fig.~\ref{fig:schematic_rCinBB}.

\subsection{Reduced CORPSE-in-BB1 (rCinBB) Sequence}

The four-pulse BB1 outer sequence \cite{WIMPERIS1994221} is written
\[
\hat{\mathcal{U}}_{\mathrm{BB1}}(\theta,\phi)=
\prod_{i=1}^{4} R(\theta_i,\phi_i),
\]
where \((\theta_1,\phi_1)\)–\((\theta_3,\phi_3)\) are given in
Table~\ref{tab:S2_theta_phi_CCCP} and \((\theta_4,\phi_4)=(\theta,\phi)\).
The three-pulse CORPSE inner block \cite{PhysRevA.67.042308},
\(\hat{\mathcal{U}}_{\mathrm{CORPSE}}(\theta,\phi)
 =R(\theta_4,\phi_4)R(\theta_5,\phi_5)R(\theta_6,\phi_6)\),
cancels ORE and is REP-PLE.
Because the first three BB1 rotations already form a trivial
\(\pi\text{–}2\pi\text{–}\pi\) block that is REP-ORE, only the
\emph{fourth} BB1 rotation must be replaced by CORPSE.
The resulting six-pulse reduced CinBB operator is therefore
\[
\hat{\mathcal{U}}_{\text{rCinBB}}(\theta,\phi)=
\prod_{i=1}^{6} R(\theta_i,\phi_i),
\]
with parameters collected in Table~\ref{tab:S2_theta_phi_CCCP}.
Compared to replacing every BB1 rotation, the reduced sequence halves the gate duration while
retaining simultaneous first-order cancellation of PLE and ORE. \cite{Bando2012Concatenated}

\begingroup
  \renewcommand{\thetable}{S\arabic{table}}

  \begin{table}
    \centering
    \begin{tabular}{|>{\centering\arraybackslash}p{0.15\textwidth}|
                    >{\centering\arraybackslash}p{0.15\textwidth}|}
      \hline
      $\theta_1 = \pi$                          & $\phi_1 = \phi + s$   \\[1ex]
      $\theta_2 = 2\pi$                        & $\phi_2 = \phi - 2\,s$ \\[1ex]
      $\theta_3 = \pi$                          & $\phi_3 = \phi + s$   \\[1ex]
      $\theta_4 = 2\pi + \tfrac{\theta}{2} - k$ & $\phi_4 = \phi$       \\[1ex]
      $\theta_5 = 2\pi - 2\,k$                  & $\phi_5 = \phi + \pi$ \\[1ex]
      $\theta_6 = \tfrac{\theta}{2} - k$        & $\phi_6 = \phi$       \\ \hline
      \multicolumn{2}{|c|}{%
        \begin{tabular}{c}
          $k = \arcsin\!\displaystyle\Bigl(\tfrac{\sin(\theta/2)}{2}\Bigr)$ \\[0.75ex]
          $s = \arccos\!\displaystyle\Bigl(-\tfrac{\theta}{4\pi}\Bigr)$
        \end{tabular}
      } \\ \hline
    \end{tabular}
    \caption{\textbf{Rotation parameters for the six-pulse rCinBB sequence.} The angles $\theta_i$ and phases $\phi_i$ define the elementary rotations $R(\theta_i,\phi_i)$ that compose the target gate $\hat{\mathcal{U}}_{\mathrm{ideal}}(\theta,\phi)$.}
    \label{tab:S2_theta_phi_CCCP}
  \end{table}
\endgroup

\subsection{SAFE-GRAPE Parameters}

Table \ref{tab:GRAPE_parameters} contains the parameters for the SAFE-GRAPE algorithm (adapted from \cite{khaneja_optimal_2005}) used in the main text. The set of trainable parameters $\{t_i , \phi_i \}_{i\in [1...N_p]}$ is initialized based on the rCinBB pulse sequence. All $t_i$ are set equal to the fraction of the rCinBB total pulse duration for chosen $\Omega$ over the number of time-bins $N_p$. Each $\phi_i$ is assigned the corresponding rCinBB phase $\in\{\phi_1, ...,\phi_6\}$ for time-bin $i$.  

\begingroup
    \renewcommand{\thetable}{S\arabic{table}}
    \begin{table}[]
        \begin{tabular}{|c|c|c|c|}
        \hline
        \textbf{General}   & \textbf{\begin{tabular}[c]{@{}c@{}}Discretised composite\\[-1ex] search space\end{tabular}} & \textbf{\begin{tabular}[c]{@{}c@{}}Gaussian\\[-1ex] weight factor\end{tabular}} & \textbf{\begin{tabular}[c]{@{}c@{}}Sigmoid\\[-1ex] reparametrization\end{tabular}} \\ \hline
        $\theta=\pi$       & $N_{\epsilon} = N_f=11$         & $\mathcal{N}=1.90$                & $t_{min}=10^{-9}$ s                \\
        $\phi=0$           & $\epsilon_{min} = f_{min}=-0.3$ & $\bar{\epsilon}=\bar{f}=0$        & $t_{max}=10^{-8}$ s                \\
        $\Omega=200$ Mrad/s & $\epsilon_{max} = f_{max}=0.3$  & $\sigma_{\epsilon}=\sigma_f=0.22$ &                                    \\
        $N_p=100$          &                                 &                                   &                                    \\ \hline
        \end{tabular}
        \caption{\textbf{SAFE-GRAPE on rCinBB optimization parameters}}
        \label{tab:GRAPE_parameters}
    \end{table}

\section{Fidelity Calculation}
In order to estimate the fidelity of the links, we divide the process into four steps:\\
(A) Single-Photon Entanglement Protocol \cite{Hermans_2023}\\
(B-D) Dynamical Decoupling Sequence as a Dephasing Channel \cite{PhysRev.94.630, meiboom_modified_1958}\\
(E) Composing the Entanglement Protocol with Dynamical Decoupling\\
(F) Including the thermal budget\\

\begin{figure*}
    \centering
    \includegraphics[width=0.9\linewidth]{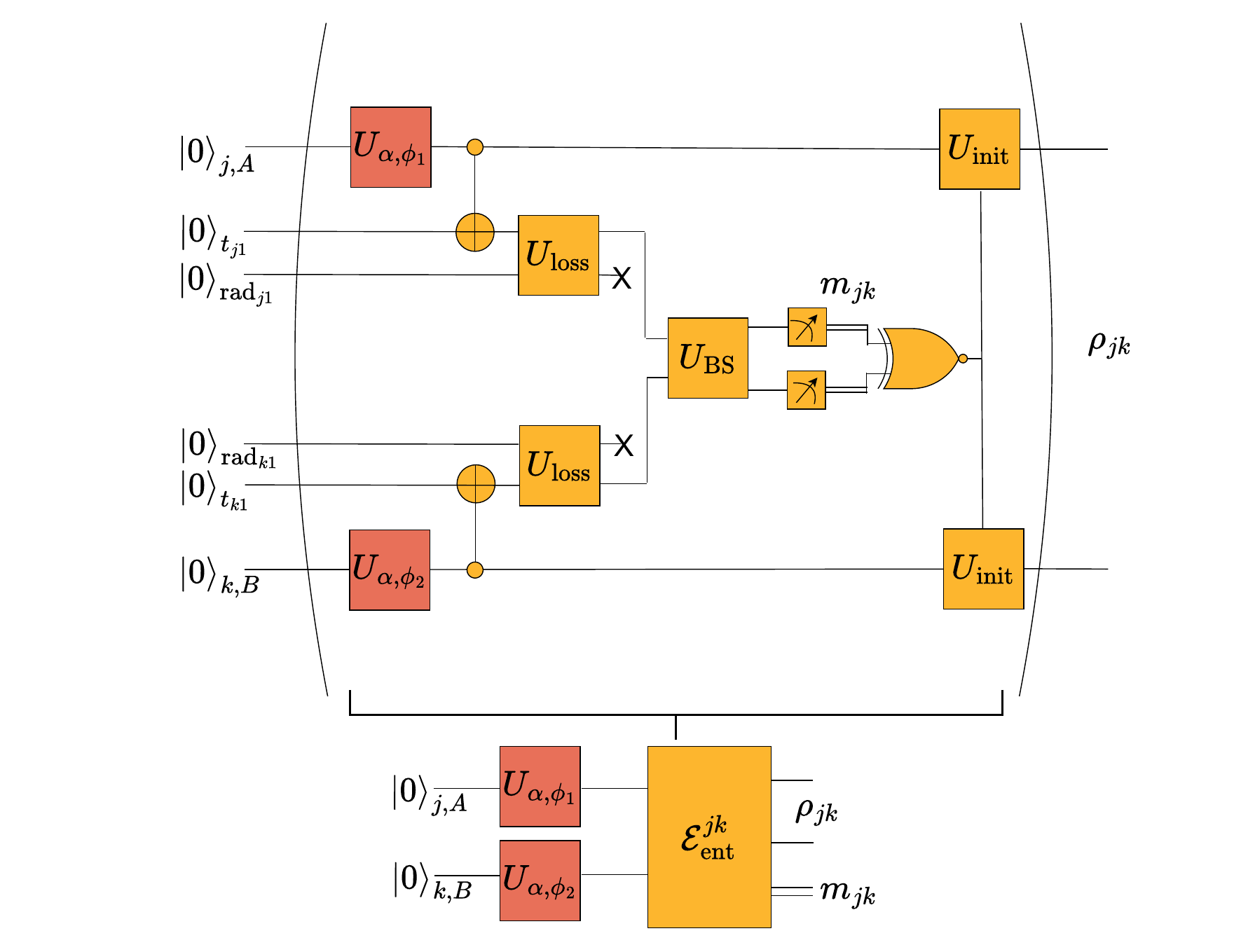}
    \caption{\textbf{Single-Photon Entanglement Protocol.} Quantum Circuit Representation to perform remote entanglement between spin qubits $\ket{0}_{j,A}$ and $\ket{0}_{k,B}$. $\ket{0}_{t}$ and $\ket{0}_{\text{rad}}$ are the vacuum modes for the time-bin and radiative mode respectively. Inverted CNOT is physically realized by laser based spin selective excitation of the spin ground states. Following are the representation of the blocks - $U_{\alpha,\phi}$: Initialization unitary, $U_{\text{loss}}$: Loss channel, $U_{\text{BS}}$: Beam Splitter, $U_{\text{init}}$: measurement controlled laser based spin initialization. $m_{jk}$ is a two bit click result, where 0(1) means absence (presence) of detector click.}
    \label{fig:SPE}
\end{figure*}

\subsection{Single-Photon Entanglement Protocol}
Fig.~\ref{fig:SPE} shows the quantum circuit representation \cite{Nielsen_Chuang_2010} of the N-cycle single-photon entanglement protocol. We start the protocol between qubits $j$ and $k$, from system A and B respectively, in their respective ground states $\ket{0}_{j,A}$ and $\ket{0}_{k,B}$. The states $\ket{0}_{t_{jm}}$ and $\ket{0}_{t_{km}}$ represents the zero photon in the time-bins $t_{jm}$ and $t_{km}$ respectively, for the $m^{\text{th}}$ attempt. Similarly, the states $\ket{0}_{\text{rad}_{jm}}$ and $\ket{0}_{\text{rad}_{km}}$ represents the zero radiative photons for the qubits $j$ and $k$ respectively, for the $m^{\text{th}}$ attempt. Thus, the initial state at the first attempt is given by:
\begin{equation}
    \ket{\psi}_{in} = \ket{0}_{j,A}\otimes\ket{0}_{t_{j1}}\otimes\ket{0}_{\text{rad}_{j1}}\otimes\ket{0}_{k,B}\otimes\ket{0}_{t_{k1}}\otimes\ket{0}_{\text{rad}_{k1}}
\end{equation}
We use the following two compact notations:
\begin{subequations}
\begin{align}
    \ket{\psi}_{in} &= \ket{00}_{jk}\otimes\ket{00}_{t}\otimes \ket{00}_{\text{rad}}\\
    &= \ket{000}_{j,t,\text{rad}}\otimes\ket{000}_{k,t,\text{rad}}
\end{align}    
\end{subequations}
where in the first notation: the three different Hilbert spaces correspond to two qubits, two time bins, and two radiative modes respectively and in the second notation: the two different Hilbert spaces correspond to the two systems. We apply the arbitrary unitary $U_{\alpha,\phi}$ on $\ket{\psi}_{in}$, whose action on the ground state is defined as:
\begin{equation}
    U_{\alpha,\phi}:\ket{0}_{j}\to \sqrt{\alpha}\ket{0}_{j} + \text{e}^{i\phi}\sqrt{1-\alpha}\ket{1}_{j}
\end{equation}
This results in the following:
\begin{subequations}
\begin{align}
     &(U_{\alpha,\phi_{1}}\otimes U_{\alpha,\phi_{2}})\ket{\psi}_{in} \nonumber \\
     &= (\sqrt{\alpha}\ket{0}_{j} + \text{e}^{i\phi_{1}}\sqrt{1-\alpha}\ket{1}_{j})\otimes(\sqrt{\alpha}\ket{0}_{k} + \text{e}^{i\phi_{2}}\sqrt{1-\alpha}\ket{1}_{k})\otimes\ket{00}_{t}\otimes\ket{00}_{\text{rad}}\\
     &= (\sqrt{\alpha}\ket{000}_{j,t,\text{rad}} + \text{e}^{i\phi_{1}}\sqrt{1-\alpha}\ket{100}_{j,t,\text{rad}})\otimes (\sqrt{\alpha}\ket{000}_{k,t,\text{rad}} + \text{e}^{i\phi_{2}}\sqrt{1-\alpha}\ket{100}_{k,t,\text{rad}})
\end{align}
\end{subequations}
Next, we apply a CNOT gate on the two systems A and B. This corresponds to laser mediated state selective excitation.  We call the resulting state $\ket{\psi}_{2}$. Each CNOT acts between the spin qubit and time-bin qubit, which gives the following:
\begin{subequations}
    \begin{align}
        \ket{\psi}_{2} =& (\text{CNOT}_{A})\otimes (\text{CNOT}_{B})(U_{\alpha,\phi_{1}}\otimes U_{\alpha,\phi_{2}})\ket{\psi}_{in}\\
        =& (\text{CNOT}_{A})(\sqrt{\alpha}\ket{000}_{j,t,\text{rad}} + \text{e}^{i\phi_{1}}\sqrt{1-\alpha}\ket{100}_{j,t,\text{rad}})\nonumber\\
        &\otimes (\text{CNOT}_{B})(\sqrt{\alpha}\ket{000}_{k,t,\text{rad}} + \text{e}^{i\phi_{2}}\sqrt{1-\alpha}\ket{100}_{k,t,\text{rad}})\\
        =& (\sqrt{\alpha}\ket{010}_{j,t,\text{rad}} + \text{e}^{i\phi_{1}}\sqrt{1-\alpha}\ket{100}_{j,t,\text{rad}})\nonumber\\
        &\otimes(\sqrt{\alpha}\ket{010}_{k,t,\text{rad}} + \text{e}^{i\phi_{2}}\sqrt{1-\alpha}\ket{100}_{k,t,\text{rad}})
    \end{align}
\end{subequations}
Next, the photon loss \(1-\eta\) is modeled using a beamsplitter model acting on the qubits labeled $t$ and $\text{rad}$ as follows:
\begin{subequations}
\begin{align}
    U_{\text{loss}}\ket{00}_{t,\text{rad}} &= \ket{00}_{t,\text{rad}}\\
     U_{\text{loss}}\ket{10}_{t,\text{rad}} &= \sqrt{\eta}\ket{10}_{t,\text{rad}} + \sqrt{1-\eta}\ket{01}_{t,\text{rad}}
\end{align}
\end{subequations}
Let $\ket{\psi}_{3}=U_{\text{loss,A}}\otimes U_{\text{loss,B}}\ket{\psi}_2$. With the beamsplitter model defined above, this gives:
\begin{subequations}
    \begin{align}
        \ket{\psi}_{3}=& U_{\text{loss,A}}(\sqrt{\alpha}\ket{010}_{j,t,\text{rad}} + \text{e}^{i\phi_{1}}\sqrt{1-\alpha}\ket{100}_{j,t,\text{rad}})\nonumber\\
        &\otimes U_{\text{loss,B}}(\sqrt{\alpha}\ket{010}_{k,t,\text{rad}} + \text{e}^{i\phi_{2}}\sqrt{1-\alpha}\ket{100}_{k,t,\text{rad}})\\
        =& (\sqrt{\alpha}\ket{0}_{j}U_{\text{loss,A}}\ket{10}_{t,\text{rad}} + \text{e}^{i\phi_{1}}\sqrt{1-\alpha}\ket{1}_{j}U_{\text{loss,A}}\ket{00}_{t,\text{rad}})\nonumber\\
        &\otimes(\sqrt{\alpha}\ket{0}_{k}U_{\text{loss,B}}\ket{10}_{t,\text{rad}} + \text{e}^{i\phi_{2}}\sqrt{1-\alpha}\ket{1}_{k}U_{\text{loss,B}}\ket{00}_{t,\text{rad}}) \\
        =& [\sqrt{\alpha}\ket{0}_{j}(\sqrt{\eta}\ket{10}_{t,\text{rad}} + \sqrt{1-\eta}\ket{01}_{t,\text{rad}}) + \text{e}^{i\phi_{1}}\sqrt{1-\alpha}\ket{1}_{j}\ket{00}_{t,\text{rad}}]\nonumber\\
        &\otimes[\sqrt{\alpha}\ket{0}_{k}(\sqrt{\eta}\ket{10}_{t,\text{rad}} + \sqrt{1-\eta}\ket{01}_{t,\text{rad}}) + \text{e}^{i\phi_{2}}\sqrt{1-\alpha}\ket{1}_{k}\ket{00}_{t,\text{rad}}] \\
        =& (\sqrt{\alpha\eta}\ket{010}_{j,t,\text{rad}} + \sqrt{\alpha(1-\eta)}\ket{001}_{j,t,\text{rad}} + \text{e}^{i\phi_{1}}\sqrt{1-\alpha}\ket{100}_{j,t,\text{rad}}) \nonumber\\
        &\otimes (\sqrt{\alpha\eta}\ket{010}_{k,t,\text{rad}} + \sqrt{\alpha(1-\eta)}\ket{001}_{k,t,\text{rad}} + \text{e}^{i\phi_{2}}\sqrt{1-\alpha}\ket{100}_{k,t,\text{rad}})\\
        =& \alpha\eta\ket{0011}_{j,k,t,t}\ket{00}_{\text{rad}} + \alpha\sqrt{\eta(1-\eta)}\ket{0010}_{j,k,t,t}\ket{01}_{\text{rad}}\nonumber\\
        & + \text{e}^{i\phi_{2}}\sqrt{\eta\alpha(1-\alpha)}\ket{0110}_{j,k,t,t}\ket{00}_{\text{rad}} + \alpha\sqrt{\eta(1-\eta)}\ket{0001}_{j,k,t,t}\ket{10}_{\text{rad}}\nonumber\\
        & + \alpha(1-\eta)\ket{0000}_{j,k,t,t}\ket{11}_{\text{rad}}+ \text{e}^{i\phi_{2}}\sqrt{\alpha(1-\alpha)(1-\eta)}\ket{0100}_{j,k,t,t}\ket{10}_{\text{rad}}\nonumber\\
        & + \text{e}^{i\phi_{1}}\sqrt{\eta\alpha(1-\alpha)}\ket{1001}_{j,k,t,t}\ket{00}_{\text{rad}}+ \text{e}^{i\phi_{1}}\sqrt{\alpha(1-\alpha)(1-\eta)}\ket{1000}_{j,k,t,t}\ket{01}_{\text{rad}}\nonumber\\
        & + \text{e}^{i(\phi_{1}+\phi_{2})}(1-\alpha)\ket{1100}_{j,k,t,t}\ket{00}_{\text{rad}}
    \end{align}
\end{subequations}
We now trace over the radiative degree of freedom:
\begin{subequations}
    \begin{align}
        \rho_{4} =& \text{Tr}_{\text{rad}}(\ket{\psi}_{3}\bra{\psi}_{3})\\
        =& \bra{00}_{\text{rad}}\ket{\psi}_{3}\bra{\psi}_{3}\ket{00}_{\text{rad}} + \bra{01}_{\text{rad}}\ket{\psi}_{3}\bra{\psi}_{3}\ket{01}_{\text{rad}}\nonumber\\
        &+ \bra{10}_{\text{rad}}\ket{\psi}_{3}\bra{\psi}_{3}\ket{10}_{\text{rad}} + \bra{11}_{\text{rad}}\ket{\psi}_{3}\bra{\psi}_{3}\ket{11}_{\text{rad}}
    \end{align}
\end{subequations}
From above equations, we get the following terms (here we omit subscript $j,k,t,t$):
\begin{subequations}
    \begin{align}
        \bra{00}_\text{rad}\ket{\psi}_{3} &= \alpha\eta\ket{0011} + \text{e}^{i\phi_{2}}\sqrt{\eta\alpha(1-\alpha)}\ket{0110} + \text{e}^{i\phi_{1}}\sqrt{\eta\alpha(1-\alpha)}\ket{1001} + \text{e}^{i(\phi_{1}+\phi_{2})}(1-\alpha)\ket{1100}\\
        \bra{01}_\text{rad}\ket{\psi}_{3} &=
        \alpha\sqrt{\eta(1-\eta)}\ket{0010} + \text{e}^{i\phi_{1}}\sqrt{\alpha(1-\alpha)(1-\eta)}\ket{1000}\\
        \bra{10}_\text{rad}\ket{\psi}_{3} &= \alpha\sqrt{\eta(1-\eta)}\ket{0001} + \text{e}^{i\phi_{2}}\sqrt{\alpha(1-\alpha)(1-\eta)}\ket{0100}\\
        \bra{11}_\text{rad}\ket{\psi}_{3} &=
        \alpha(1-\eta)\ket{0000}
    \end{align}
\end{subequations}
We use the following short-hand notation:
\begin{equation}
        \ket{c_{d(j)}}\equiv\ket{j}
\end{equation}
where $j$ is a 4-bit binary number, and $d(j)$ is decimal representation of $j$. We can rewrite Eq.12 as follows:
\begin{subequations}
    \begin{align}
        \bra{00}_\text{rad}\ket{\psi}_{3} =& \alpha\eta\ket{c_{3}} + \text{e}^{i\phi_{2}}\sqrt{\eta\alpha(1-\alpha)}\ket{c_{6}} + \text{e}^{i\phi_{1}}\sqrt{\eta\alpha(1-\alpha)}\ket{c_{9}} \nonumber\\
        &+ \text{e}^{i(\phi_{1}+\phi_{2})}(1-\alpha)\ket{c_{12}}\\
        \bra{01}_\text{rad}\ket{\psi}_{3} =&
        \alpha\sqrt{\eta(1-\eta)}\ket{c_{2}} + \text{e}^{i\phi_{1}}\sqrt{\alpha(1-\alpha)(1-\eta)}\ket{c_{8}}\\
        \bra{10}_\text{rad}\ket{\psi}_{3} =& \alpha\sqrt{\eta(1-\eta)}\ket{c_{1}} + \text{e}^{i\phi_{2}}\sqrt{\alpha(1-\alpha)(1-\eta)}\ket{c_{4}}\\
        \bra{11}_\text{rad}\ket{\psi}_{3} =&
        \alpha(1-\eta)\ket{c_{0}}
    \end{align}
\end{subequations}
Using the above notations, and plugging in all the terms in Eq.11, we get the following:
\begin{align}
    \rho_{4} =& \alpha^2\eta^2\ket{c_3}\bra{c_3} + \eta\alpha(1-\alpha)\ket{c_6}\bra{c_6} + \eta\alpha(1-\alpha)\ket{c_9}\bra{c_9} + (1-\alpha)^2\ket{c_{12}}\bra{c_{12}}\nonumber\\
    &+ \sqrt{\alpha^3\eta^{3}(1-\alpha)}(\text{e}^{-i\phi_{2}}\ket{c_3}\bra{c_6} + \text{e}^{i\phi_{2}}\ket{c_6}\bra{c_3}) \nonumber\\
    &+ \sqrt{\alpha^3\eta^{3}(1-\alpha)}(\text{e}^{-i\phi_{1}}\ket{c_3}\bra{c_9} + \text{e}^{i\phi_{1}}\ket{c_9}\bra{c_3})\nonumber\\
    &+ \eta\alpha(1-\alpha)(\text{e}^{i(\phi_{2}-\phi_{1})}\ket{c_6}\bra{c_9} + \text{e}^{i(\phi_{1}-\phi_{2})}\ket{c_9}\bra{c_6}) \nonumber\\
    &+ \sqrt{\eta\alpha(1-\alpha)^3}(\text{e}^{-i\phi_{1}}\ket{c_6}\bra{c_{12}} + \text{e}^{i\phi_{1}}\ket{c_{12}}\bra{c_6})\nonumber\\
    &+ \sqrt{\eta\alpha(1-\alpha)^3}(\text{e}^{-i\phi_{2}}\ket{c_9}\bra{c_{12}} + \text{e}^{i\phi_{2}}\ket{c_{12}}\bra{c_9}) \nonumber\\
    &+ \eta\alpha(1-\alpha)(\text{e}^{-i(\phi_{1}+\phi_{2})}\ket{c_3}\bra{c_{12}} + \text{e}^{i(\phi_{1}+\phi_{2})}\ket{c_{12}}\bra{c_3})\nonumber\\
    &+ \alpha^2\eta(1-\eta)\ket{c_2}\bra{c_2} + \alpha(1-\alpha)(1-\eta)\ket{c_8}\bra{c_8} \nonumber\\
    &+ \sqrt{\alpha^3\eta(1-\alpha)(1-\eta)^2}(\text{e}^{-i\phi_{1}}\ket{c_2}\bra{c_8} + \text{e}^{i\phi_{1}}\ket{c_8}\bra{c_2})\nonumber\\
    &+ \alpha^2\eta(1-\eta)\ket{c_1}\bra{c_1} + \alpha(1-\alpha)(1-\eta)\ket{c_4}\bra{c_4} \nonumber\\
    &+ \sqrt{\alpha^3\eta(1-\alpha)(1-\eta)^2}(\text{e}^{-i\phi_{2}}\ket{c_1}\bra{c_4} + \text{e}^{i\phi_{2}}\ket{c_4}\bra{c_1})\nonumber\\
    &+ \alpha^2(1-\eta)^2\ket{c_0}\bra{c_0}
\end{align}
After this stage, the two time bin qubits experience the following transformation due to the beam splitter:
\begin{subequations}
    \begin{align}
        &U_{\text{BS}}:\ket{00}^{A,B}_{t,t}\to \ket{00}^{C,D}_{t,t}\\
        &U_{\text{BS}}:\ket{01}^{A,B}_{t,t}\to \frac{\ket{10}^{C,D}_{t,t}-\ket{01}^{C,D}_{t,t}}{\sqrt{2}}\\
        &U_{\text{BS}}:\ket{10}^{A,B}_{t,t}\to \frac{\ket{10}^{C,D}_{t,t}+\ket{01}^{C,D}_{t,t}}{\sqrt{2}}\\
        &U_{\text{BS}}:\ket{11}^{A,B}_{t,t}\to \frac{\ket{20}^{C,D}_{t,t}-\ket{02}^{C,D}_{t,t}}{2}\\
        &(\text{assuming perfect indistinguishability of the two incoming photons})\nonumber
    \end{align}
\end{subequations}
Here, $A, B$ are the incoming ports and $C,D$ are the outgoing ports of the beamsplitter. The state after the beamsplitter is:
\begin{equation}
    \rho_{5} = U_{\text{BS}}\rho_{4}U_{\text{BS}}^{\dagger}
\end{equation}
We define the following measurement operators for getting statistics of clicks:
\begin{subequations}
    \begin{align}
        \text{M}_{C1} &= \ket{10}^{C,D}_{t,t} \bra{10}^{C,D}_{t,t}: \text{single click in port C and no click in port D}\\
        \text{M}_{D1} &= \ket{01}^{C,D}_{t,t} \bra{01}^{C,D}_{t,t}: \text{single click in port D and no click in port C}\\
        \text{M}_{0} &= \ket{00}^{C,D}_{t,t} \bra{00 }^{C,D}_{t,t}: \text{no clicks in both ports}
    \end{align}
\end{subequations}
The probabilities for the events above are given as follows:
\begin{subequations}
    \begin{align}
        p_{C1} &= \text{Tr}(M_{C1}\rho_{5}) = \text{Tr}(\bra{10}^{C,D}_{t,t}\rho_{5}\ket{10}^{C,D}_{t,t}) = \text{Tr}(\bra{10}^{C,D}_{t,t}U_{\text{BS}}\rho_{4}U_{\text{BS}}^{\dagger}\ket{10}^{C,D}_{t,t})\\
        &= \text{Tr}\bigg(\frac{\bra{10}^{A,B}_{t,t}+\bra{01}^{A,B}_{t,t}}{\sqrt{2}} \rho_{4}        \frac{\ket{10}^{A,B}_{t,t}+\ket{01}^{A,B}_{t,t}}{\sqrt{2}} \bigg)\\
        &= \frac{ \text{Tr}(\bra{10}\rho_{4}\ket{10}) + \text{Tr}(\bra{10}\rho_{4}\ket{01}) + \text{Tr}(\bra{01}\rho_{4}\ket{10}) + \text{Tr}(\bra{01}\rho_{4}\ket{01}) }{2}
    \end{align}
\end{subequations}
Only the terms with $(c_{1}, c_{2}, c_{5}, c_{6}, c_{9}, c_{10}, c_{13}, c_{14})$ in $\rho_{4}$ lead to non-zero trace in the above expression. Let $\rho_{C1}$ be the density after the measurement $M_{C1}$. 
\begin{subequations}
    \begin{align}
        \rho_{C1} =&  \eta\alpha(1-\alpha)\ket{c_6}\bra{c_6} + \eta\alpha(1-\alpha)\ket{c_9}\bra{c_9} + \eta\alpha(1-\alpha)(\text{e}^{i(\phi_{2}-\phi_{1})}\ket{c_6}\bra{c_9} \nonumber\\
        &+ \text{e}^{i(\phi_{1}-\phi_{2})}\ket{c_9}\bra{c_6}) + \alpha^2\eta(1-\eta)\ket{c_2}\bra{c_2} + \alpha^2\eta(1-\eta)\ket{c_1}\bra{c_1}\\
        =& \frac{\eta\alpha(1-\alpha)}{2}\ket{01}_{j,k}\bra{01} + \frac{\eta\alpha(1-\alpha)}{2}\ket{10}_{j,k}\bra{10} + \frac{\eta\alpha(1-\alpha)}{2}\text{e}^{i(\phi_{2}-\phi_{1})}\ket{01}_{j,k}\bra{10}\nonumber\\
        &+ \frac{\eta\alpha(1-\alpha)}{2}\text{e}^{i(\phi_{1}-\phi_{2})}\ket{10}_{j,k}\bra{01} + \alpha^2\eta(1-\eta)\ket{00}_{j,k}\bra{00}\\
        =& \frac{\eta\alpha(1-\alpha)}{2}(\ket{01}_{j,k}\bra{01} + \ket{10}_{j,k}\bra{10} + \text{e}^{-i\Delta\phi}\ket{01}_{j,k}\bra{10} + \text{e}^{i\Delta\phi}\ket{10}_{j,k}\bra{01})\nonumber\\
        &+ \alpha^2\eta(1-\eta)\ket{00}_{j,k}\bra{00}\\
        =& \eta\alpha(1-\alpha)\ket{\Phi^{\Delta\phi}_{+}}\bra{\Phi^{{\Delta\phi}}_{+}}+ \alpha^2\eta(1-\eta)\ket{00}_{j,k}\bra{00}
    \end{align}
\end{subequations}
Here, $\ket{\Phi^{\Delta\phi}_{\pm}} = \frac{1}{\sqrt{2}}(\ket{01}_{j,k} \pm \text{e}^{i\Delta\phi}\ket{10}_{j,k})$, and $\Delta\phi = (\phi_{1}-\phi_{2})$. Hence we find:
\begin{equation}
    p_{C1} = \text{Tr}(\rho_{C1}) = p_{\text{click}} = 2\alpha\eta - 2\alpha^2\eta^2
\end{equation}
The normalized density matrix is given by:
\begin{align}
    \Tilde{\rho}_{C1} = \frac{1-\alpha}{1-\alpha\eta}\ket{\Phi^{\Delta\phi}_{+}}\bra{\Phi^{\Delta\phi}_{+}} + \frac{\alpha(1-\eta)}{1-\alpha\eta}\ket{00}_{j,k}\bra{00}
\end{align}
Analogously, we get the following expression for $\Tilde{\rho}_{D1}$:
\begin{align}
    \Tilde{\rho}_{D1} = \frac{1-\alpha}{1-\alpha\eta}\ket{\Phi^{\Delta\phi}_{-}}\bra{\Phi^{\Delta\phi}_{-}} + \frac{\alpha(1-\eta)}{1-\alpha\eta}\ket{00}_{j,k}\bra{00}
\end{align}
Thus, the fidelity of the heralded state is given by:
\begin{equation}
    F = \bra{\Phi^{\Delta\phi}_{+}}\Tilde{\rho}_{C1}\ket{\Phi^{\Delta\phi}_{+}}=\bra{\Phi^{\Delta\phi}_{-}}\Tilde{\rho}_{D1}\ket{\Phi^{\Delta\phi}_{-}} = F =\frac{1-\alpha}{1-\alpha\eta}
\end{equation}
In reality because of the dephasing in individual systems, phases of qubits in system A and B have noise contributions. Suppose for a measurement run, the phases are given by:
\begin{subequations}
    \begin{align}
        \Tilde{\phi}_{1} &= \phi_{1} + n_{\phi_{1}}\\
        \Tilde{\phi}_{2} &= \phi_{2} + n_{\phi_{2}}\\
        \Delta\Tilde{\phi} &= \Tilde{\phi}_{1}- \Tilde{\phi}_{2} = \Delta\phi + (n_{\phi_{1}}-n_{\phi_{2}}) = \Delta\phi + \Delta n_{\phi}
    \end{align}
\end{subequations}
Here, $n_{\phi_{1(2)}}$ is the random phase noise for a measurement run in system A(B). Let $\Tilde{\rho}_{C1,exp}$ be the density matrix that we expect experimentally, given by:
\begin{align}
    \Tilde{\rho}_{C1,exp} = \frac{1-\alpha}{1-\alpha\eta}\ket{\Phi^{\Delta\Tilde{\phi}}_{+}}\bra{\Phi^{\Delta\Tilde{\phi}}_{+}} + \frac{\alpha(1-\eta)}{1-\alpha\eta}\ket{00}_{j,k}\bra{00}
\end{align}
In that case, the fidelity we expect experimentally is given by:
\begin{subequations}
   \begin{align}
        F_{exp} &= \bra{\Phi^{\Delta\phi}_{+}}\Tilde{\rho}_{C1,exp}\ket{\Phi^{\Delta\phi}_{+}}=\bra{\Phi^{\Delta\phi}_{-}}\Tilde{\rho}_{D1,exp}\ket{\Phi^{\Delta\phi}_{-}}\\
        &= \text{cos}^{2}\bigg(\frac{\Delta n_{\phi}}{2}\bigg)\frac{1-\alpha}{1-\alpha\eta}
    \end{align} 
\end{subequations}
Similarly, the state we get after the measurement $M_{0}$ is given by:
\begin{subequations}
    \begin{align}
       \rho_{0} =& \bra{00}^{C,D}_{t,t}\rho_{5}\ket{00}^{C,D}_{t,t}\\ 
       =& (1-\alpha)^2\ket{11}_{j,k}\bra{11} + \alpha(1-\alpha)(1-\eta)(\ket{01}_{j,k}\bra{01} + \ket{10}_{j,k}\bra{10})\nonumber\\
       &+ \alpha^2(1-\eta)^2\ket{00}_{j,k}\bra{00}
    \end{align}
\end{subequations}
and the corresponding probability $p_0$ by:
\begin{equation}
    p_{0} = \text{Tr}(\rho_{0}) = (1-\alpha\eta)^2
\end{equation}
So the normalized density matrix becomes:
\begin{subequations}
    \begin{align}
        \Tilde{\rho}_{0} =& \frac{\rho_{0}}{\text{Tr}(\rho_{0})}\\
        =& \bigg(\frac{\alpha-\alpha\eta}{1-\alpha\eta}\bigg)^2\ket{00}_{j,k}\bra{00} + \bigg(\frac{1-\alpha}{1-\alpha\eta}\bigg)^2\ket{11}_{j,k}\bra{11}\nonumber\\
        &+ \frac{\alpha(1-\alpha)(1-\eta)}{(1-\alpha\eta)^2}\bigg(\ket{01}_{j,k}\bra{01} + \ket{10}_{j,k}\bra{10}\bigg)
    \end{align}
\end{subequations}
\begin{figure}
    \centering
    \includegraphics[width=0.8\linewidth]{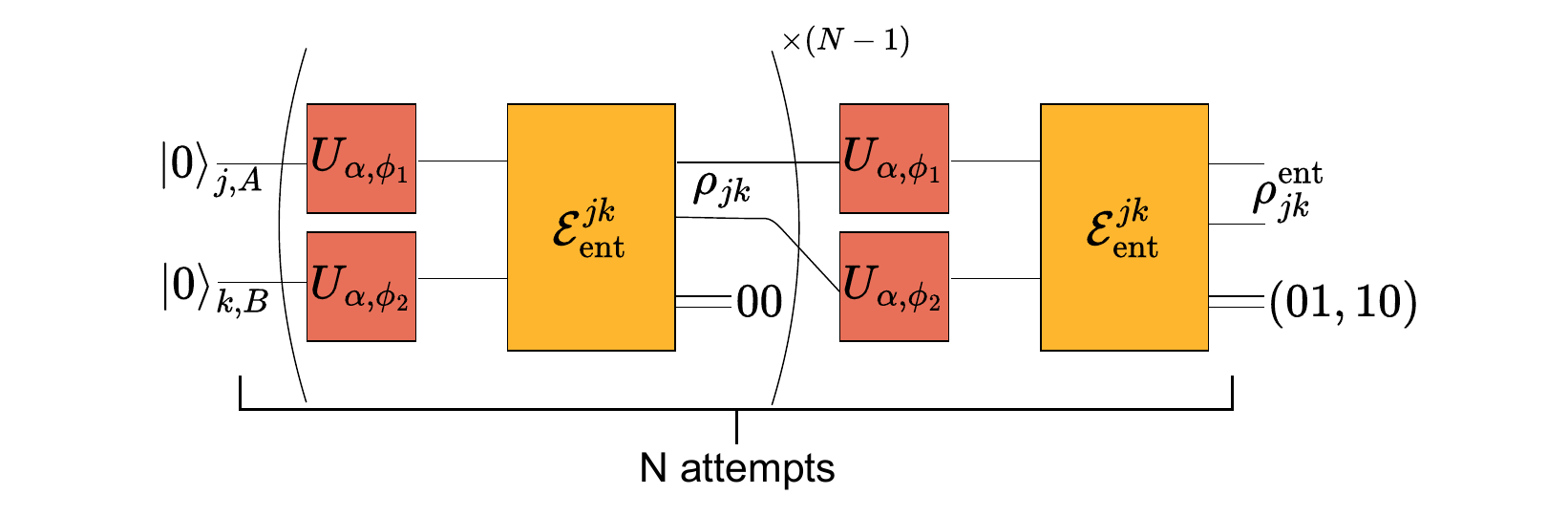}
    \caption{\textbf{Stochastic Nature of the Entanglement Protocol.} First ($M$-1) attempts are unsuccessful as $m_{jk}=00$, while $m_{jk}=(01,10)$ at the $M^{\text{th}}$ attempt heralds a successful attempt. An important aspect here that we omit is the swap with nuclear spins $\ket{n}_{j,A}$ and $\ket{n}_{k,B}$, after a successful entanglement is established between electron spin qubits $\ket{0}_{j,A}$ and $\ket{0}_{k,B}$.}
    \label{fig:Si_5}
\end{figure}
The above expression suggests that heralding on zero clicks on both ports leads to a mixed state between qubits $j$ and $k$. Thus, in order to re-utilize it for the next attempt, we propose a controlled laser based spin-initialization depending on if there is a click or not, $U_{\text{init}}$ which performs the following transformation:
\begin{align}
        U_{\text{init}}:\rho_{jk}\to \ket{00}_{j,k}\bra{00}
\end{align}
Now, we repeat this protocol until the measurement outcome is either 01 or 10, as in Figure~\ref{fig:Si_5}. An important aspect here that we omit is the respective swap with nuclear spins $\ket{n}_{j,A}$ and $\ket{n}_{k,B}$, after a successful entanglement is established between electron spin qubits $\ket{0}_{j,A}$ and $\ket{0}_{k,B}$. For our analysis we also assume that the electron-nuclear SWAP gate error is negligible and constant wrt to entanglement error. This assumption allows us to focus largely on entanglement errors arising due to electron spin. This gives the following fidelity and probability of getting clicks only in the $M^{th}$ attempt (assuming $\eta<<1$):
\begin{subequations}
    \begin{align}
        F_{jk} &= \text{cos}^{2}\bigg(\frac{\Delta n_{\phi}}{2}\bigg)\frac{1-\alpha}{1-\alpha\eta}\\
        P_{jk}(M) &= 2\alpha\eta(1-2\alpha\eta)^{N-1}
    \end{align}
\end{subequations} 
Let, $n_{jk}$ be the attempt number which leads to first successful entanglement. As discussed in the main text, this leads to a distribution in $n$:
\begin{equation}
\mathcal{D}_{n} = \{n_{jk}\}_{\substack{1\leq i \leq N_{a} \\ 1 \leq j \leq N_{b}}}
\end{equation}
where each $n_{jk}$ is sampled from the geometric distribution in Eq.(32b), and $N_{a}$ and $N_{b}$ are the number of qubits in system A and B respectively.
\subsection{Thermal Decoherence}
The thermal induced quantum noise acting on the state $\rho$, can be modeled as a generalized amplitude damping channel, $\mathcal{E}_{\text{th}}(\rho)$ given by \cite{Nielsen_Chuang_2010}:
\begin{align}
    &\mathcal{E}_{\text{th}}(\rho) = 
\begin{bmatrix}
(1 - \gamma)\rho_{00} + \gamma p_{\text{th}} & \sqrt{1 - \gamma} \, \rho_{01} \\
\sqrt{1 - \gamma} \, \rho_{01}^* & 1 - (1 - \gamma)\rho_{00} - \gamma p_{\text{th}}
\end{bmatrix}\\
&\text{for a general state }
\rho = 
\begin{bmatrix}
\rho_{00}& \rho_{01} \\
\rho_{01}^* & (1 - \rho_{00})
\end{bmatrix}
\end{align}
where, $\gamma = (1- \text{e}^{-t/T_{1}})$, and $p_{\text{th}}$ is the thermodynamic steady-state probability given by the Boltzmann distribution:
\begin{equation}
    p_{\text{th}} = \frac{\text{e}^{-E_0 / (k_B T)}}{\text{e}^{-E_0 / (k_B T)} + \text{e}^{-E_1 / (k_B T)}} 
= \frac{1}{1 + \text{e}^{-\hbar\omega / (k_B T)}},
\end{equation}
where $E_{1(0)}$ is the energy of the qubit $\ket{1}(\ket{0})$, and $E_1 - E_0 = \hbar\omega$, and $T$ is the local temperature of the qubit environment.
If we start from a state, $\ket{\psi}_{\alpha,\phi} = \sqrt{\alpha}\ket{0} + \text{e}^{i\phi}\sqrt{1-\alpha}\ket{1}$, which corresponds to $\rho_{\alpha,\phi}$:
\begin{align}    \rho_{\alpha,\phi} &= 
    \begin{bmatrix} 
    \alpha & \text{e}^{-i\phi} \, \sqrt{\alpha(1-\alpha)}\\\\
    \text{e}^{i\phi} \, \sqrt{\alpha(1-\alpha)} & 1 - \alpha
    \end{bmatrix} 
\end{align}
Let us look at the expectation value of $\sigma_y$:
\begin{align}
    \langle\sigma_{y}\rangle_{\alpha,\phi} &= 
    2 \, \sqrt{\alpha(1-\alpha)} \, \sin(\phi)
\end{align}
In the presence of a thermal decoherence channel, $\alpha$ is a stochastic variable, hence the averaged state is given by:
\begin{align}
    \overline{\rho}_{\alpha,\phi}(t) &= 
    \begin{bmatrix} 
    \overline{\alpha} & \text{e}^{-i\phi} \, \overline{\sqrt{\alpha(1-\alpha}} \\
    \text{e}^{i\phi} \, \overline{\sqrt{\alpha(1-\alpha)}} & 1 - \overline{\alpha}
    \end{bmatrix}
\end{align}
So we find the following average expectation value of $\sigma_y$:
\begin{align}
    \overline{\langle\sigma_{y}\rangle}_{\alpha,\phi} &= 
    2 \, \overline{\sqrt{\alpha(1-\alpha)}} \, \sin(\phi)
\end{align}

where the overline refers to the average over $\alpha$. Equating Eq.~(40) with Eq.~(34), we find the following:
\begin{subequations}
\begin{align}
    \overline{\sqrt{\alpha(1-\alpha)}} &\approx \sqrt{1-\gamma}\sqrt{\alpha(1-\alpha)} = \text{e}^{-\frac{t}{2T_{1}}}\sqrt{\alpha(1-\alpha)}\\
    \overline{\alpha} &= \text{e}^{-\frac{t}{T_{1}}}\alpha + (1- \text{e}^{-\frac{t}{T_{1}}})p_{\text{th}}
\end{align}    
\end{subequations}
Thus, we get following expression for thermal-induced decoherence:
\begin{equation}
\overline{\langle\sigma_{y}\rangle}_{\alpha,\phi} = 2\text{e}^{-\frac{t}{2T_{1}}}~\sqrt{\alpha(1-\alpha)}~\text{sin}(\phi) 
\end{equation}

\subsection{Dephasing Channel}
Let $\mathcal{E}_{\text{dep}}(\rho)$ be a pure dephasing channel given by \cite{Nielsen_Chuang_2010}:
\begin{equation}
    \mathcal{E}_{\text{dep}}:\rho\to (1-p_{\text{dep}})\rho + p_{\text{dep}}\sigma_{\text{z}}\rho\sigma_{\text{z}}
\end{equation}
Similar to the previous case, we start from a state, $\ket{\psi}_{\alpha,\phi} = \sqrt{\alpha}\ket{0} + \text{e}^{i\phi}\sqrt{1-\alpha}\ket{1}$. In the presence of dephasing channel, $\phi$ is a stochastic variable, hence the ensemble average state is given by:
\begin{align}
    \dashbar{\rho_{\alpha,\phi}} &= \begin{bmatrix} 
    \alpha & \dashbar{\text{e}^{-i\phi}}\sqrt{\alpha(1-\alpha)} \\\\
    \dashbar{\text{e}^{i\phi}}\sqrt{\alpha(1-\alpha)} & 1 -\alpha
\end{bmatrix}
\end{align}
This gives the following ensemble average expectation value of $\sigma_y$:
\begin{align}
\dashbar{\langle\sigma_{y}\rangle}_{\alpha,\phi} &= 2\sqrt{\alpha(1-\alpha)}~\dashbar{\text{sin}(\phi)}
\end{align}

Here the angled bracket corresponds to the expectation value for the quantum evolution of $\rho$ (accounts for the pure dephasing), and tilde corresponds to the ensemble averaging (accounts for inhomogeneous broadening based dephasing), which averages on the stochastic phase noise $n_{\phi}$ of the control or environment from one experimental run to another. 

Suppose, $\phi$ is centered around $\phi_{0}$ such that, $\phi=\phi_{0} + n_{\phi}$, where $n_{\phi}$ is the stochastic phase variable.
\begin{equation}
     |\dashbar{\langle\sigma_{y}\rangle}_{\alpha,\phi_{0}}| = 2\sqrt{\alpha(1-\alpha)}\text{sin}(\phi_{0})|~\dashbar{\text{cos}(n_{\phi})}| + \text{cos}(\phi_{0})~|\dashbar{\text{sin}(n_{\phi})}|
\end{equation}
For a system characterized for the initial state  $\ket{Y_{+}}=\frac{\ket{0}+i\ket{1}}{\sqrt{2}}$ \cite{Biercuk_2011}:
\begin{equation}
    |\dashbar{\langle\sigma_{y}\rangle}| = |\dashbar{\text{cos}(n_{\phi})}| = \text{e}^{-\chi_{\text{dep}}(\tau)}
\end{equation}
Here the decoherence function $\chi_{\text{dep}}(\tau)$ is given by:
\begin{equation}
    \chi_{\text{dep}}(\tau) = \bigg(\frac{\tau}{T_{\phi}}\bigg)^{z_{\phi}} + \bigg(\frac{\tau}{T_{\text{inh}}}\bigg)^{z_{\text{inh}}}
\end{equation}
Here $\tau$ is the experimental duration of dephasing, $1/T_{\phi}$ is the pure-dephasing rate and $1/T_{\text{inh}}$ is the inhomogeneous-broadening rate. Further, depending on the noise spectra of these individual sources, we assume a scaling factor of $z_{\phi}$ and $z_{\text{inh}}$ respectively.

\subsection{Composing Thermal Decoherence with a Dephasing Channel}
In the presence of thermal decoherence and dephasing, both $\alpha$ and $\phi$ are stochastic variables, which means the coherence function needs to be averaged over both, yielding the following:
\begin{equation}    \dashbar{\overline{\rho_{\alpha,\phi}}} = 
    \begin{bmatrix} 
    \overline{\alpha} & \dashbar{\text{e}^{-i\phi}} \, \overline{\sqrt{\alpha(1-\alpha)}} \\\\
    \dashbar{\text{e}^{i\phi}} \, \overline{\sqrt{\alpha(1-\alpha)}} & 1 - \overline{\alpha}
    \end{bmatrix}
\end{equation}
\begin{subequations}
    \begin{align}
     \dashbar{\overline{\langle\sigma_{y}\rangle}}_{\alpha,\phi} &= 
    2 \, \overline{\sqrt{\alpha(1-\alpha)}} \, \dashbar{\sin(\phi)}\\
    &= 2~\sqrt{\alpha(1-\alpha)}~\text{e}^{-(\frac{\tau}{2T_{1}}+\chi_{\text{dep}}(\tau))}\\
    &= 2~\sqrt{\alpha(1-\alpha)}~\text{e}^{-\chi^{*}(\tau)}
    \end{align}
\end{subequations}
where we introduced the effective decoherence function $\chi^{*}(\tau)$:
\begin{subequations}
\begin{align}
   \chi^{*}(\tau) &= \frac{\tau}{2T_{1}}+\chi_{\text{dep}}(\tau)\\
   &= \frac{\tau}{2T_{1}}+\bigg(\frac{\tau}{T_{\phi}}\bigg)^{z_{\phi}} + \bigg(\frac{\tau}{T_{\text{inh}}}\bigg)^{z_{\text{inh}}} \simeq \bigg(\frac{\tau}{T^{*}_{2}}\bigg)^{z^{*}}
\end{align}    
\end{subequations}
The scaling $z^{*}$ is the effective asymptotic scaling factor depending on $\tau$ and individual scalings $z_{\phi}$ and $z_{\text{inh}}$. In the main text we characterize a dynamical decoupling sequence for initial state along the bloch-axis y (i.e. $\alpha=1/2, \phi= \pi/2$), which gives:
\begin{equation}
    \chi_{\text{dds}}(\tau) = \bigg(\frac{\tau}{T_{2}}\bigg)^{z_{\text{dds}}}
\end{equation}
Since the dynamical decoupling sequence mainly reduces the inhomogeneous phase noise, we assume that $T_{\text{inh}}$ increases to $\kappa_{\text{dds}}T_{\text{inh}}$, where $\kappa_{\text{dds}}>1$ is a factor of improvement. From Eq. (51b) and Eq.(52), we get the following:
\begin{subequations}
    \begin{align}
      \bigg(\frac{\tau}{T_{2}}\bigg)^{z_{\text{dds}}} &= \frac{\tau}{2T_{1}}+\bigg(\frac{\tau}{T_{\phi}}\bigg)^{z_{\phi}} + \bigg(\frac{\tau}{\kappa_{\text{dds}}T_{\text{inh}}}\bigg)^{z_{\text{inh}}}\\
      &= \frac{\tau}{2T_{1}} + \chi^{\prime}_{\text{dep}}(\tau)
    \end{align}
\end{subequations}
Assuming $T_{1}>>T_{2}$ this becomes:
\begin{equation}
     \bigg(\frac{\tau}{T_{2}}\bigg)^{z_{\text{dds}}} \approx \chi^{\prime}_{\text{dep}}(\tau)
\end{equation}
Thus, in the presence of thermal decoherence, phase noise and a dynamical decoupling sequence, $\chi^{\prime}_{\text{dep}}(\tau)$ quantifies the effective decoherence rate of the phase.
\subsection{Combining the Entanglement Protocol and Dynamical Decoupling Sequence}
The operation order for a single entanglement attempt after performing dynamical decoupling is shown in Fig. \ref{fig:singleentanglementblock}.
\begin{figure}
    \centering
    \includegraphics[width=0.9\linewidth]{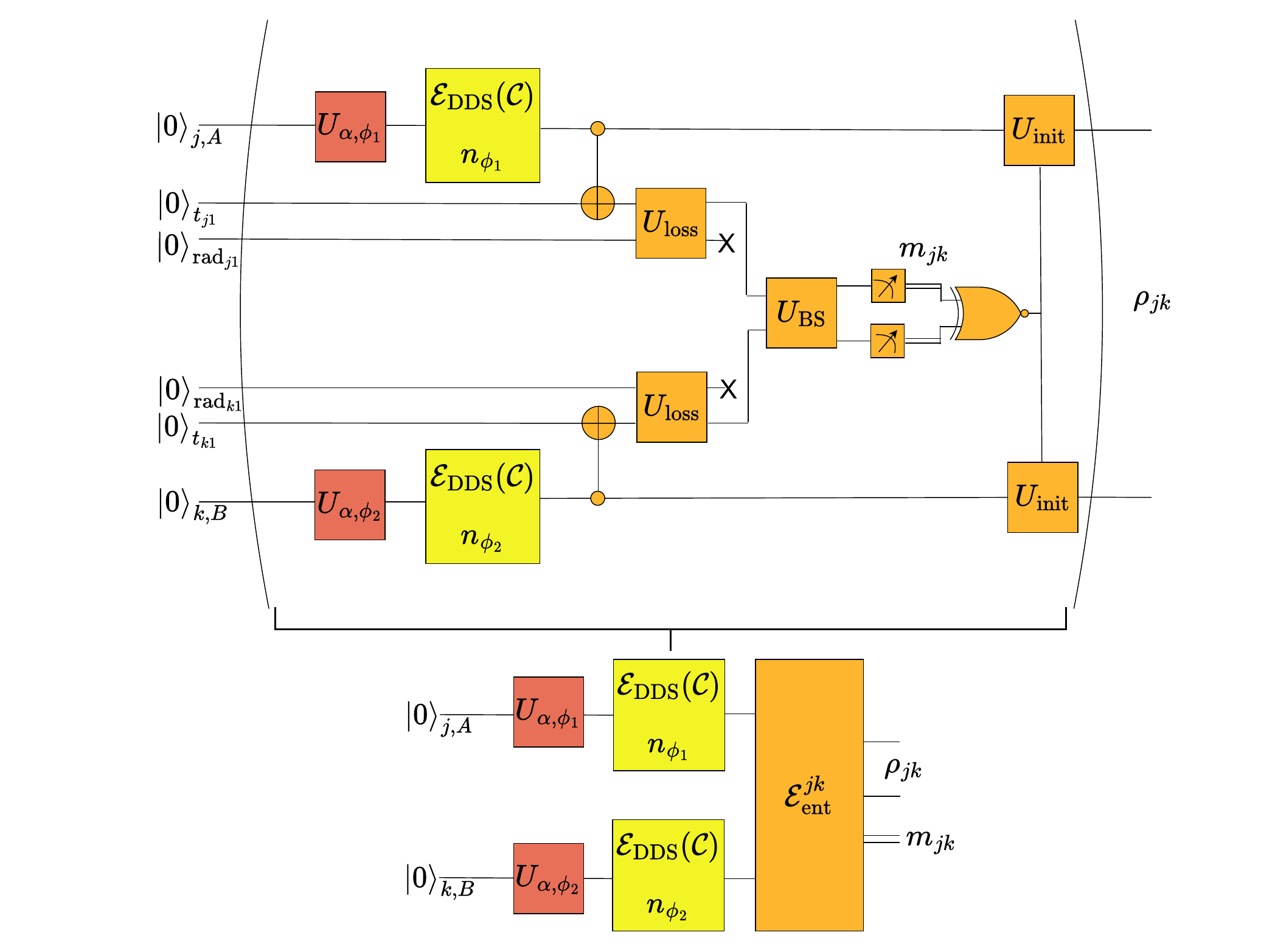}
    \caption{A single attempt of combination of Dynamical Decoupling Channel and Single Photon Entanglement for a qubit pair $(j_{A},k_{B})$}
    \label{fig:singleentanglementblock}
\end{figure}
In Eq.40, we estimate the entanglement fidelity $F_{exp}$ for a single run, but in order to make it closer to experiments, one has to take an ensemble average for both thermal and dephasing channels:
\begin{equation}
    \dashbar{\overline{F}}_{exp} = \overline{\bigg(\frac{1-\alpha}{1-\alpha\eta}\bigg)}\dashbar{\bigg(\text{cos}^2\bigg(\frac{\Delta n_{\phi}}{2}  \bigg)\bigg)}
\end{equation}
Assuming that $\eta<<1$, we get the following:
\begin{equation}
   \dashbar{\overline{F}}_{\text{exp}}= (1-\overline{\alpha})\bigg(\frac{1 + \dashbar{\text{cos}(\Delta n_{\phi})}}{2}\bigg)
\end{equation}
Let us break down the ensemble averaging:
    \begin{align}
        \dashbar{\text{cos}(\Delta n_{\phi})} = \dashbar{\text{cos}(n_{\phi_{1}} - n_{\phi_{2}})} = \dashbar{\text{cos}(n_{\phi_{1}})}~\dashbar{\text{cos}(n_{\phi_{2}})} + \dashbar{\text{sin}(n_{\phi_{1}})}~\dashbar{\text{sin}(n_{\phi_{2}})}
    \end{align}
Since $n_{\phi_{1}}$ and $n_{\phi_{2}}$ correspond to noise spectra on two spatially separated identical systems, we assume them to be identical and independent. The dynamical decoupling sequence improves the phase coherence for the entanglement protocol $(T_2^*\rightarrow T_2)$, therefore:
\begin{subequations}
    \begin{align}
        |\dashbar{\text{cos}(n_{\phi_{1}})}| &= \text{e}^{-\chi^{\prime}_{1,\text{dep}}(\tau)}\\
        |\dashbar{\text{cos}(n_{\phi_{2}})}| &= \text{e}^{-\chi^{\prime}_{2,\text{dep}}(\tau)}
    \end{align}
\end{subequations}
Here, $\chi^{\prime}_{1(2),\text{dep}}$ is given by Eq.(54).
Without loss of generality, let's assume the following convention:
\begin{subequations}
    \begin{align}
        &\dashbar{\text{cos}(n_{\phi_{1}})} = p; ~\dashbar{\text{sin}(n_{\phi_{1}})} = q\\
        &\dashbar{\text{cos}(n_{\phi_{2}})} = r; ~\dashbar{\text{sin}(n_{\phi_{2}})} = s\\
        &P = |\dashbar{\text{cos}(\Delta n_{\phi_{}})}| = |pr + qs|
    \end{align}
\end{subequations}
By triangle inequality of distance measures, we have the following:
\begin{equation}
   |p||r| - |q||s|  \le P \le |p||r| + |q||s| 
\end{equation}
We further have the following inequalities:
\begin{subequations}
\begin{align}
    &\dashbar{\text{cos}^{2}(n_{\phi_{1}})} = 1 - \dashbar{\text{sin}^{2}(n_{\phi_{1}})} \ge \dashbar{\text{cos}(n_{\phi_{1}})}~^{2} = p^{2}\\
    &\implies 1 - p^2 \ge \dashbar{\text{sin}^{2}(n_{\phi_{1}})} \ge \dashbar{\text{sin}(n_{\phi_{1}})}~^{2} = q^{2} \\
    &\implies |q| \le \sqrt{1-p^2} 
\end{align}
\end{subequations}
Similarly, we have: $|s| \le \sqrt{1-r^2}$.

From this we get:
\begin{equation}
    P \ge |p||r| - |q||s| \ge |p||r| - \sqrt{1-p^2}\sqrt{1-q^2}
\end{equation}
Since $|p|=\text{e}^{-\chi^{\prime}_{1,\text{dep}}(\tau)}$ and $|r|=\text{e}^{-\chi^{\prime}_{2,\text{dep}}(\tau)}$, we get:
\begin{equation}
    \dashbar{\text{cos}(\Delta n_{\phi})} \ge \text{e}^{-(\chi^{\prime}_{1,\text{dep}}(\tau)+\chi^{\prime}_{2,\text{dep}}(\tau))} - \sqrt{1 - \text{e}^{-2\chi^{\prime}_{1,\text{dep}}(\tau)}}\sqrt{1 - \text{e}^{-2\chi^{\prime}_{2,\text{dep}}(\tau)}}
\end{equation}
Thus, by combining Eq.~(41b), (56) and (63), we get the following:
\begin{tcolorbox}[breakable, colback=white]
\begin{align}
\dashbar{\overline{F_{\text{exp}}(\tau)}} 
&\ge \bigg(1 - e^{-\frac{\tau}{T_1}} \alpha - (1 - e^{-\frac{\tau}{T_1}}) p_{\text{th}} \bigg) \nonumber \\
&\quad \times \bigg( \frac{1 + e^{-(\chi'_{1,\text{dep}}(\tau) + \chi'_{2,\text{dep}}(\tau))} 
- \sqrt{1 - e^{-2\chi'_{1,\text{dep}}(\tau)}} \sqrt{1 - e^{-2\chi'_{2,\text{dep}}(\tau)}}}{2} \bigg)
\end{align}
\end{tcolorbox}

Here, we assumed that the $T_{1}$ process and $T_{1}$ timescale for both systems are identical. In order to estimate the error of this protocol $\epsilon$, we use the following definition of distance measure $\mathcal{D}$ between two states $\rho$ and $\sigma$:
\begin{equation}
    \mathcal{D}(\rho,\sigma) = \frac{1}{2}\text{Tr}|\rho-\sigma|
\end{equation}
We use the following property:
\begin{subequations}
\begin{align}
    &1 - F(\rho,\sigma) \le \mathcal{D}(\rho,\sigma) \le \sqrt{1 - F(\rho,\sigma)^2}\\
    &1 - \overline{F(\rho,\sigma)} \le \overline{\mathcal{D}(\rho,\sigma)} \le \sqrt{1 - \overline{F(\rho,\sigma)
    }^2}~: \text{By Jensen's Inequality}
\end{align}
\end{subequations}
This gives:
\begin{equation}
    \dashbar{\overline{\mathcal{D}_{jk}(\tau)}}_{\text{min}} = 1 - \dashbar{\overline{{F}_{exp}(\tau)}}
\end{equation}
Using the universality principle that an arbitrary 2-qubit gate can implemented by a 1-qubit unitary and a CNOT gate, we get the following expression for the average 2-qubit error between qubits $j$ and $k$:
\begin{equation}
    \epsilon_{jk}(\tau) = \text{max}(\dashbar{\overline{\mathcal{D}_{jk}(\tau)}}_{\text{min}} + \epsilon_{\text{1-qubit}})
\end{equation}
Including the $N$-pulse CPMG unitary errors modifies the above error as follows:
\begin{equation}
   \epsilon_{jk}(\tau,N) = \text{max}(\dashbar{\overline{\mathcal{D}_{jk}(\tau)}}_{\text{min}}) + (2N+1)\epsilon_{1-\text{qubit}} 
\end{equation}

Combining Eq. Eq.~(64), (67) and (69) gives us our final expression for the average 2-qubit error between qubits $j$ and $k$:

\begin{tcolorbox}[breakable, colback=white]
\begin{align}
&\epsilon_{jk}(\tau,N, T_{1}, T_{2j}, T_{2k}, z_{j,\text{dds}}, z_{k,\text{dds}} ) \nonumber\\&=
 1 - \left(1 - \text{e}^{-\frac{\tau}{T_{1}}}\alpha 
- (1 - \text{e}^{-\frac{\tau}{T_{1}}})\,p_{\text{th}}\right) \notag \\
&\quad \left. \times 
\left(
\frac{1 +\text{e}^{-(\chi^{\prime}_{j,\text{dep}}(\tau)+\chi^{\prime}_{k,\text{dep}}(\tau))} 
- \sqrt{1 - \text{e}^{-2\chi^{\prime}_{j,\text{dep}}(\tau)}}
\sqrt{1 - \text{e}^{-2\chi^{\prime}_{k,\text{dep}}(\tau)}}}{2}
\right) \right. \notag \\
&\quad  + (2N+1)\epsilon_{1\text{-qubit}} 
\end{align}
\end{tcolorbox}

The $M$-attempts composition of the global decoupling channel with pairwise entanglement blocks is summarized in Fig. \ref{fig:globaldecouplingchannel}.

\begin{figure}
    \centering
    \includegraphics[width=0.9
    \linewidth]{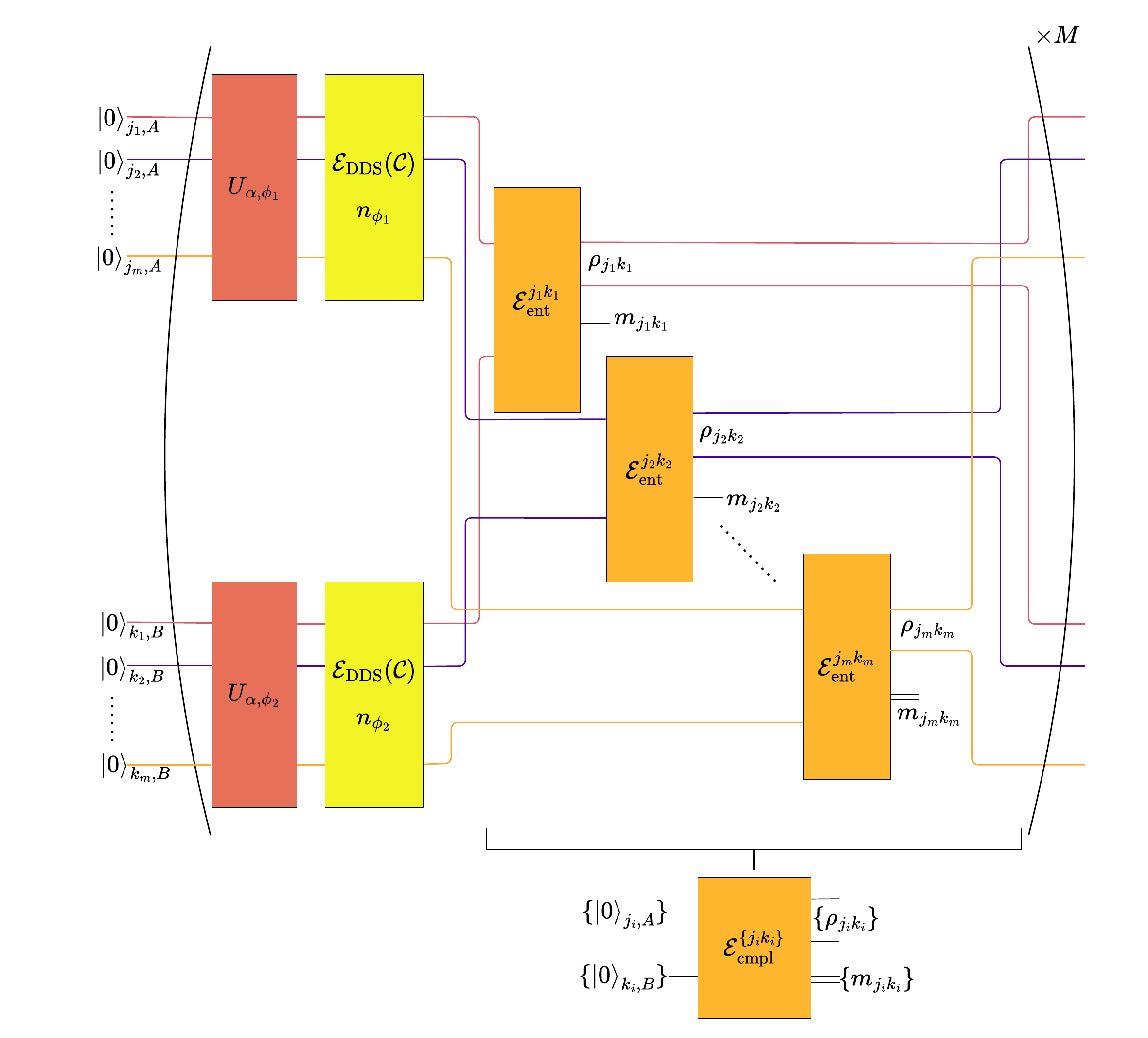}
    \caption{$M$ attempts of composing the global decoupling channel $\mathcal{E}_{\text{DDS}}$ and pairwise entanglement blocks $\{\mathcal{E}^{j_{i}k_{i}}_{\text{ent}} \}$ given by the compilation strategy.}
    \label{fig:globaldecouplingchannel}
\end{figure}

For completeness, the full end-to-end protocol, executed repeatedly over $M$ attempts, is shown in the block diagram of Fig. \ref{fig:fullblockdiagram}.

\begin{figure}
    \centering
    \includegraphics[width=0.6\linewidth]{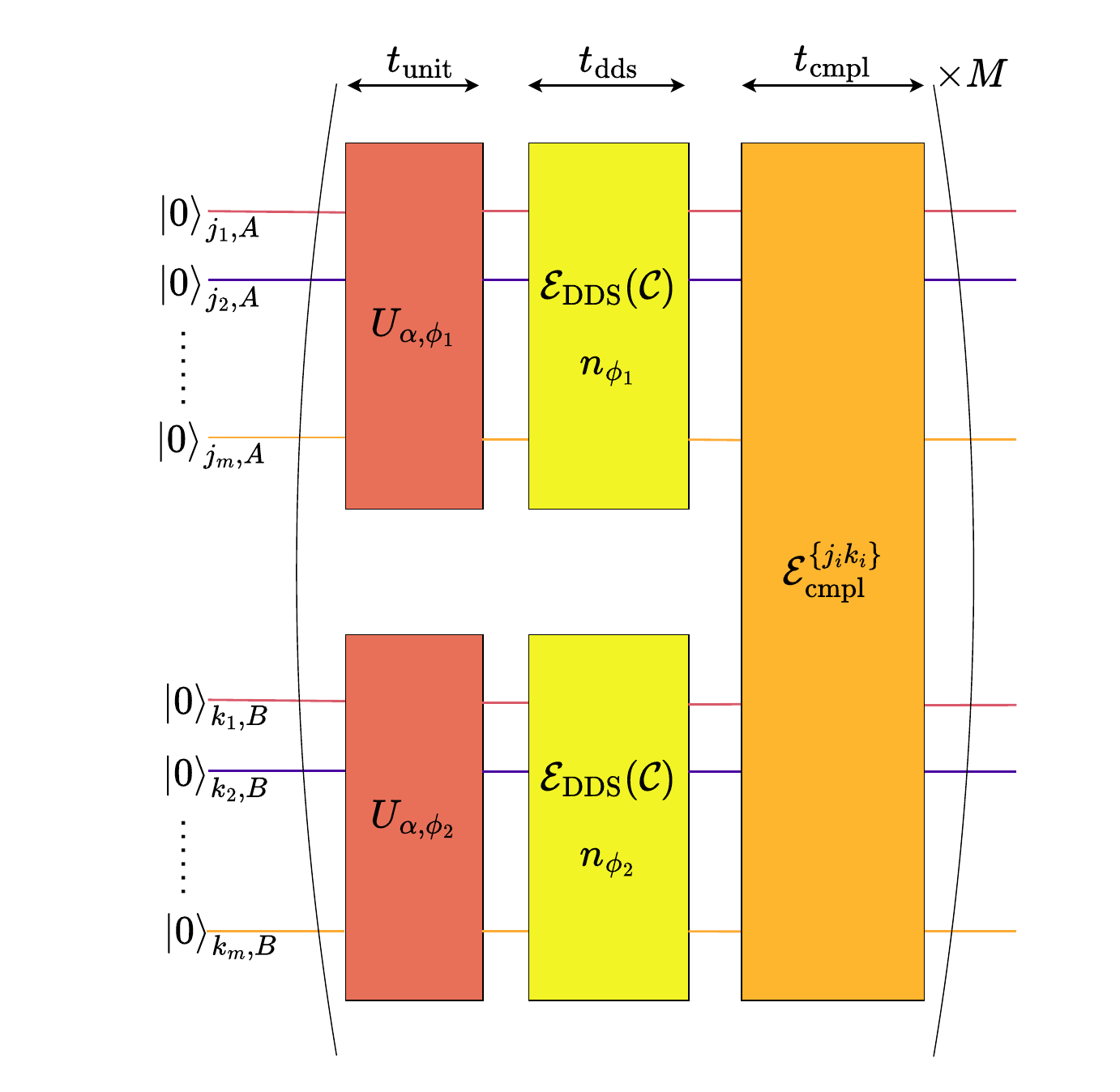}
    \caption{Block diagram of the full protocol repeated $M$ times.}
    \label{fig:fullblockdiagram}
\end{figure}

\subsection{Thermal Budget}
\subsubsection{Cold-Plate Stage}
The $T_{1}$ and $T_{2}$ parameters in the previous section also depend on the local temperature of the SiV$^-$ qubit. This section discusses about modeling the thermal budget of the system.
We assume a train of N pulses, each containing an energy of $E_{\text{pulse}}$, having pulse length $t_{\text{pul}}$, and inter-pulse duration $t_{\text{ip}}$. This gives measurement time $t_{\text{dds}} = \text{N}(t_{\text{pul}} + t_{\text{ip}})$. Suppose the heat-capacity of the system is $C_{\text{sys}}$, the maximum cooling power available in the cryostat is $P_{\text{cool}}$ and the heat-load at the qubit stage is $P_{\text{stage}}$. We assume that the qubit is at the cold-plate stage of the dilution refrigerator.

\begin{figure}
    \centering
    \includegraphics[width=0.9\linewidth]{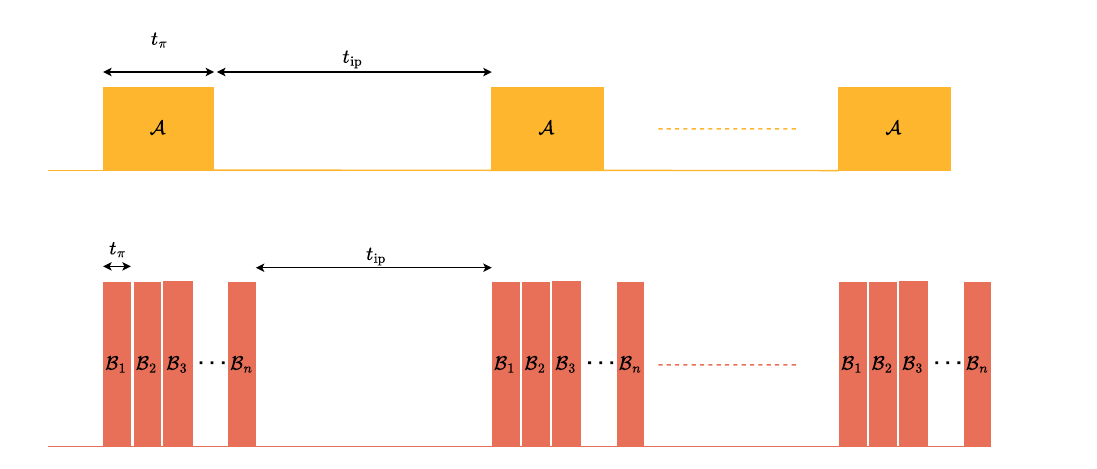}
    \caption{Pulse level diagram}
    \label{fig:enter-label}
\end{figure}

Since the qubit interacts via strain coupling, the approximate energy-density contained per unit mechanical pulse is given by:
\begin{equation}
    u_{\text{mech}}  = \frac{1}{2}\mathrm{Y}_{\text{dia}}\epsilon_{\text{str}}^{2}
\end{equation}
Here, $\mathrm{Y}_{\text{dia}}$ is the Young's modulus of diamond, $\epsilon
_{\text{str}}$ is the strain amplitude for the mechanical wave. With $c_{\text{dia}}$ the speed of sound in diamond, the intensity of the propagating pulse is given by:
\begin{equation}
    \text{I}_{\text{mech}} =  u_{\text{mech}} c_{\text{dia}}  = \frac{1}{2}\mathrm{Y}_{\text{dia}}\epsilon_{\text{str}}^{2}c_{\text{dia}}
\end{equation}
Further, suppose the wave is confined within an aperture area given by $\sim\lambda_{\text{mech}}^{2}$, where $\lambda_{\text{mech}}$ is the wavelength of the mechanical wave. Then the power contained in the pulse is given by:
\begin{equation}
    P_{\text{mech}} = \frac{1}{2}\mathrm{Y}_{\text{dia}}\epsilon_{\text{str}}^{2}c_{\text{dia}}\lambda_{\text{mech}}^{2}
\end{equation}
Given the transduction efficiency $\eta_{\text{tr}}$ from microwave mode to mechanical mode, the microwave power delivered to the sample near the cold-plate stage is given by:
\begin{equation}
    P_{\text{MW}} = \frac{1}{2\eta_{\text{tr}}}\mathrm{Y}_{\text{dia}}\epsilon_{\text{str}}^{2}c_{\text{dia}}\lambda_{\text{mech}}^{2} \propto \epsilon_{\text{str}}^{2}
\end{equation}
We plug in typical values from the literature: $\eta_{\text{tr}}=0.35$ \cite{mirhosseini2020superconducting}, $\mathrm{Y}_{\text{dia}}\sim1000$ GPa \cite{brisDiamondProperties}, $c_{\text{dia}}\sim 12000$ m/s \cite{brisDiamondProperties}, $\lambda_{\text{mech}}\sim 2.4~\mu$m (for 5 GHz). This gives the following:
\begin{equation}
    P_{\text{MW}} \approx 9.87\times10^{4}\epsilon_{\text{str}}^{2} \text{W}
\end{equation}

\begin{figure}
    \centering
    \includegraphics[width=0.5\linewidth]{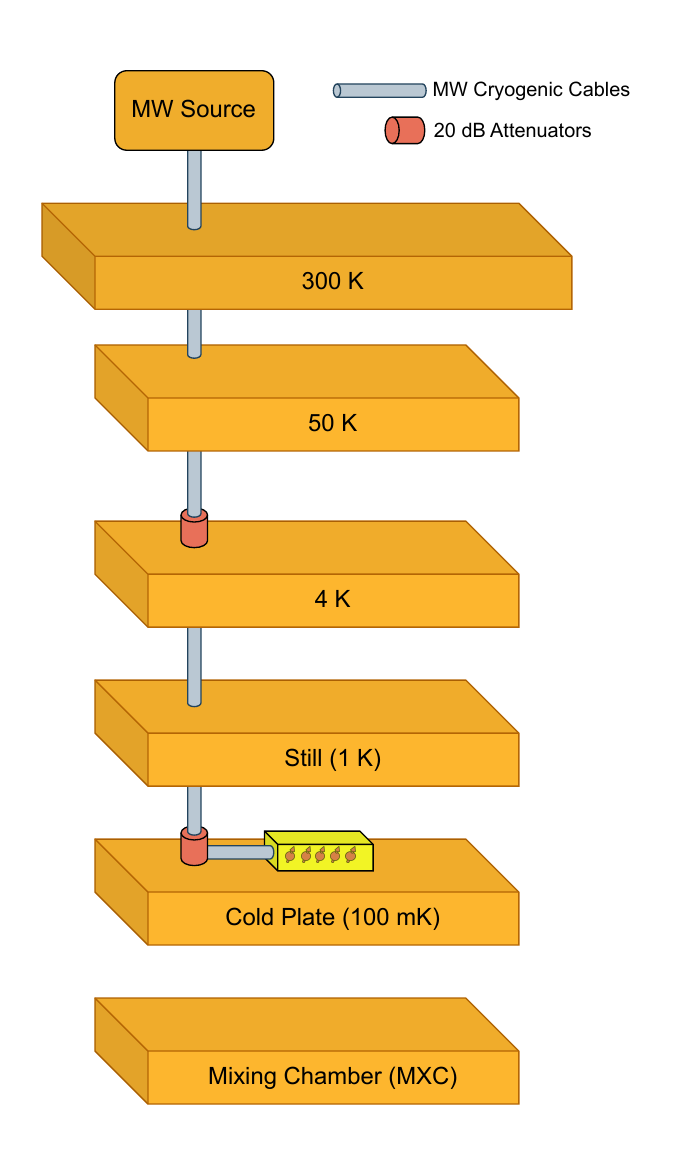}
    \caption{Dilution Fridge Configuration \cite{krinner2019engineering}}
    \label{fig:dil_fridge}
\end{figure}
For $\epsilon_{\text{str}}= 1.56\times 10^{-6}$ (See SI Sec. \ref{sec:straindriving}), we get $P_{\text{MW}}\approx0.24~\mu$W. We assume the stages of dilution fridge as in Figure \ref{fig:dil_fridge} (adapted from \cite{krinner2019engineering}), which has the following stages: Room temperature 300 K - 50 K - 4 K - 1 K - 100 mK (Cold Plate, CP) - MXC (Mixing Chamber) with 20 dB attenuators at 4 K, 100 mK, and MXC. The major sources of heat dissipation at the CP stage are:
\begin{itemize}
    \item Active Heat Load: Heat dissipated by the attenuators at the respective stages. Suppose $P^{\text{in/out}}_{\text{T}}$ represents the incoming or outgoing microwave power at stage with temperature T. Further, suppose $L_{j},~\alpha_{j}$ correspond to the length and attenuation per unit length of the cable reaching the stage $j+1$. Here $j$ varies from 1 to 5, and is the index for the set: \{300K, 50K, 4K, Still, CP, MXC\}. We assume that cables connecting all the stages are identical in material, i.e. $\alpha_{\text{T}} = \alpha_{\text{th}}$ Then we have following relations:
    \begin{subequations}
    \begin{align}
    P^{\text{in/out}}_{\text{300K}} &= P_{\text{source}}\\
    P^{\text{out}}_{\text{50K}} &= P^{\text{in}}_{\text{50K}} = P^{\text{out}}_{\text{300K}}10^{-\alpha_{\text{th}}L_{1}}\\
    P^{\text{in}}_{\text{4K}} &= P^{\text{out}}_{\text{50K}}10^{-\alpha_{\text{th}}L_{2}}\\
    P^{\text{out}}_{\text{4K}} &= P^{\text{in}}_{\text{4K}}/100\\
    P^{\text{out}}_{\text{1K}} &= P^{\text{in}}_{\text{1K}} = P^{\text{out}}_{\text{4K}}10^{-\alpha_{\text{th}}L_{3}}\\
     P^{\text{in}}_{\text{CP}} &= P^{\text{out}}_{\text{1K}}10^{-\alpha_{\text{th}}L_{4}}\\
     P^{\text{out}}_{\text{CP}} &= P^{\text{in}}_{\text{CP}}/100
    \end{align}
    \end{subequations}

This gives the following:
\begin{equation}
    P^{\text{out}}_{\text{CP}} = 10^{-(\alpha_{\text{th}}(L_{1}+L_{2}+L_{3}+L_{4})+4)}P_{\text{source}}
\end{equation}
We take the following values: $\alpha_{\text{th}}=2$ dB/m \cite{krinner2019engineering}, $L_{j}=\{200, 290, 250, 170, 140\}$ mm \cite{krinner2019engineering}, which gives $ P^{\text{out}}_{\text{CP}}\approx6.58\times 10^{-5} P_{\text{source}}$. Thus, in order to have $P^{\text{out}}_{\text{CP}}\approx 0.24~\mu$W, we need $P_{\text{source}}\approx 3.65$ mW. Thus, the active heat load ($h_{\text{act}}$) at the CP stage is given by:
\begin{subequations}
    \begin{align}
      h_{\text{act}} &= P^{\text{out}}_{1K} - P^{\text{out}}_{CP} = 10^{-(\alpha_{\text{th}}(L_{1}+L_{2}+L_{3})+2)}P_{\text{source}} - P^{\text{out}}_{CP}\\
      h_{\text{act}} &\approx 25.6~\mu W
    \end{align}
\end{subequations}
\item Passive Heat Load: This is the heat load due to the finite thermal conductivity of the coaxial cables connecting stages at different temperature. Typical passive heat load for stainless steel coaxial cable at the CP is $h_{\text{pass}}\sim 0.5~\mu$W \cite{krinner2019engineering}.
\item Impedance Mismatch heat load: This occurs due to inefficiency in the MW to phonon transduction and other impedance mismatch reflections that occur along the transmission line, which can be given by, $h_{\text{imp}}\sim (1-\eta_{\text{tr}})P^{\text{out}}_{\text{CP}} = 0.156~\mu$W. 

\item Sample Losses: This occurs due to the mechanical dissipation within the sample, which can be given by $h_{\text{sample}}=\gamma_{\text{th}}\eta_{\text{tr}}P_{\text{CP}}^{\text{out}} < 0.084~\mu$W. We assume a $\gamma_{\text{th}}=10^{-3}$ implying a modest quality factor of $\sim10^{3}$ 
\end{itemize}
Thus, total heat-load $h_{\text{tot}}$ is given by:
\begin{subequations}
    \begin{align}
      h_{\text{tot}} &= h_{\text{act}} + h_{\text{pass}} + h_{\text{imp}} + h_{\text{sample}}\\
  h_{\text{tot}} &\approx 25.6~\mu W + 0.5~\mu W + 0.156~\mu W + 0.084~\mu W = 26.34~\mu W
    \end{align}
\end{subequations}
We compare the two approaches $\mathcal{A}$ and $\mathcal{B}$ as seen in Fig 9. Suppose both the sequences implement CPMG-N over a timescale $t_{\text{dds}}$, then inter-pulse duration for the two approaches are given by:
\begin{subequations}
    \begin{align}
        t_{\text{ip,A}} &= \frac{t_{\text{dds}}}{N} - t_{\pi,\text{A}}\\
        t_{\text{ip,B}} &= \frac{t_{\text{dds}}}{N} - m ~t_{\pi,\text{B}}
    \end{align}
\end{subequations}
Plugging in numbers from our simulations, $t_{\text{dds}}=5$ ms, $t_{\pi,\text{A}}\approx150$ ns, $t_{\pi,\text{B}}\approx10$ ns, $N=\{2, 4, 8, 16\}$, this gives 0.31 ms $\le t_{\text{ip,A}} \le$ 2.5 ms, and (0.31 - $m\cdot10^{-5}$) ms $\le t_{\text{ip,B}} \le$ (2.5 - $m\cdot10^{-5}$) ms. The thermalization timescale for the sample is $\tau_{\text{th},\text{SiV}}\approx2-10~\mu$s \cite{nguyen2019integrated}, whereas the same for the cold-plate stage is of the order of $\tau_{\text{th,CP}}\approx100$ s. For sequence $\mathcal{A}$, $t_{\text{ip,A}}>>\tau_{\text{th,SiV}}$, whereas for sequence $\mathcal{B}$ the inequality: $t_{\text{ip,B}}>>\tau_{\text{th,SiV}}$ holds as long as $m<<3\cdot10^{4}$. Since, for our simulations, we consider an ensemble with only 121 qubits, we can assume that both the inequalities hold. This means that by the time subsequent pulse arrives, the SiV$^-$ sample already thermalizes with the cold plate stage, thus we can assume that the temperature of SiV$^-$ is the same as that of the cold plate, i.e. $\Theta_{\text{SiV}}$ $\approx$ $\Theta_{\text{CP}}$. Further, since $\tau_{\text{th,CP}}>>t_{\text{dds}}$, the effective heat-load observed by the cold-plate stage needs to be corrected by the duty-cycle (dc) of the pulse-sequences, given by:
\begin{subequations}
    \begin{align}
        \text{dc}_{A} = \frac{t_{\pi,\text{A}}}{t_{\text{dds}}}N \approx 3\cdot10^{-5}N\\
        \text{dc}_{B} = \frac{t_{\pi,\text{B}}}{t_{\text{dds}}}mN \approx 2\cdot10^{-6}mN
    \end{align}
\end{subequations}

\begin{figure}
    \centering
    \includegraphics[width=0.7\linewidth]{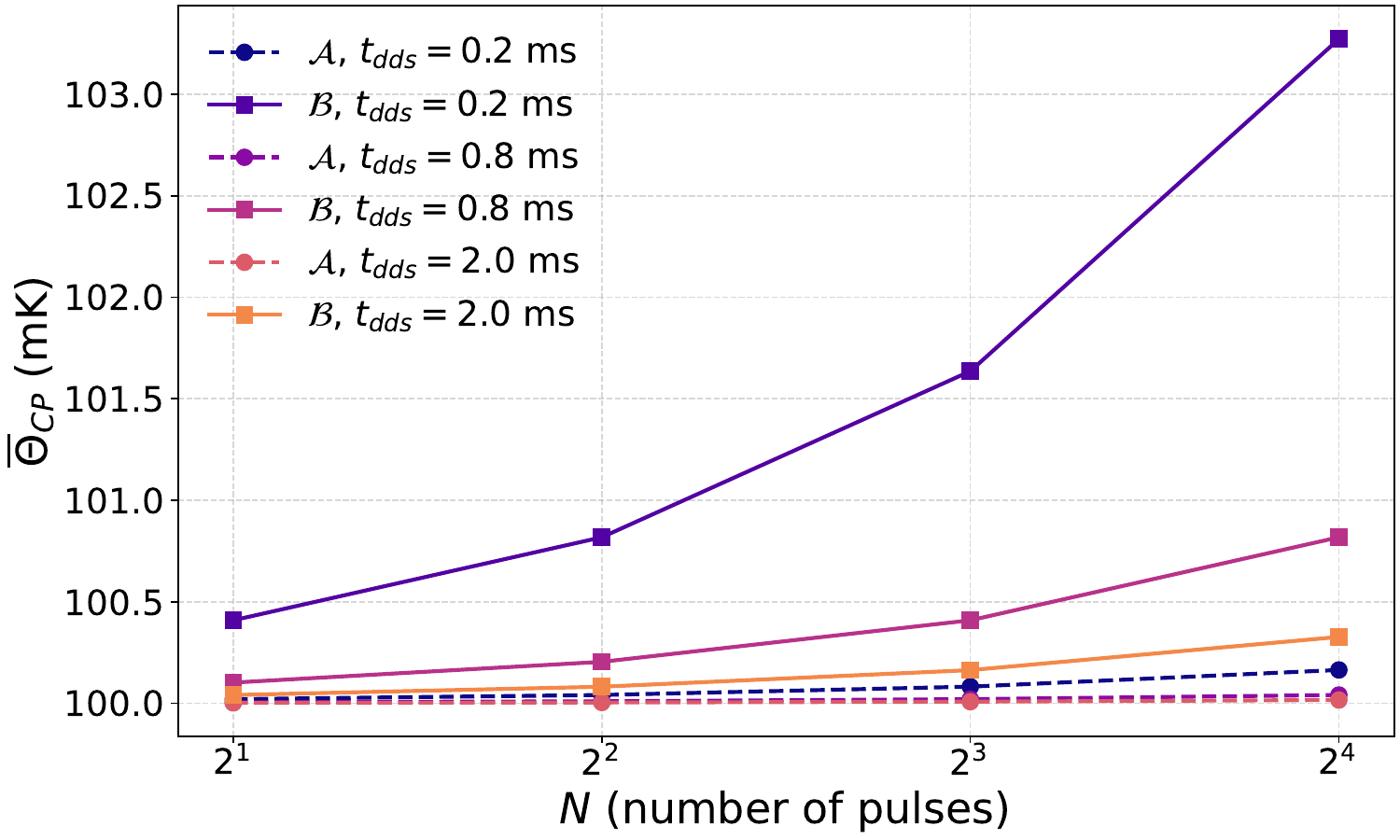}
    \caption{Variation of average cold-plate stage temperature $\overline{\Theta}_{\text{CP}}$ with the number of pulses of CPMG sequence.}
    \label{fig:CPtemp}
\end{figure}

From, Eq. (80b), we get that the total heat load for unit pulse $\mathcal{A}$ driven at $\Omega_{\text{rabi,A}}=200$ M-rad/s, is $h_{A}\approx26.34~\mu$W. Since, pulse sequence $\mathcal{B}$ is an unoptimized pulse sequence, we get: $\Omega_{\text{rabi,B}}=\pi/t_{\pi,\text{B}}\approx 314.16$ M-rad/s. Since $P\propto\Omega^{2}$, we estimate that $h_{B}\approx65~\mu$W. After incorporating the duty cycle we get the following effective heat-loads for sequence $\mathcal{A}_{N}$ and $\mathcal{B}_{N}$ implementing a $N$-pulse CPMG:
\begin{subequations}
    \begin{align}
        \Tilde{h}_{\text{A}}(N) &= \text{dc}_{A}h_{A} \approx (0.79 N)~\text{nW}\\
        \Tilde{h}_{\text{B}}(N) &= \text{dc}_{B}h_{B} \approx (0.13 mN)~\text{nW}
    \end{align}
\end{subequations}
Given the effective heat-loads at the cold-plate, we use the calibration data from this reference \cite{krinner2019engineering}, to get an estimate of approximate rise in the temperature of the cold-plate stage. From this source \cite{krinner2019engineering}, we see that for heat load under 50 $\mu$W, the relative rise in cold-plate temperature is roughly linear, and is $\approx5.2\cdot10^{-3}/(\mu\text{W})$. Thus, the relative rise $\delta \overline{\Theta}/\overline{\Theta}_{\text{CP}}$ in average cold-plate temperature due to $\mathcal{A}_{N}$ and $\mathcal{B}_{N}$ is given by:
\begin{subequations}
    \begin{align}
        \frac{\delta \overline{\Theta}_{A}(N)} {\overline{\Theta}_{\text{CP}}} &\approx 4.11\cdot10^{-6} N\\
          \frac{\delta \overline{\Theta} _{B}(N)} {\overline{\Theta}_{\text{CP}}} &\approx 0.68\cdot10^{-6} m
        N
    \end{align}
\end{subequations}
Here, the overline corresponds to the ensemble average, over different cycles of the entanglement. Figure \ref{fig:CPtemp} plots the resulting average cold-plate temperature versus the number of pulses $N$ for multiple $t_{\text{dds}}$.

\subsubsection{Thermal Environment of Silicon Vacancy}
From the previous section, we estimated the heat-load in the sample is roughly $h_{\text{sample}}\approx\gamma_{\text{th}}0.084~\mu$W. We assume that the dynamic temperature of SiV$^{-}$ due to pulse incident at time $t_{0}$ is given by \cite{nguyen2019integrated}:
\begin{equation}
    \Theta_{\text{SiV}}(t,t_{0   }) = \Theta_{\text{CP}} + P_{\text{th}} (\text{e}^{-(t-t_{0})/\tau_{\text{th,SiV}}} - \text{e}^{-9(t-t_{0})/\tau_{\text{th,SiV}}})~\text{ReLU}(t-t_{0})
\end{equation}
Here, $P_{\text{th}}$ is a normalization constant, $\text{ReLU}$ is the rectified linear unit function. In the above expression, first term is the baseline temperature of the cold-plate, second term arises due to the slow thermalization of sample with the bath, and third term is due to the fast heating process due to the active heat-load on the sample. Suppose the volume of the diamond chip is $V_{\text{chip}}$. Since the Debye temperature of diamond is $\Theta_{\text{deb}}\approx 2230$ K \cite{mukherjee1967debye} 
and the operating temperature is within 1K, we can assume that the heat-capacity $C_{v}(\Theta)$ for diamond at temperature $\Theta$ is given by the Debye model:
\begin{align} 
    C_{v}(\Theta) &= \frac{12\pi^{4}}{5}\bigg(\frac{\Theta}{\Theta_{\text{deb}}}\bigg)^{3}N_{\text{atoms}}\text{k}_{\text{B}}
\end{align}
Here $N_{\text{atoms}}$ is the number of atoms of diamond in a chip of volume $V_{\text{chip}}$, and $\text{k}_{\text{B}}$ is the Boltzmann constant. For a typical chip footprint of $5\times5\times0.5$ mm$^{3}$, we get $C_{v}(\Theta=0.1\text{K})=6.42\cdot10^{-13}$ J/K. Suppose, the time $t_{0}$ at which pulse is applied the temperature of the SiV is $\Theta_{\text{SiV}}(t_{0})$, giving a heat-capacity of $C_{v}(\Theta_{\text{SiV}}(t_{0}))\equiv C_{v}(t_{0})$. We use the following approximation for estimating normalization constant $P_{\text{th}}$. We assume that the maximum temperature rise achieved by the sample is due to the total heat absorbed by the sample from the MW line, meaning: $h_{\text{sample}}t_{\pi}=C_{v}(\Theta^{\text{max}}_{\text{SiV}}-\Theta_{\text{CP}})$. After solving the above, we get the following: 
\begin{equation}
    P_{\text{th}}(t_{0}) \approx \frac{9~h_{\text{sample}}t_{\pi}}{8~\beta~C_{v}(t_{0})}
\end{equation}
$\beta = 9^{-1/8}$ arises by solving the derivative of Eq.(85). This gives the following expression for $\Theta_{\text{SiV}}$ for a single pulse:
\begin{equation}
    \Theta_{\text{SiV}}(t, t_{0}) = \Theta_{\text{CP}} + \frac{9~h_{\text{sample}}t_{\pi}}{8~\beta~C_{v}(t_{0})} (\text{e}^{-(t-t_{0})/\tau_{\text{th,SiV}}} - \text{e}^{-9(t-t_{0})/\tau_{\text{th,SiV}}})~\text{ReLU}(t-t_{0})
\end{equation}
Now, if instead of a single pulse, it is a pulse train incoming at integer multiples of $t_{0}$ i.e. \{$jt_{0}$\}, where $j$ is a natural number, then the temperature is given by:
\begin{equation}
    \Theta_{\text{SiV}}(t) = \Theta_{\text{CP}} + \sum_{j}\bigg(\Theta_{\text{SiV}}(t,jt_{0})-\Theta_{\text{CP}}\bigg)
\end{equation}

\begin{figure}
    \centering
    \includegraphics[width= \linewidth]{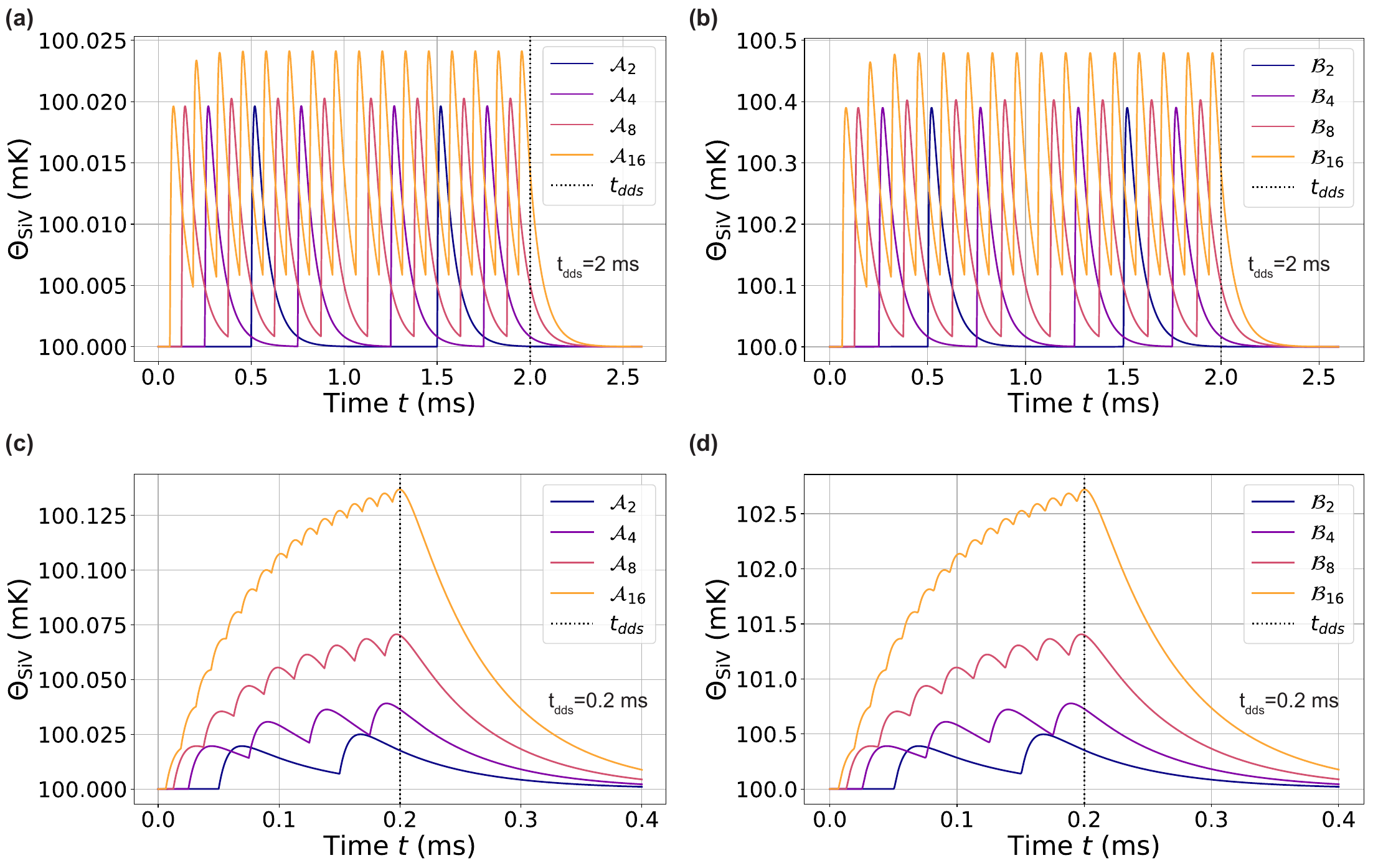}
    \caption{\textbf{Variation of $\Theta_{\text{SiV}}$ with time.} Vertical dotted line shows the $t_{\text{dds}}$ mark where the dynamical decoupling sequence stops. \textbf{(a-b)} For large values of $t_{\text{dds}}=2$ ms, the incoming pulses are separated far enough for SiV to thermalize, that's why the temperature stays bounded. Sequence $\mathcal{B}$ leads to higher temperatures than $\mathcal{A}$ due to larger heat-load. \textbf{(c-d)} When the $t_{\text{dds}}$ is reduced to 0.2 ms, pulses are coming at a rate faster than the thermalization rate of SiV which leads to unbounded rise in average temperature with time until $t<t_{\text{dds}}$, after which the decoupling sequence ends and SiV starts to thermalize again. In this case also sequence $\mathcal{B}$ leads to higher temperatures than $\mathcal{A}$ due to larger heat-load.}
    \label{fig:sivtemperature}
\end{figure}

For sequence $\mathcal{A}_{N}$, pulse incoming times are $t_{j,A}=\{ \frac{(2j-1)t_{\text{dds}}}{2N} \}$, where $1\le j \le N$. For sequence $\mathcal{B}_{N}(m)$, the incoming times are $t_{jk,B}=\{ \frac{(2j-1)t_{\text{dds}}}{2N} + (k-1)t_{\pi,B}\}$, where $1\le j \le N$ and $1\le k \le m$. Thus, we get following SiV$^-$ temperatures for implementing both sequences:
\begin{subequations}
\begin{align}
    \Theta_{\text{SiV}}(t) &= \overline{\Theta}_{\text{CP}} + \sum^{N}_{j=1}\bigg(\Theta_{\text{SiV}}(t,t_{j,A})-\overline{\Theta}_{\text{CP}} \bigg)\hspace{35pt}: \text{For}~\mathcal{A}_{N}\\
    \Theta_{\text{SiV}}(t) &= \overline{\Theta}_{\text{CP}} + \sum^{N}_{j=1}\sum^{m}_{k=1}\bigg(\Theta_{\text{SiV}}(t,t_{jk,B}) - \overline{\Theta}_{\text{CP}}\bigg)\hspace{10pt}: \text{For}~\mathcal{B}_{N}(m)
\end{align}    
\end{subequations}

 Here, $\overline{\Theta}_{\text{CP}}$ takes into account the slow temperature rise in the cold-plate as in Eq.(84). In Fig. \ref{fig:sivtemperature}, we plot fast rise in $\Theta_{\text{SiV}}$ for $\mathcal{A}_{N}$ and $\mathcal{B}_{N}$, for different values of $N$, where we take $\overline{\Theta}_{\text{CP}}=100$ mK, in order to decouple the contribution from slow and fast rise in temperature. When the measurement window is large ($t_{\text{dds}}=2 $ ms), the SiV$^-$ gets time to re-thermalize between pulses, hence the temperature has a saw-tooth trend, and once the decoupling stops, the temperature thermalizes to $T_{\text{CP}}$. On the other hand, for low measurement time windows ($t_{\text{dds}}=0.2$ ms), there is less time for the SiV$^-$ to thermalize before the next pulse arrives, leading to the accumulation of heat and a rise in temperature with the incoming pulses. We measure the $T_{2}$, $T_{1}$ times of the silicon vacancy at the end of the measurement window (i.e. $t=t_{\text{dds}}$). We refer to this paper \cite{Sukachev_2017}, which measured $T^{*}_{2}\approx10~\mu$s  and $T_{1}\gtrsim1s$ (for $\Theta_{\text{SiV}} = 100$ mK). Further, we assume a linear dependence of $1/T^{*}_{2}$ ($\sim3~\text{MHz/K}$) and $1/T_{1}$ ($\sim2.4~\text{MHz/K}$) w.r.t. $\Theta_{\text{SiV}}$ \cite{Jahnke_2015}. If we assume that $z_{\text{dds}}\approx z_{\phi}\approx z_{\text{inh}}$, then we can further assume that $1/T_{2}$ also has a linear dependence on $\Theta_{\text{SiV}}$. Since, $T_{1}>>T^{*}_{2}$, we assume that $1/T_{2}$ has a similar scaling as $1/T^{*}_{2}$, which gives the following:
\begin{subequations}
    \begin{align}
       \frac{1}{T_{2}(\Theta_{\text{SiV}})} \left[1/\text{s}\right] &= \frac{1}{T_{2}(0.1~\text{K})} \left[1/\text{s}\right] + (\Theta_{\text{SiV}}-0.1)\cdot 3\cdot10^{6} \left[1/\text{K-s}\right]\\
       T_{2}(\Theta_{\text{SiV}}) &= \frac{T_{2}(0.1~\text{K})}{1 + T_{2}(0.1~\text{K})\cdot(\Theta_{\text{SiV}}-0.1)\cdot3\cdot10^{6}}
    \end{align}
\end{subequations}
Thus, we get the following effective $T_{2}$ times for a measurement window $t_{\text{dds}}$ for both sequences:
\begin{subequations}
    \begin{align}
       T_{2}(\Theta_{\text{SiV}}(t_{\text{dds}},\mathcal{A}_{N})) &= \frac{T_{2}(0.1~\text{K},\mathcal{A}_{N})}{1 + T_{2}(0.1~\text{K},\mathcal{A}_{N})\cdot(\Theta_{\text{SiV}}(t_{\text{dds}},\mathcal{A}_{N})-0.1)\cdot3\cdot10^{6}}\\[1.5ex]
       T_{2}(\Theta_{\text{SiV}}(t_{\text{dds}},\mathcal{B}_{N})) &= \frac{T_{2}(0.1~\text{K},\mathcal{B}_{N})}{1 + T_{2}(0.1~\text{K},\mathcal{B}_{N})\cdot(\Theta_{\text{SiV}}(t_{\text{dds}},\mathcal{B}_{N})-0.1)\cdot3\cdot10^{6}}
    \end{align}
\end{subequations}
Based on the slopes, we can write a similar expression for $T_{1}$:
\begin{subequations}
    \begin{align}
       T_{1}(\Theta_{\text{SiV}}(t_{\text{dds}},\mathcal{A}_{N})) &= \frac{T_{1}(0.1~\text{K},\mathcal{A}_{N})}{1 + T_{1}(0.1~\text{K},\mathcal{A}_{N})\cdot(\Theta_{\text{SiV}}(t_{\text{dds}},\mathcal{A}_{N})-0.1)\cdot2.4\cdot10^{6}}\\[1.5ex]
       T_{1}(\Theta_{\text{SiV}}(t_{\text{dds}},\mathcal{B}_{N})) &= \frac{T_{1}(0.1~\text{K},\mathcal{B}_{N})}{1 + T_{1}(0.1~\text{K},\mathcal{B}_{N})\cdot(\Theta_{\text{SiV}}(t_{\text{dds}},\mathcal{B}_{N})-0.1)\cdot2.4\cdot10^{6}}
    \end{align}
\end{subequations}
\newline
Since we now have a relation for the time-dependence of $T_{1}, T_{2}$, from Eq. (70), we can write the following:
\begin{equation}
    \epsilon_{jk}(\tau,N, T_{1}(\tau), T_{2j}(\tau), T_{2k}(\tau), z_{j,\text{dds}}, z_{k,\text{dds}} ) \equiv \epsilon_{jk}(\tau,N, z_{j,\text{dds}}, z_{k,\text{dds}})
\end{equation}
Since the entanglement process occurs in the time-window $t=(t_{\text{dds}}, t_{\text{dds}}+t_{\text{cmpl}})$, where $t_{\text{cmpl}}$ is the total time allocated for entanglement, we take the temporal average of Eq.(94), resulting in:

\begin{tcolorbox}[breakable, colback=white]
\begin{equation} 
    \overline{\epsilon_{jk}}(t_{\text{dds}}, t_{\text{cmpl}}, N, z_{j,\text{dds}}, z_{k,\text{dds}}) = \frac{1}{t_{\text{cmpl}}}\int^{t_{\text{dds}}+t_{\text{cmpl}}}_{t_{\text{dds}}}\epsilon_{jk}(\tau,N, z_{j,\text{dds}}, z_{k,\text{dds}})~\text{d}\tau
 \end{equation}
\end{tcolorbox}
We further change the number of qubits $m$ in sequence $\mathcal{B}(\tau,m)$ to see the impact on $\Theta_{\text{SiV}}$ and $T_{2}$ as seen in Fig.~\ref{fig:scaling}. As seen from the Fig.~\ref{fig:scaling}b, we observe that for larger value of $m$, the $T_{2}$ plot for $\mathcal{B}$ is completely submerged in the orange shaded region (signifying $T_{2}<t_{\text{dds}}$). This suggests that the pulse sequence stops to perform dynamical decoupling as we scale the number of qubits to be addressed.  

\begin{figure}
    \centering
    \includegraphics[width=0.8\linewidth]{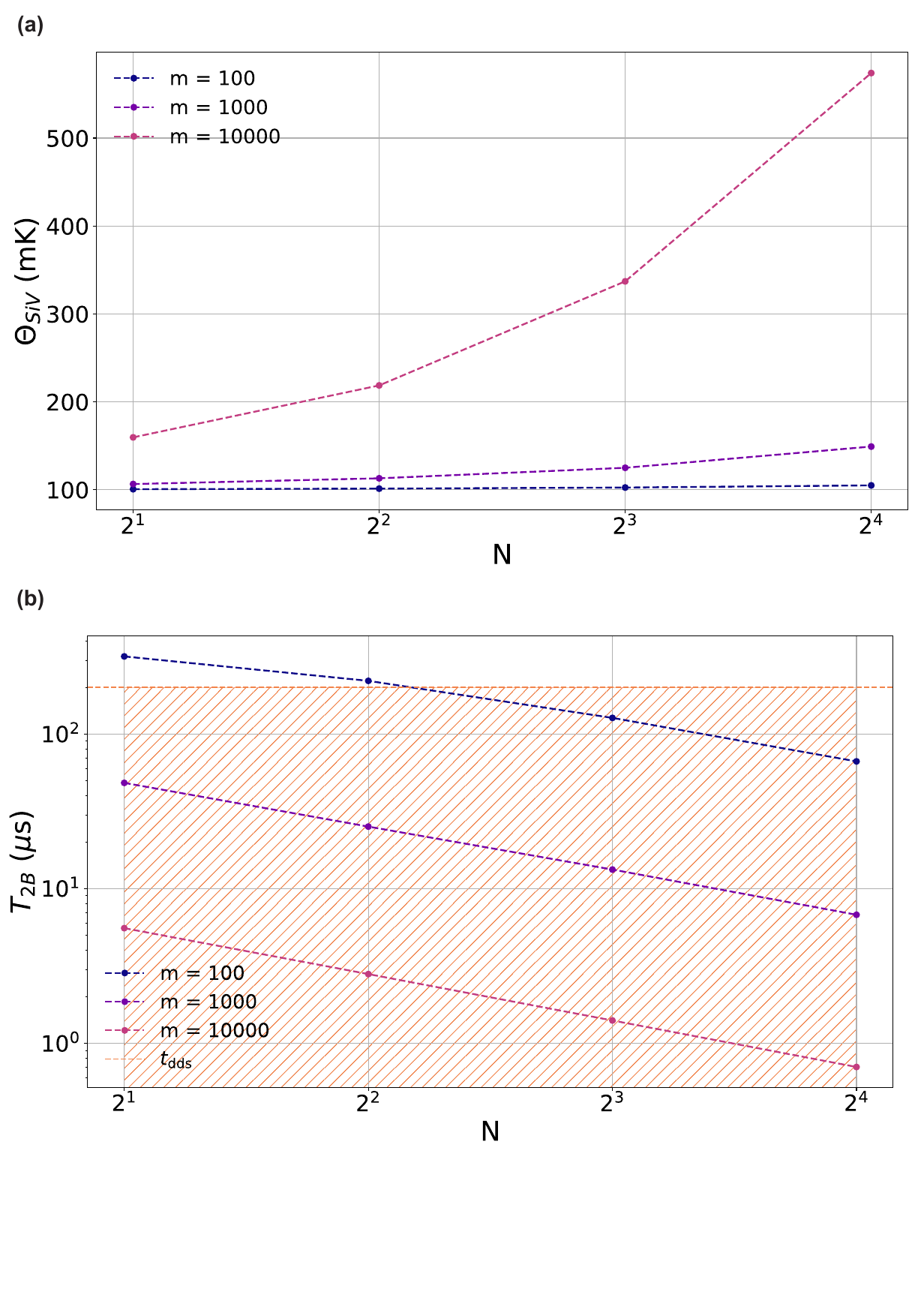}
    \caption{\textbf{Increasing the number of qubits}. \textbf{(a)} As the number of qubits $m$ increases from 100 to 10000, the heat-load due to pulse $\mathcal{B}$ increases, as seen from $\Theta_{\text{SiV}}$ due to $\mathcal{B}$ increasing by almost 4 times for $m=10000$ as $N$ changes from 2 to 16. \textbf{(b)} $T_{2}$ simulations for $\mathcal{B}$ for m = 1000 and 10000 are completely submerged in the orange shaded region which signifies the region $T_{2B} < t_{\text{dds}}$.
    }
    \label{fig:scaling}
\end{figure}

\newpage
\section{Derivation of the Fundamental Period $T$}

This derivation supports \textbf{Algorithm 2} from the main text by calculating the value of $\mathcal{J}$, which determines the minimum measurement window required to achieve the highest possible number of unique entanglement links. The calculation finds the fundamental period $T$ of the combined system, after which the pattern of entanglement attempts repeats. The value $\mathcal{J}$ is then determined from this period, setting the necessary number of attempts, $j_{\text{max}}$, to ensure all possible qubit pairs are brought into temporal coincidence at least once.\\

\textit{Setup.}

Device A has a sawtooth pattern of period $2T_a$:
\begin{align}
  \forall\,k\in\mathbb{Z}
  :\quad A(k + 2T_a) \;=\; A(k)
\end{align}

Device B has a sawtooth pattern of period $2T_b$ and runs at speed $d = \tfrac{p}{q}$, with $\gcd(p,q) = 1$:
\begin{gather}
\begin{cases}
    \forall\,k \in \mathbb{Z}:\; B(k + 2T_b) \;=\; B(k)\\
    \forall\,k\in\mathbb{Z}:\; B\!\bigl(\lfloor d\,k\rfloor \bigr) \;=\;
    B\!\Bigl(\left\lfloor \tfrac{p}{q}\,k \right\rfloor\Bigr) 
\end{cases}
\end{gather}

\textit{Definition of $T$.}

A positive integer $T$ is called a \emph{fundamental period} if
\begin{align}
   \forall\,k \in \mathbb{Z}
   :\quad
   A(k+T) = A(k)
   \;\;\;\wedge\;\;\;
   B\!\Bigl(\left\lfloor \tfrac{p}{q}(k+T)\right\rfloor\Bigr)
   \;=\;
   B\!\Bigl(\left\lfloor \tfrac{p}{q}\,k\right\rfloor\Bigr)
   \label{eq:fundamentalperiod}
\end{align}

We seek the minimal such $T$.

\textit{Condition for $A$.}

Since $A$ has period $2T_A$, we require
\begin{align}
   \forall\,k\in \mathbb{Z}
   :\quad
   A(k+T) = A(k)
   \quad\Longrightarrow\quad
   2T_a \;\mid\; T
\end{align}

\textit{Condition for $B$.}

We also require
\begin{align}
  \forall\,k \in \mathbb{Z}
  :\quad
  B\!\Bigl(\left\lfloor \tfrac{p}{q}(k+T)\right\rfloor\Bigr)
  \;=\;
  B\!\Bigl(\left\lfloor \tfrac{p}{q}\,k\right\rfloor\Bigr)
\end{align}

Since $B(\cdot)$ is $2T_b$--periodic in its integer input, it suffices that:

\begin{align}
    \forall\,k \in \mathbb{Z}: \quad
    \left\lfloor \tfrac{p}{q}(k+T)\right\rfloor 
    \;\equiv\;
    \left\lfloor \tfrac{p}{q}\,k\right\rfloor
    \pmod{2T_b}
\end{align}

A sufficient (and necessary) condition is:
\begin{align}
  \exists\,z \in \mathbb{Z}
  :\quad
  \frac{p}{q}\,T 
  \;=\;
  2T_b\,z.
\end{align}

\textit{Combining both conditions.}

$T$ must satisfy:
\begin{align}
   2T_a \;\mid\; T
   \quad\wedge\quad
   \exists\,z \in \mathbb{Z}
   :\;
   \frac{p\,T}{q} = 2T_b\,z
\end{align}

Equivalently, $2T_b\,q \,\mid\, p\,T$. Let $\gamma = \gcd\!\bigl(p,2T_b\,q\bigr)$, and write $p = \gamma\,p'$, $2T_b\,q = \gamma\,w$, with $\gcd(p',w)=1$. Then $2T_b\,q \mid p\,T$ is equivalent to $w \mid p'\,T$. Hence $T$ must be a multiple of $w$, and simultaneously a multiple of $2T_a$. The minimal positive $T$ satisfying both is:
\begin{align}
    T 
    \;=\;
    \mathrm{lcm}\!\Bigl(
      2T_a,
      \;\frac{2T_b\,q}{\gcd(p,\,2T_b\,q)}
    \Bigr).
\end{align}

This $T$ is the fundamental period, ensuring the condition of Eq. \ref{eq:fundamentalperiod}.



\section{Quantum Sensing}

Color centers in diamond have been proven a promising platform for quantum sensing applications due to their remarkable sensitivity to external perturbations, such as magnetic and electric fields, temperature, and strain. Particularly, the negatively charged nitrogen vacancy (NV$^-$) center in diamond has been interesting for magnetometry due to its long spin coherence times at room temperature. One of the core concepts of this paper, resonantly applying a global control pulse simultaneously performing dynamical decoupling on an ensemble of color centers, can be easily extended from group-IV color centers to the NV$^-$ center. While the SAFE-GRAPE algorithm and dynamical decoupling sequence remain the same, the implementation of the control pulse is different. Instead of using strain driving, the control pulse can be implemented by microwave signals via an electromagnetic resonator.

By using an ensemble of $N_q$ color centers instead of a single one, the magnetometer's sensitivity can be enhanced by a factor $\sqrt{N_q}$. Current state-of-the-art ensemble-NV$^-$ magnetometers exhibit sensitivities down to the pT/$\sqrt{\text{Hz}}$ range. Consider the CPMG measurement sequence as our AC magnetometry protocol. The sensitivity $\eta$ in the presence of this dynamical decoupling is then given by \cite{RevModPhys.92.015004}: 

\begin{equation}
    \eta \approx \underbrace{\frac{\pi}{2}\frac{\hbar}{\Delta m_s g_e \mu_B} \frac{1}{\sqrt{N_q \tau}}}_{\text{spin projection limit}} \underbrace{\frac{1}{e^{-(\tau/T_2)^p}}}_{\text{spin decoherence}} \kappa_{\text{readout}} \kappa_{\text{duty}}
\end{equation}

Here $\Delta m_s=1$ as we use the effective $S=\frac{1}{2}$ NV$^-$ subspace, $g_e=2.003$ is the NV$^-$ electronic $g$ factor, $\mu_B$ is the Bohr magneton, $\tau$ is the interrogation time per measurement, $T_2$ is the coherence time and the stretched exponential parameter $p$ depends on the type of dephasing. There are two additional corrections: $\kappa_{\text{readout}}=\frac{1}{\mathcal{F}_{\text{readout}}}$ deals with the effect of imperfect readout fidelity $\mathcal{F}_{\text{readout}}$ and $\kappa_{\text{duty}}=\sqrt{\frac{t_I+\tau+t_R}{\tau}}$ takes into account the duty cycle, i.e. the fraction of interrogation time $\tau$ w.r.t. the total cycle time which also includes the initialization time $t_I$ and readout time $t_R$ of a measurement. 

Since $\eta \propto \frac{1}{\sqrt{N_q \tau}}$ for $\tau \lesssim T_2$, the sensitivity can be drastically improved by using the SAFE-GRAPE based composite pulse sequence for dynamical decoupling. This allows to achieve large $T_2$ for a large amount $N_q$ of color centers, which in turn lets you select a larger interrogation time $\tau$.

\endgroup

\end{document}